\documentclass{jfm}

\usepackage{xcolor}
\usepackage{graphicx}
\usepackage{newtxtext}
\usepackage{newtxmath}
\usepackage{natbib}
\usepackage[final]{changes}
\usepackage{hyperref}
\hypersetup{
    colorlinks = true,
    urlcolor   = blue,
    citecolor  = black,
}

\newcommand\Bon{\mbox{\textit{Bo}}}
\newcommand\Web{\mbox{\textit{We}}}
\newcommand\Kap{\mbox{\textit{Ka}}}
\newcommand{\RomanNumeralCaps}[1]
\linenumbers
\usepackage{setspace}

\title{Modeling and dynamics of axisymmetric thin liquid film flow along a conical surface
}

\author{Longmin Tang\aff{1} and
  Guangzhao Zhou\aff{1, 2}\corresp{\email{zgz@ucas.ac.cn}}}

\affiliation{\aff{1}School of Engineering Science, University of Chinese Academy of Sciences, Beijing 101408, China
\aff{2}State Key Laboratory of Nonlinear Mechanics, IMECH \& UCAS,
Beijing 100190, China}

\begin{document}
\maketitle

\begin{abstract}
This study focuses on the modeling and dynamics of gravity-driven, axisymmetric thin liquid film flow along a conical surface.
Spatial linear stability analysis is performed on the basis of a Benney-type equation derived for the present configuration.
In particular, streamwise curvature of the free surface is found to exert a crucial influence on the stability threshold.
For simulations of surface waves, a second-order low-dimensional model is developed under the long-wave assumption, achieving accuracy comparable to direct numerical simulations at far lower cost.
With this model, the characteristics of both linear and nonlinear waves are examined.
A key difference from the flow over a flat plate is the dependence of wave dynamics on radial distance from the cone apex.
At relatively high flow rates, a transition from solitary to sinusoidal waves is observed, with the transition position correlating closely with the linear stability threshold.
Within the parameter range investigated, quantitative results of the conical film flow are almost identical to those in the flat-plate case when local parameters are substituted, indicating that inertial effects of the conical geometry are negligible.
The models and findings presented in this paper may aid the design and optimization of industrial processes such as film coating and liquid-film-based heat and mass transfer on conical surfaces. 
\end{abstract}

\section{Introduction} \label{sec:intro}
Thin liquid film flows along solid surfaces are ubiquitous in daily life.
They are also widely employed in industrial heat and mass transfer applications because of their large surface-to-volume ratios.
Examples include gas absorption \citep{Hu2014CES}, desalination \citep{Dai2022Falling}, and distillation \citep{2005Valluri}.
Since the pioneering experimental work of \citet{kapitza1948wave} and \citet{kapitsa1949wave}, the study of liquid film flow has remained an active area of both fundamental and applied research, attracting sustained attention for decades.

The physical understanding and mathematical description of gravity-driven liquid film flow over a flat plate have been extensively studied~\citep{1994chang, chang2002complex, 2004Weinstein, 2009matar, 2011kalliadasis, bandi2018reduced}.
For a given volumetric flow rate per unit width, $\tilde Q_\mathrm{2D}$, the thickness of an undisturbed film, known as the Nusselt film thickness \citep{nusselt1916oberflachenkondensation}, is
\begin{equation}
\tilde{h}_{N\mathrm{2D}} = \left( \frac{3 \nu \tilde Q_\mathrm{2D}}{g \sin\tilde \varphi} \right)^{1/3},
\label{eq:NusseltThickness}
\end{equation}
where $\nu$ is the kinematic viscosity of the liquid, $g$ is the gravitational acceleration, and $\tilde \varphi$ is the plate's inclination angle relative to the horizontal plane.
Linear stability analysis shows that when the Reynolds number, defined as
\begin{equation}
\Rey_\mathrm{2D} = \frac{\tilde Q_\mathrm{2D}}{\nu},
\label{eq:defineRef}
\end{equation}
exceeds $5 \cot \left( \tilde \varphi\right) / 6$ \citep{benjamin1957wave, yih1963stability}, the film becomes unstable to low-wavenumber disturbances.
The cut-off wavenumber $\tilde k_{c\mathrm{2D}}$ below which the flow is unstable satisfies \citep[see, e.g.,][p. 196]{2011kalliadasis}
\begin{equation}
\tilde k_{c\mathrm{2D}} \tilde h_{N\mathrm{2D}} = \frac{1}{\sqrt{\Web}} \left( \frac{6}{5} \Rey_\mathrm{2D} - \cot \tilde \varphi \right)^{1/2},
\label{eq:flatPlateKC}
\end{equation}
where the Weber number defined in their context is $\Web = \sigma / (\rho g \tilde{h}_{N\mathrm{2D}}^2 \sin \tilde \varphi)$, which is essentially an inverse Bond number.
$\rho$ and $\sigma$ are density and surface tension coefficient of the liquid, respectively.
Eq.~\eqref{eq:flatPlateKC} shows that the linear stability is influenced by the combined effects of inertia, viscosity, and surface tension force.
Specifically, the surface tension plays a role of stabilization.

When the flow is unstable, small-amplitude disturbances on the free surface first grow according to linear theory and eventually saturate as finite-amplitude nonlinear waves.
If the dynamics is dominated by a single wavelength--e.g., when a monochromatic disturbance is imposed at the inlet--the free-surface profile is spatially periodic.
The resulting shape depends on the wavelength $\tilde \lambda$ \citep{liu1994solitary, 2002Malamataris}: short waves are nearly sinusoidal, whereas long waves form solitary humps preceded by small capillary ripples within each wave unit \citep{joo1991long,1994chang, chang2002complex, rohlfs2015phase, 2020zhouprf}.
Considering the critical role of the wavy structures in heat and mass transfer \citep{Miyara1999, sisoev2005absorption, 2019CHAR, 2019Dietze, 2020zhou}, much effort has been devoted to examine the dynamics of these surface waves both experimentally \citep[e.g.,][]{1985ALEKSEENKO, liu1993measurements, liu1994solitary, nosoko1996characteristics, meza2008modeling, 2009DIETZE, 2014Kofman} and numerically \citep[e.g.,][]{2002Malamataris, mudunuri2006solitary, 2014Dietze, 2014Chakraborty, denner2016self}.

In numerical simulations, since the streamwise dimension of the liquid film is usually much larger than the film thickness, with a moderate cell aspect ratio, a very large total number of mesh cells is required, making the computation expensive.
Two main strategies have been used to reduce the cost.
The first exploits periodicity: a single wave unit is simulated with periodic boundary conditions at both ends of it \citep[see, e.g.,][]{salamon1994traveling, 1996Ramaswamy, aktershev2010heat}.
The second relies on the separation of streamwise and cross-stream scales, which allows development of a variety of low-dimensional models \citep[see, e.g.,][]{nguyen2000modeling, ruyer2000improved, scheid2006wave, mudunuri2006solitary, samanta2023elliptic}.
In derivation of these models, a certain type of integration with respect to the cross-stream coordinate is performed through the film thickness.
Therefore, for two-dimensional liquid films, the flow field is explicitly dependent on the streamwise coordinate only.
The computational efficiency is significantly improved compared with direct numerical simulations (DNS) that require discretization in both directions \citep[see, e.g.,][]{2002Malamataris, gao2003numerical, Denner2018JFM, 2020zhou}.

\begin{figure}
\centering
\includegraphics[width=7cm]{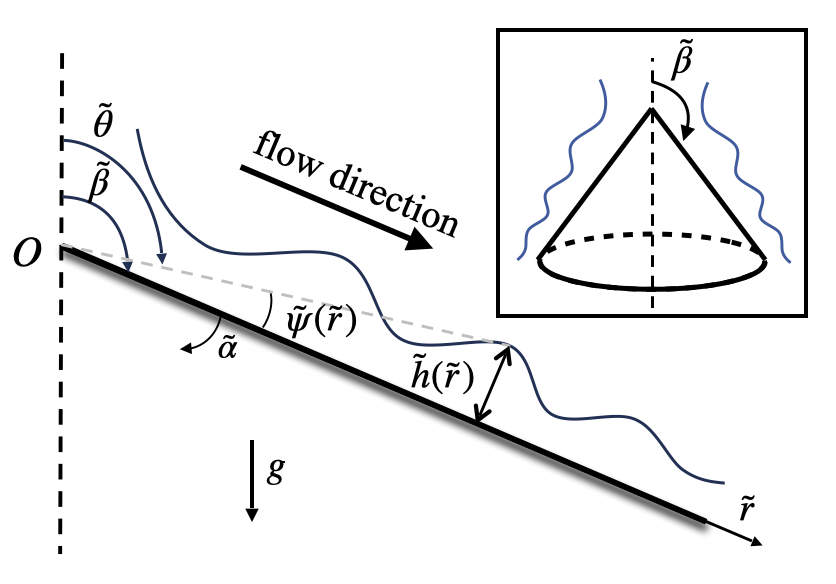}
\caption{Geometry and coordinate system.
}
\label{fig:geometry}
\end{figure}

A distinct subset of liquid film flow occurs over conical geometries, as when rainwater runs down a cone-shaped roof.
Research on this configuration is much less extensive than the well-studied flat-plate case.
In this paper, we consider a standard situation in which the cone axis is aligned with gravity, so that the flow along the surface is axisymmetric under a properly imposed upstream condition (figure~\ref{fig:geometry}).
Industrial examples of this type of flow include the Centritherm evaporator and spinning-cone distillation columns \citep{tanguy2015concentration, puglisi2022evaluation}, where the cones may be either rotating or stationary. 
These systems place high requirements for heat and mass transfer efficiency. 
As mentioned earlier, the transport processes are significantly influenced by surface wave dynamics.
Therefore, investigating the behavior of the liquid film flow on conical surfaces and the accompanying surface waves holds considerable importance.

The dynamics of liquid film flow on a cone differ fundamentally from those on a flat plate.
The key feature is the increase in circumference with the radial distance $\tilde r$ from the cone apex, which results in an $\tilde r$-dependent local volumetric flow rate per unit width
\begin{equation}
\tilde Q_\mathrm{2D}(\tilde{r}) = \frac{\tilde Q}{2\pi \tilde{r} \sin \tilde \beta},
\label{eq:Q2D-cone}
\end{equation}
where $\tilde Q$ is the total volumetric flow rate, $90^\circ < \tilde \beta < 180^\circ$ is the supplementary of the half-apex angle of the cone (figure~\ref{fig:geometry}).
A leading-order estimate of the local thickness $\tilde h_{N\ell}$ for a steady, gravity-driven liquid film flow can be obtained by substituting Eq.~\eqref{eq:Q2D-cone} into Eq.~\eqref{eq:NusseltThickness} (with $\tilde \varphi$ replaced by $\tilde \beta - \pi/2$), giving  \citep{1980Zollars, zhou2022hydraulic}
\begin{equation}
\tilde{h}_{N\ell}(\tilde{r}) = -\left( \frac{3 \nu \tilde Q}{2 \pi g \tilde{r} \sin \tilde \beta \cos \tilde \beta }  \right)^{1/3},
\label{eq:local_hN}
\end{equation}
i.e., in contrast to the Nusselt thickness that is independent of the streamwise coordinate, the undisturbed film thickness on a cone is a function of the radial distance following approximately $\tilde{h}_{N\ell} \propto \tilde{r}^{-1/3}$.

With Eqs.~\eqref{eq:defineRef} and \eqref{eq:Q2D-cone}, the local Reynolds number for the conical liquid film flow is expressed as
\begin{equation}
\Rey_\ell (\tilde{r}) = \frac{\tilde Q}{2\pi \nu \tilde{r} \sin\tilde\beta},
\label{eq:defRel}
\end{equation}
which is proportional to $\tilde{r}^{-1}$.
Since both the linear and nonlinear behaviors of liquid film flow have a strong dependence on the Reynolds number in the flat-plate situation~\citep[see, e.g.,][]{liu1993measurements, denner2016self, rohlfs2017hydrodynamic}, these characteristics are expected to vary significantly along a conical surface.
Indeed, by doing linear stability analysis, \citet{1980Zollars} noted that the phase speed and wavelength of surface waves on a cone decrease with $\tilde{r}$.
They also concluded that unstable disturbances eventually stabilize far from the apex.

This work centers on the dynamics of gravity-driven, axisymmetric liquid film flow on a stationary cone, and develops a model for its numerical simulation.
The paper is organized as follows.
Section~\ref{sec:gov-eq} gives the mathematical formulation, where the Navier-Stokes equations and boundary conditions are normalized and truncated.
Section~\ref{sec:linear} derives a Benney-type equation and revisits the linear stability problem, improving the results of~\citet{1980Zollars} to a second-order accuracy in the small parameter $\varepsilon$ (to be introduced in section~\ref{sec:gov-eq}).
Section~\ref{sec:model} presents a second-order low-dimensional model for efficient simulation of the flow dynamics, which is shown to yield more accurate results compared with the ``first-order model'' developed earlier by \citet{zhou2022hydraulic}.
Using this model, for the first time to our knowledge, the spatio-temporal evolution of surface waves on the conical liquid film, including their transition into linear waves, is examined in section~\ref{sec:nonlinear}.
Section~\ref{sec:funnel} briefly discusses flow on the upper surface of an inverted cone.
Further considerations on the long-wave expansion are provided in section~\ref{sec:furtherCons}.
Finally, conclusions are given in section~\ref{sec:conclusions}.

\section{Mathematical formulation} \label{sec:gov-eq}

As noted in section~\ref{sec:intro}, thin liquid film flows exhibit a clear scale separation between the streamwise and cross-stream directions.
Referred to as the long-wave approximation \citep{2011kalliadasis}, this property has been used to simplify the full Navier-Stokes equations for flow over a flat plate \citep{1966benney, ruyer2000improved, 2008ruyer, mudunuri2006solitary, kim2024low}.
In this section, we adopt the same strategy to establish the basic framework for subsequent model development.
The procedure closely follows \citet{zhou2022hydraulic}, but the resultant equations are carried to a higher order with respect to the small parameter to be introduced below.

A spherical coordinate system is employed with the polar axis aligned with the cone axis and the origin $O$ at the apex, see figure~\ref{fig:geometry}.
This coordinate system is different from that used in~\citet{1980Zollars}, who actually employ a cylindrical coordinate system with its radial axis rotated to align with the cone surface.
In the present coordinate system, the polar angle $\tilde{\theta}$ is measured from the upward direction.
The cone surface is expressed as $\tilde{\theta} = \tilde \beta$; and the radial coordinate $\tilde{r}$ represents the distance from the apex along the wall.
For convenience, we introduce the local coordinate $\tilde{\alpha} = \tilde{\theta} - \tilde \beta$.
The free surface is denoted by $\tilde{\alpha} = \tilde{\psi}$, where $\tilde{\psi}$ is a function of the radial distance $\tilde{r}$ and time $\tilde{t}$.
For the configuration shown in figure~\ref{fig:geometry}, $\tilde{\psi}$ is negative.
The liquid occupies the region $\tilde{\psi}(\tilde{r}, \tilde{t}) \le \tilde{\alpha} \le 0$.
The local film thickness $\tilde{h} (\tilde{r}, \tilde{t})$ can be approximated by $- \tilde{r} \tilde{\psi}(\tilde{r},\tilde{t})$.

Because the flow is axisymmetric, the Navier-Stokes equations in spherical coordinates can be readily simplified.
The continuity equation reads
\begin{equation} \label{eq:continuity}
	\frac{1}{\tilde{r}^2}\frac{\partial}{\partial \tilde{r}} \left(\tilde{r}^2 \tilde{u}\right) + \frac{1}{\tilde{r} \sin\tilde{\theta}} \frac{\partial}{\partial \tilde{\theta}} \left(\tilde{v} \sin\tilde{\theta}\right) = 0.
\end{equation}
The momentum equation in the radial direction is
\begin{equation} \label{eq:momentum-u}
    \begin{split}
        \frac{\partial \tilde{u}}{\partial \tilde{t}} & + \tilde{u} \frac{\partial \tilde{u}}{\partial \tilde{r}} + \frac{\tilde{v}}{\tilde{r}} \frac{\partial \tilde{u}}{\partial \tilde{\theta}} - \frac{\tilde{v}^2}{\tilde{r}} = - \frac{1}{\rho} \frac{\partial \tilde{p}}{\partial \tilde{r}} - g \cos\tilde{\theta} \\
	    & + \nu \left[\frac{1}{\tilde{r}} \frac{\partial^2}{\partial \tilde{r}^2} \left(\tilde{r} \tilde{u}\right) + \frac{1}{\tilde{r}^2} \frac{\partial^2 \tilde{u}}{\partial \tilde{\theta}^2} + \frac{\cot \tilde \theta}{\tilde{r}^2} \frac{\partial \tilde{u}}{\partial \tilde{\theta}} - \frac{2}{\tilde{r}^2} \frac{\partial \tilde{v}}{\partial \tilde{\theta}} - \frac{2 \tilde{u}}{\tilde{r}^2}  - \frac{2 \cot\tilde{\theta}}{\tilde{r}^2} \tilde{v}\right];
    \end{split}
\end{equation}
and that in the cross-stream direction is
\begin{equation} \label{eq:momentum-theta}
	\begin{split}
		&\frac{\partial \tilde{v}}{\partial \tilde{t}} + \tilde{u} \frac{\partial \tilde{v}}{\partial \tilde{r}} + \frac{\tilde{v}}{\tilde{r}} \frac{\partial \tilde{v}}{\partial \tilde{\theta}} + \frac{\tilde{u} \tilde{v}}{\tilde{r}} = - \frac{1}{\rho \tilde{r}} \frac{\partial \tilde{p}}{\partial \tilde{\theta}} + g \sin\tilde{\theta} \\
		& + \nu \left[\frac{1}{\tilde{r}} \frac{\partial^2}{\partial \tilde{r}^2} \left(\tilde{r} \tilde{v}\right) + \frac{1}{\tilde{r}^2} \frac{\partial^2 \tilde{v}}{\partial \tilde{\theta}^2} + \frac{\cot\tilde{\theta}}{\tilde{r}^2} \frac{\partial \tilde{v}}{\partial \tilde{\theta}} + \frac{2}{\tilde{r}^2} \frac{\partial \tilde{u}}{\partial \tilde{\theta}} - \frac{\tilde{v}}{\tilde{r}^2 \sin^2\tilde{\theta}} \right],
	\end{split}
\end{equation}
where $\tilde{u}$ and $\tilde{v}$ are the velocity components in the above two directions, respectively.
$\tilde{p}$ is pressure;
$g$ is gravitational acceleration;
$\rho$ and $\nu$ represent liquid density and kinematic viscosity, respectively.

On the wall of the cone where $\tilde \alpha = 0$, the no-slip condition requires
\begin{equation} \label{eq:bc-nonslip}
    \tilde{u} = \tilde{v} = 0.
\end{equation}
At the free surface of the liquid film, $\tilde{\alpha} = \tilde{\psi}$, the kinematic boundary condition is
\begin{equation} \label{eq:bc-kinematic}
	\frac{\partial \tilde{\psi}}{\partial \tilde{t}} + \tilde{u} \frac{\partial \tilde{\psi}}{\partial \tilde{r}} - \frac{\tilde{v}}{\tilde{r}} = 0\,;
\end{equation}
the tangential- and normal-stress balances read
\begin{equation} \label{eq:bc-tangential}
   \tilde{ \boldsymbol n}  \cdot \tilde{ \mathsfbi{T} } \cdot \tilde{ \boldsymbol  \tau }  = 0,
\end{equation}
and
\begin{equation} \label{eq:bc-normal}
\tilde {\boldsymbol n } \cdot \tilde{\mathsfbi{T}}  \cdot \tilde{\boldsymbol  n}  + \sigma \tilde{\mathcal{C}} = 0,
\end{equation}
where $\tilde{\mathsfbi{T}}$ is the stress tensor, $\tilde {\boldsymbol{n}}$ and $\tilde{\boldsymbol{\tau}}$ are the unit vectors normal and tangential to the free surface, respectively. 
$\sigma$ is the surface tension coefficient;
$\tilde{\mathcal C}$ is the free-surface curvature.
In writing Eqs.~\eqref{eq:bc-tangential} and \eqref{eq:bc-normal}, the dynamic effect of the gas phase has been neglected.

Multiplying Eq.~\eqref{eq:continuity} by $\tilde{r}^2 \sin\tilde{\theta}$ and integrating from $\tilde{\alpha} = 0$ to $\tilde{\alpha} = \tilde{\psi}$ gives
\begin{equation}
	\frac{\partial}{\partial \tilde{r}} \left(\int_{0}^{\tilde\psi} \tilde{r}^2 \tilde{u} \sin\tilde{\theta} d\tilde{\alpha}\right) - \tilde{r}^2 \sin\tilde \theta \left[\left.\tilde{u}\right|_{\tilde{\alpha} = \tilde{\psi}} \frac{\partial \tilde{\psi}}{\partial \tilde{r}} - \frac{1}{\tilde{r}} \left.\tilde{v}\right|_{\tilde{\alpha} = \tilde{\psi}}\right] = 0,
\label{eq:integrate}
\end{equation}
where the boundary condition Eq.~\eqref{eq:bc-nonslip} has been used.
The first term in Eq.~\eqref{eq:integrate} can be expressed with the volumetric flow rate $\tilde Q$ or $\tilde{q} \equiv \tilde Q/(2\pi)$, which, by definition, is
\begin{equation} \label{eq:q-definition}
	\tilde{q} = \frac{\tilde Q}{2\pi} =   \int^{0}_{\tilde \psi} \tilde{r}^2 \tilde{u} \sin\tilde{\theta} d\tilde{\alpha}.
\end{equation}
Invoking Eq.~\eqref{eq:bc-kinematic}, Eq.~\eqref{eq:integrate} can be rewritten as
\begin{equation} \label{eq:psi}
	\frac{\partial \tilde{\psi}}{\partial \tilde{t}} - \frac{1}{\tilde{r}^2 \sin(\tilde \beta + \tilde{\psi})} \frac{\partial \tilde{q}}{\partial \tilde{r}} = 0.
\end{equation}

Following \citet{zhou2022hydraulic}, we introduce two length scales, $H$ and $L$, to normalize the film thickness $\tilde{h}$ and radial coordinate $\tilde{r}$, respectively.
The ratio between these two scales is denoted by $\varepsilon$, which is expected to be small:
\begin{equation}
\varepsilon \equiv \frac{H}{L} \ll 1.
\end{equation}
It is noteworthy that $|\tilde \psi| \approx \tilde{h}/\tilde{r} \sim H/L = \varepsilon$.

Similarly, we denote the streamwise and cross-stream velocity scales by $U$ and $V$, respectively.
Inserting them into Eq.~\eqref{eq:continuity} and noting $\partial/\partial \tilde\theta = \partial/\partial \tilde \alpha \sim \varepsilon^{-1}$ gives $V/U \sim \varepsilon $.
The time scale is expressed as the combination $L/U$, indicating that the temporal variation occurs together with the liquid transportation in the streamwise direction.
For the gravity-driven flow considered currently, the gravitational acceleration is responsible for both the pressure distribution inside the liquid film and the streamwise flow velocity.
Therefore, the pressure is normalized by the hydrostatic value $\rho g H$.
The velocity scale is expressed as $U = g H^2/\nu$, the form of which corresponds to the Nusselt velocity profile~\citep{2008Kalliadasis} for a flat-plate film and, as will be seen, also well represents the order of magnitude of the radial velocity for an undisturbed liquid film flow on a cone.

With the above preparations, we normalize the governing equations \eqref{eq:continuity}-\eqref{eq:momentum-theta} by writing
\begin{equation} \label{eq:scaling}
	\tilde{r} = L r, \quad \tilde{\alpha} = \varepsilon \alpha,  \quad \tilde{u} = U u, \quad \tilde{v} = \varepsilon U v, \quad \tilde{t} = \frac{L t}{U}, \quad \tilde{p} = \rho g H p.
\end{equation}
The constant angle $\tilde \beta = \beta$ requires no re-scaling.

The results are
\begin{equation} \label{eq:nondim-continuity}
	\frac{\partial}{\partial r} \left(r^2 u\right) + \frac{\partial}{\partial \alpha} \left(r v\right) + \varepsilon r v \left(\cot\beta - \varepsilon \alpha \csc^2 \beta\right) = 0,
\end{equation}
\begin{equation} \label{eq:r-f}
	\begin{split}
		&\varepsilon \Rey \left(\frac{\partial u}{\partial t} + u \frac{\partial u}{\partial r} + \frac{v}{r} \frac{\partial u}{\partial \alpha}\right) = -\varepsilon \frac{\partial p}{\partial r} - \cos\beta + \varepsilon \alpha \sin\beta + \frac{1}{2} \varepsilon^2 \alpha^2 \cos\beta \\
          + & \varepsilon^2 \frac{1}{r^2} \frac{\partial}{\partial r} \left(r^2 \frac{\partial u}{\partial r}\right) + \frac{1}{r^2} \frac{\partial^2 u}{\partial \alpha^2} + \varepsilon \frac{\cot\beta}{r^2} \frac{\partial u}{\partial \alpha} - \varepsilon^2 \frac{\alpha}{r^2 \sin^2\beta} \frac{\partial u}{\partial \alpha} - 2 \frac{\varepsilon^2}{r^2}\left(\frac{\partial v}{\partial \alpha} + u\right),
	\end{split}
\end{equation}
and
\begin{equation} \label{eq:theta-f}
		\frac{1}{r} \frac{\partial p}{\partial \alpha} =  \alpha \varepsilon \cos\beta + \sin\beta + \varepsilon \left(\frac{1}{r^2} \frac{\partial^2 v}{\partial \alpha^2}  +  \frac{2}{r^2} \frac{\partial u}{\partial \alpha}\right),
\end{equation}
respectively.
In order to achieve a second-order accuracy, Eqs.~\eqref{eq:nondim-continuity} and \eqref{eq:r-f} are truncated at $O(\varepsilon^2)$.
Whereas it is sufficient to retain terms up to $O(\varepsilon)$ in Eq.~\eqref{eq:theta-f} in all subsequent derivations.
We have implicitly assumed $\sin \beta \sim O(1)$, which holds unless $\beta$ is very close to $0$ or $\pi$.
This restriction renders the equations inapplicable to limiting cases where the liquid films are on the inner or outer surface of a vertical round tube, which demands special treatments \citep[see, e.g.,][]{2008ruyer, 2020Dietze, 2024Dietze}.

The Reynolds number appearing in Eq.~\eqref{eq:r-f} is defined as $\Rey = UH/\nu = gH^3/\nu^2$, whose order of magnitude is assigned to be $O(1)$.
As discussed in \citet{2011kalliadasis}, this constraint can be relaxed.
The present derivations are expected to be valid for moderate Reynolds numbers with $\Rey \gtrsim O(1)$.

The boundary conditions \eqref{eq:bc-nonslip}-\eqref{eq:bc-normal} are normalized in the same manner.
Eqs.~\eqref{eq:bc-nonslip} and \eqref{eq:bc-kinematic} become
\begin{equation}
    u = v = 0,
    \label{eq:noSlipBC_dimless}
\end{equation}
and
\begin{equation}\label{eq:kineBC_dimless}
    \frac{\partial \psi}{\partial t} + u \frac{\partial \psi}{\partial r} - \frac{v}{r} = 0.
\end{equation}

The balance of tangential stress at the free surface $\alpha = \psi$, truncated at $O(\varepsilon^2)$, reads
\begin{equation} \label{eq:bc-tangential-nondim}
	\frac{\partial u}{\partial \alpha} = 2 \varepsilon^2 r^2 \frac{\partial \psi}{\partial r} \frac{\partial u}{\partial r} - \varepsilon^2 r \frac{\partial v}{\partial r} + \varepsilon^2 v - 2 \varepsilon^2 r \frac{\partial \psi}{\partial r} \frac{\partial v}{\partial \alpha} - 2 \varepsilon^2 r \frac{\partial \psi}{\partial r} u.
\end{equation}
The stress balance in the normal direction, accurate to $O(\varepsilon)$, which is sufficient for later use, is
\begin{equation} \label{bc:normal-nondim}
	p = 2 \varepsilon \left(\frac{1}{r} \frac{\partial v}{\partial \alpha} + \frac{u}{r}\right) - \left(\frac{\varepsilon}{\Bon}\right) \mathcal{C},
\end{equation}
where $ \Bon = \rho g H^2 / \sigma$ is the Bond number.

The normalized curvature of the free surface, given by
\begin{equation}
    \label{eq:curv}
	\mathcal C = \frac{\cot\beta}{r} + \varepsilon \frac{h \cot^2\beta}{r^2} + \varepsilon \frac{1}{r} \frac{\partial h}{\partial r} + \varepsilon \frac{\partial^2 h}{\partial r^2},
\end{equation}
is retained to $O (\varepsilon)$
(A detailed derivation of $\mathcal{C}$ is provided in Appendix~\ref{ap:curv}).
The first term in Eq.~\eqref{eq:curv} accounts for the lateral curvature induced by the conical geometry.
While the last three terms arise from streamwise fluctuations of the free surface. 

According to Eq.~\eqref{eq:theta-f} in which the highest order is $O(1)$, the order of $\Bon$ in Eq.~\eqref{bc:normal-nondim} cannot be higher than $O(\varepsilon)$.
Considering the surface tension forces are typically large for common liquids~\citep{2011kalliadasis}, we assign $\Bon \sim O(\varepsilon)$.
Keeping this in mind, in the following text wherever $\Bon$ appears, we enclose the combination $\varepsilon/\Bon$ in parentheses and treat it as a single entity of order 1.

Finally, it is noteworthy that the difference between $\tilde{\psi}$ and $- \tilde{h} / \tilde{r}$ is $\tilde{\psi} + \tilde{h} / \tilde{r} = \tilde{\psi} - \tan\tilde{\psi} = O(\tilde{\psi}^3) \sim O(\varepsilon^3)$, which is negligibly small.
Replacing $\tilde{\psi}$ with $- \tilde{h} / \tilde{r}$ in Eq.~\eqref{eq:psi} yields, in dimensionless form,
\begin{equation} \label{eq:h-nondim}
	\frac{\partial h}{\partial t} = - \frac{1}{r \sin\left(\beta - \varepsilon h/r\right)} \frac{\partial q}{\partial r},
\end{equation}
where $q = \tilde q  / (U H L)$.

\section{Spatial linear stability analysis} \label{sec:linear}
\subsection{Stability threshold} \label{sec:stability}
Unlike the flat-plate scenario, the liquid film flow on a conical surface is non-parallel even when the disturbances are absent.
This dependence of the base flow on the streamwise coordinate usually demands special consideration in linear stability analysis \citep{herbert1997parabolized,govindarajan1995stability, govindarajan1997low, bertolotti1997response}.
For the presently studied conical liquid film flow, \citet{1980Zollars} derived an asymptotic base-flow solution from the Navier-Stokes equations, and performed a spatial linear stability analysis with respect to axisymmetric disturbances.
However, their results only contained the effect of lateral curvature of the gas-liquid interface.
The streamwise curvature was not included due to its smaller order of magnitude (see Eq.~\eqref{eq:curv}) in their first-order model.

In this section, we revisit the spatial linear stability problem with a similar approach to that of \citet{1980Zollars}, extending it to second order and paying special attention to the influence of the streamwise curvature.
Instead of employing the stream function as is done in \citet{1980Zollars}, however, the present analysis is based on a Benney-type equation derived below.
For liquid film flow on a flat plate, the cut-off wavenumber obtained from the Benney equation~\citep{1966benney} coincides with that from the long-wave expansion of the Orr-Sommerfeld equation \citep[][section 5.1.5]{2011kalliadasis}.

We expand the variables $u(r, \alpha, t)$, $v(r, \alpha, t)$, and $p(r, \alpha, t)$ in powers of $\varepsilon$:
\begin{subequations}  \label{eq:expansion}
\begin{align}
   	u & = u^{(0)} + \varepsilon u^{(1)} + \varepsilon^2 u^{(2)} + O(\varepsilon^3),\label{eq:expansion_a} \\
	v & = v^{(0)} + \varepsilon v^{(1)} + \varepsilon^2 v^{(2)} + O(\varepsilon^3), \\
	p & = p^{(0)} + \varepsilon p^{(1)} + \varepsilon^2 p^{(2)} + O(\varepsilon^3).
\end{align}
\end{subequations}

Eqs.~\eqref{eq:expansion} are substituted into Eqs.~\eqref{eq:nondim-continuity}-\eqref{eq:theta-f}.
The terms with superscripts $(0)$, $(1)$, and $(2)$ can be solved sequentially.
Expressions of these terms are given in Eqs.~\eqref{eq:v0}-\eqref{eq:u2} in Appendix~\ref{ap:linear-results} for reference.
Specifically, 
\begin{equation} \label{eq:u0}
	u^{(0)} = -  r^2 \psi^2  \left(\frac{\alpha}{\psi} - \frac{1}{2} \frac{\alpha^2}{\psi^2}\right) \cos\beta.
\end{equation}
In dimensional form, $\tilde{u}^{(0)} = - g \tilde{h}^2 \cos\tilde{\beta}/(2\nu) $ at the free surface $\tilde \alpha=\tilde \psi$.
This justifies the choice of the velocity scale $U=gH^2/\nu$ in section~\ref{sec:gov-eq}.

The solution \eqref{eq:expansion_a} is substituted into Eq.~\eqref{eq:h-nondim} through the definition of $\tilde q$ in Eq.~\eqref{eq:q-definition}, i.e.,
\begin{equation} \label{eq:be2}
	\frac{\partial h}{\partial t} = \frac{1}{r \sin(\beta - \varepsilon h / r)} \frac{\partial}{\partial r} \int_{0}^{\psi} r^2 \left(u^{(0)} + \varepsilon u^{(1)} + \varepsilon^2 u^{(2)}\right) \sin(\beta + \varepsilon \alpha) d\alpha.
\end{equation}
The integral and derivative operations in the right-hand side (RHS) are straightforward after performing Taylor expansion for the sinusoidal function.
The final result, being a little lengthy, is given by Eqs.~\eqref{eq:be2-final}-\eqref{eq:be2-q2} in Appendix~\ref{ap:linear-results} (some of the symbolic manipulations in this paper were carried out using the Python library SymPy 1.13~\citep{SymPy}).
Eq.~\eqref{eq:be2} is the second-order Benney-type equation for liquid film flow along a conical surface.

Next, we seek the steady-state film thickness $\bar h$ at fixed flow rate $q=q_s$ by writing
\begin{equation} \label{eq:hbar}
	\bar h = \bar h^{(0)} + \varepsilon \bar h^{(1)} + \varepsilon^2 \bar h^{(2)} + O ( \varepsilon^3 ).
\end{equation}
After replacing $h$ with $\bar h$ and setting $\partial \bar h/\partial t=0$ in Eq.~\eqref{eq:be2}, $\bar h^{(0)}$,  $\bar h^{(1)}$, and  $\bar h^{(2)}$ are solved successively as
\begin{equation} \label{eq:hbar-leading}
	\bar h^{(0)} = K (q_s / r)^{1/3},
\end{equation}
\begin{equation}
	\bar h^{(1)} = - \frac{4 K^{2} \Rey q_s^{5/3}}{35 r^{8/3} \sin\beta} + \frac{K^{2} q_s^{2/3} \sin\beta}{9 r^{5/3} \cos\beta} + \frac{K^{2} q_s^{2/3} \cos\beta}{3 r^{5/3} \sin\beta} - \left(\frac{\varepsilon}{\Bon}\right) \frac{K q_s^{1/3}}{3 r^{7/3} \sin\beta},
\end{equation}
and
\begin{equation} \label{eq:hbar-2nd}
	\begin{split}
		\bar h^{(2)} = & \frac{19648 \Rey^{2} q_s^{3}}{121275 r^{5} \sin^{3}\beta \cos\beta} + \frac{32 \Rey q_s^{2}}{105 r^{4} \sin\beta \cos^{2}\beta} + \frac{83 \Rey q_s^{2}}{140 r^{4} \sin^{3}\beta} - \frac{7 q_s \sin\beta}{27 r^{3} \cos^{3}\beta} \\
		& - \frac{67 q_s}{24 r^{3} \sin\beta \cos\beta} - \frac{27 q_s \cos\beta}{40 r^{3} \sin^{3}\beta} - \frac{7 q_s}{40 r^{3} \sin^{3}{\beta} \cos{\beta}} \\ 
		& + \left(\frac{\varepsilon}{\Bon}\right) \left[ - \frac{4 K^{2} \Rey q_s^{5/3}}{105 r^{\frac{14}{3}} \sin^{2}\beta} - \frac{40 K^{2} q_s^{2/3}}{81 r^{11/3} \cos\beta} - \frac{K^{2} q_s^{2/3} \cos\beta}{r^{11/3} \sin^{2}\beta} + \left(\frac{\varepsilon}{\Bon}\right) \frac{2 K q_s^{1/3}}{9 r^{13/3} \sin^{2}\beta}\right],
	\end{split}
\end{equation}
where the constant $K = - \sqrt[3]{3 / (\sin\beta \cos\beta)}$ is positive if $\beta > \pi/2$.
This base-flow solution is used in the subsequent linear stability analysis.
\citet{1976Zollars} also derived a first-order asymptotic expression for the steady-state film thickness by applying a perturbation expansion to the stream function. 
Truncating our result at the same order, i.e., $\bar h^{(0)} + \varepsilon \bar h^{(1)}$, reproduces their formula once their scaling is adopted.
These expressions for $\bar h$ are expected to be valid except in the vicinity of the cone axis (see section~\ref{subsec:valiRang} for more discussions).

Before proceeding, it might be interesting to examine the shapes of the undisturbed film flow predicted by Eq.~\eqref{eq:hbar}.
However, the dependence of the dimensionless quantities on the unspecified length scales $H$ and $L$ causes inconvenience in displaying results and describing liquid properties/working conditions.
It is noteworthy that, $H$ and $L$ (thus $\varepsilon$) are introduced to simplify the governing equations by distinguishing the orders of different terms explicitly.
Once the final results, e.g. Eq.~\eqref{eq:hbar}, are obtained, it would be useful to re-nondimensionalize the
variables with a more specific length scale such as the film thickness at the inlet, $\tilde h_{\mathrm{in}}$, or the combination $(\nu^2/g)^{1/3}$.
Actually, on substitution of $L=H=(\nu^2/g)^{1/3}$, one finds
\begin{equation}
    \Rey = \frac{g H^3}{\nu^2} = 1, \quad \Bon = \frac{\rho g H^2}{\sigma} = \frac{\rho g^{1/3} \nu^{4/3}}{\sigma} \equiv \frac{1}{\Kap_\perp}, \quad  q = \frac{\tilde q}{U H L} = \frac{\tilde q g^{1/3}}{\nu^{5/3}} \equiv N_{q\perp},
\end{equation}
i.e., $\Rey$ is always 1; the original Bond number converts to a form whose inverse is usually designated to the Kapitza number, $\Kap_\perp$; and the nondimensionalized flow rate $q$ becomes $N_{q\perp}$.
The subscript $\perp$ indicates the dimensionless groups are ``vertical'' versions that are irrelevant to the angle $\tilde \beta$.

\begin{figure}
    \centering
    \input{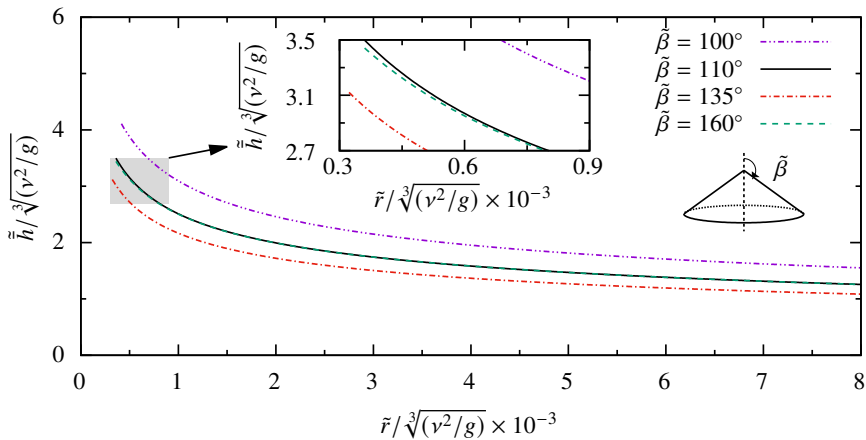}
    \caption{
    Steady-state film thickness (Eq.~\eqref{eq:hbar}) versus radial distance at different $\tilde{\beta}$.
    $\Kap_\perp = 242$, $N_{q\perp} = 1703$. }
    \label{fig:hbar-beta}
\end{figure}

Figure~\ref{fig:hbar-beta} shows $\tilde {\bar h}$ as a function of $\tilde r$ predicted by Eq.~\eqref{eq:hbar} at four different $\tilde \beta$.
The other parameters are fixed as $\Kap_\perp = 242$, and $N_{q\perp} = 1703$.
For an aqueous glycerol solution with $\nu = 5.77 \times 10^{-6}\, \mathrm{m^2} / \mathrm{s}$ (case C in \citet{Denner2018JFM}), this $N_{q\perp}$ corresponds to a dimensional volumetric flow rate of $\tilde Q = 2 \pi \tilde q = 9.28 \times 10^{-6}\,\mathrm{m}^3/\mathrm{s}$. 
Interestingly, all four curves in figure~\ref{fig:hbar-beta} are nearly parallel to each other.
With the same volumetric flow rate, the film thickness at the same radial position decreases as $\tilde \beta $ increases from $100^\circ$ to $135^\circ$.
However, further increasing $\tilde \beta $ leads to a higher elevation of the film surface.
Moreover, except for the very small region close to the apex, the result with $\tilde \beta =160^\circ$ is almost indistinguishable form that with  $\tilde \beta =110^\circ$.

The above observation results from the fact that the corrections $\bar h^{(1)}$ and $\bar h^{(2)}$ to $\bar h^{(0)}$ are negligible unless in regions where $r$ is small (the powers of $r$ in the denominators successively increase from $\bar h^{(0)}$ to $\bar h^{(2)}$, indicating faster decay with $r$ for high-order corrections).
As a consequence, the film shape is dominated by $\bar h^{(0)}$ (Eq.~\eqref{eq:hbar-leading}), which, on substitution of $K$, yields
\begin{equation}
    \bar h^{(0)} 
    = - \left( \frac{3 q_s}{r \sin\beta \cos\beta} \right)^{1/3} 
    = - \left[ \frac{6 q_s}{r \sin(2\beta)} \right]^{1/3}.
    \label{eq:h0}
\end{equation}
Noting that the dimensional form of Eq.~\eqref{eq:h0} is exactly the same as Eq.~\eqref{eq:local_hN}.
Obviously, Eq.~\eqref{eq:h0} indicates that $\bar h^{(0)}(\beta)$ is symmetric about $\beta = 3\pi/4$ or $\tilde \beta = 135^\circ$.
This explains the overlap of the film surface for $\tilde \beta =110^\circ$ and $160^\circ$ in figure~\ref{fig:hbar-beta}.

This non-monotonic behavior can be interpreted as below: as $\tilde \beta$ increases from $90^\circ$ to $180^\circ$, i.e., when the cone becomes progressively sharper, there are two competing mechanisms that can affect the film thickness at a fixed position $\tilde r$.
On the one hand, 
the local volumetric flow rate per unit width is calculated as $\tilde Q_\mathrm{2D} = \tilde Q / (2 \pi \tilde r \sin \tilde \beta) $, which increases with $\tilde{\beta}$.
Thus, the film thickness also has a tendency to increase.
On the other hand, the streamwise component of the gravitational acceleration $-g\cos\tilde\beta$ increases.
The liquid in the film is therefore accelerated to gain a higher mean velocity, resulting in a thinner film at a fixed flow rate.
Consequently, the actual film thickness is determined jointly by the geometric and dynamic factors, which are associated with $\sin\beta$ and $-\cos \beta$ in Eq.~\eqref{eq:h0}, respectively.

Notably, the existence of a minimal film thickness at a given flow rate might be helpful in designing heat and mass transfer equipment, as a thinner film has a smaller resistance for the transport process between the wall and the gas side.

In order to perform the stability analysis, we express the film thickness as the sum of the base flow and a small disturbance $\hat h \ll 1$:
\begin{equation}
h = \bar h (r) + \hat h(r,t).
\label{eq:h_disturbed}
\end{equation}
After substitution of Eq.~\eqref{eq:h_disturbed}, Eq.~\eqref{eq:be2} can be linearized by dropping all terms containing second- and higher-order powers of $\hat h$.
The result is 
\begin{equation} \label{eq:hhat-linear}
	\begin{split}
		& \frac{\partial \hat h}{\partial t} + S_0 \hat h + S_1 \frac{\partial \hat h}{\partial r} + S_2 \frac{\partial^{2} \hat h}{\partial r^{2}} + S_3 \frac{\partial^{3}  \hat h}{\partial r^{3}} + S_4 \frac{\partial^{4} \hat h}{\partial r^{4}} = 0.
	\end{split}
\end{equation}
The coefficients $S_0$-$S_4$ are given by Eqs.~\eqref{eq:S0}-\eqref{eq:S4} in Appendix~\ref{ap:linear-results}.

On the basis of Eq.~\eqref{eq:hhat-linear}, we conduct a spatial linear stability analysis.
Assuming the disturbance is periodic in time and grows/decays in space, we set
\begin{equation} \label{eq:hhatForm}
	\hat h(r, t) = R(r) e^{\mathrm j \omega t},
\end{equation}
where $\mathrm j$ is the imaginary unit,
and $\omega$, denoting the dimensionless angular frequency of the disturbance, is a real number.
Substitution of Eq.~\eqref{eq:hhatForm} into \eqref{eq:hhat-linear} leads to an equation for $R(r)$.
Following \citet{1980Zollars}, we assume $R(r)$ is expressed as
\begin{equation} \label{eq:sol-hypo}
   R(r) = e^{f_0(r) + \varepsilon f_1(r) + \varepsilon^2 f_2(r) + O (\varepsilon^3)},
\end{equation}
where the functions $f_i(r)$ ($i=0$, 1, 2) are complex in general and solved sequentially.
Their expressions are listed in Eqs.~\eqref{eq:f}-\eqref{eq:mIm} of Appendix~\ref{ap:linear-results}.
By taking the real part in the final result, $\hat h$ reads
\begin{equation} \label{eq:hhat}
    \hat{h}(r, t) = C \frac{e^{B r}}{r^{1/3}} \cos \left(\omega t - A r \right),
\end{equation}
where
\begin{equation} \label{eq:A}
	A = - \left[\Imag (f_0) + \varepsilon \Imag (f_1) + \varepsilon^2 \Imag (f_2) \right] / r,
\end{equation}
\begin{equation} \label{eq:B}
	B = \left[\varepsilon \Real (f_1) + \varepsilon^2 \Real (f_2)\right] / r,
\end{equation}
and $C$ is an integral constant.
$\Real (\cdot)$ and $\Imag (\cdot)$ denote the real and imaginary parts of a complex number, respectively.

It is noteworthy that, at large radial distance $r$, $B$ is dominated by
\begin{equation}
    - \varepsilon^2 \left(\frac{\varepsilon}{\Bon}\right) \frac{K^{2} \omega^{4} r^{7/3} \sin\beta}{270 q_s^{7/3} \cot^{2}\beta},
    \label{eq:B_large_r}
\end{equation}
which is always negative.
Noting Eq.~\eqref{eq:hhat}, this fact indicates that for any given $\omega$, the amplitude of the disturbance $\hat h$ decays to zero at large $r$, i.e., the flow ultimately stabilizes.

Within a certain range of moderate $r$, however, the disturbances can be amplified locally.
\citet{1980Zollars} classified the instability of this non-parallel flow into two types:
the absolute growth refers to the scenario in which the raw disturbance amplitude increases with $r$, i.e., $d[\exp{(B r) / r^{1/3}}] / d r > 0$; whereas the relative growth means an increase of the amplitude scaled by the local mean film thickness $\bar h$ whose value is given in Eq.~\eqref{eq:hbar}, i.e., $d[\exp{(B r) / (\bar h r^{1/3}})] / d r > 0$.
Correspondingly, two modified spatial amplification factors can be defined as
\begin{equation} \label{eq:Ga}
    G_a \equiv \frac{1}{\exp{(B r) / r^{1/3}}} \frac{d[\exp{(B r) / r^{1/3}}]}{d r} = r \frac{d B}{d r} + B - \frac{1}{3 r}
\end{equation}
for absolute growth and
\begin{equation} \label{eq:Gr}
    G_r \equiv \frac{1}{\exp{(B r) / (\bar h r^{1/3}})} \frac{d[\exp{(B r) / (\bar h r^{1/3}})]}{d r} = r \frac{d B}{d r} + B - \frac{1}{\bar h} \frac{\partial \bar h}{\partial r} - \frac{1}{3 r}
\end{equation}
for relative growth (the definition of $G_r$ differs from that in \citet{1980Zollars}).
The difference between the two expressions is the term $-(\partial \bar h/\partial r ) / \bar h$, which is positive since $\bar h$ is negatively correlated with $r$, see Eq.~\eqref{eq:hbar-leading}.
Therefore, this term has a destabilizing effect, and it is possible that at the same $r$, a flow is stable in the sense of absolute growth ($G_a<0$), yet unstable in the sense of relative growth ($G_r>0$).

\begin{figure}
    \centering
    \input{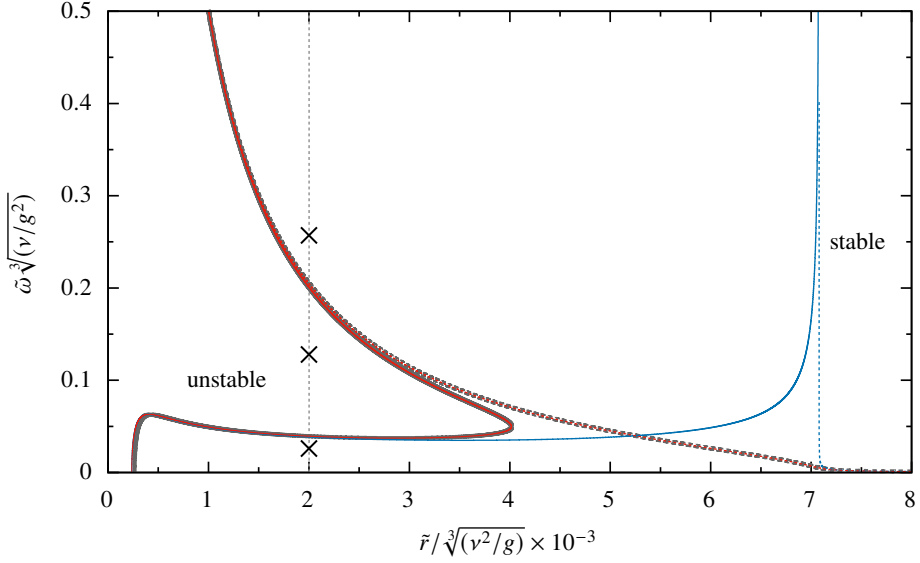}
    \caption{Neutral stability curves for absolute (solid lines) and relative (dashed lines) growth in the parameter space spanned by disturbance angular frequency $\tilde \omega$ and radial distance $\tilde r$;
    $\tilde \beta = 150^\circ$, $\Kap_\perp = 242$, and $N_{q\perp} = 1703$.
    Gray lines: second-order model;
    blue lines: first-order model;
    red lines: first-order formulae plus second-order streamwise-curvature correction.
    The three symbols correspond to the cases shown in figure~\ref{fig:hhat}.
    } 
    \label{fig:neutral_curves}
\end{figure}

Figure~\ref{fig:neutral_curves} shows the neutral stability curves in the parameter space spanned by the disturbance angular frequency $\tilde \omega$ and radial coordinate $\tilde r$.
The stability thresholds for absolute ($G_a=0$) and relative ($G_r=0$) growth are represented by solid and dashed lines, respectively.
The thick gray curves show second-order results.
The blue curves are obtained by inserting the first-order ($O(\varepsilon)$) expressions of $B$ and $\bar h$ into Eqs.~\eqref{eq:Ga} and \eqref{eq:Gr}, i.e., with the order of accuracy achieved in \citet{1980Zollars}.
For both absolute and relative growth, a dependence of the stability thresholds on the radial distance is observed.
Any disturbance will be stabilized at sufficiently large $\tilde r$, as expected.
Compared with the first-order result, the second-order neutral curves for both instabilities shift toward the small $\tilde r$ region, especially at high disturbance frequencies.

In fact, the difference between the first- and second-order results primarily originates from the term $\varepsilon \partial^2 h /\partial r^2$ in Eq.~\eqref{eq:curv}, which essentially represents the contribution of streamwise curvature of the free surface.
The expressions of $G_a$ and $G_r$ with only this additional second-order correction read
\begin{equation}
\begin{aligned}
G_a & = - \varepsilon \frac{2 \Rey \omega^{2}}{15 \cos{\beta }} - \varepsilon \frac{4 \sqrt[3]{3} \Rey q_s^{4/3}}{3 r^{10/3} \sin^{4/3}{\beta } \cos^{1/3}{\beta }} - \varepsilon \frac{\omega^{2} r  \sin^{2}{\beta }}{9 q_s \cos^{2}{\beta }} + \varepsilon \frac{8 \sqrt[3]{3} q_s^{1/3} \sin^{2/3}{\beta }}{27 r^{7/3} \cos^{4/3}{\beta }} \\
& + \varepsilon \frac{8 \sqrt[3]{3} q_s^{1/3} \cos^{2/3}{\beta }}{9 r^{7/3} \sin^{4/3}{\beta }} - \frac{1}{3 r} + \varepsilon \left(\frac{\varepsilon}{\Bon}\right) \frac{2}{3 r^{3} \sin{\beta}} - \varepsilon^{2} \left(\frac{\varepsilon}{\Bon}\right) \frac{3^{2/3} \omega^{4} r^{7/3} \sin^{7/3}{\beta}}{81 q_s^{7/3} \cos^{8/3}{\beta}},
\end{aligned}
\label{eq:Ga1st_fullCurv}
\end{equation}
and
\begin{equation}
\begin{aligned}
G_r & = - \varepsilon \frac{2 \Rey \omega^{2}}{15 \cos{\beta }} - \varepsilon \frac{16 \sqrt[3]{3} \Rey q_s^{4/3}}{15 r^{10/3} \sin^{4/3}{\beta } \cos^{1/3}{\beta }} - \varepsilon \frac{\omega^{2} r \sin^{2}{\beta }}{9 q_s \cos^{2}{\beta }} + \varepsilon \frac{\sqrt[3]{3} \sqrt[3]{q_s} \sin^{2/3}{\beta }}{9 r^{\frac{7}{3}} \cos^{4/3}{\beta }} \\
& + \varepsilon \frac{11 \sqrt[3]{3} q_s^{1/3} \cos^{2/3}{\beta }}{27 r^{7/3} \sin^{4/3}{\beta }} + \varepsilon \frac{\sqrt[3]{3} q_s^{1/3}}{27 r^{7/3} \sin^{4/3}{\beta } \cos^{4/3}{\beta }} - \varepsilon^{2} \left(\frac{\varepsilon}{\Bon}\right) \frac{3^{2/3} \omega^{4} r^{7/3} \sin^{7/3}{\beta }}{81 q_s^{7/3} \cos^{8/3}{\beta}}.
\end{aligned}
\label{eq:Gr1st_fullCurv}
\end{equation}

The stability thresholds calculated with these simplified versions of $G_a$ and $G_r$ (red lines in figure~\ref{fig:neutral_curves})
are nearly indistinguishable from the full second-order results (gray lines).
In conclusion, the streamwise curvature has a significant influence in stabilizing the liquid film flow on a conical surface, confirming the speculation of \citet{1980Zollars} who did not, however, provide the corresponding formulae or quantitative results.

Moreover, further comparison (results not shown here) reveals that the effect of the lateral curvature, i.e., the term $\cot\beta/r$ in Eq.~\eqref{eq:curv}, is negligible.
The reason why the seemingly smaller streamwise curvature (second-order) has a higher impact than the first-order lateral curvature will be further discussed in section~\ref{sec:doublee}.

A noteworthy feature of the absolute-growth threshold is its multi-valued dependence on $\tilde r$: over a finite radial interval, the flow is stable at both low and high disturbance frequencies, and unstable only at intermediate values.
Another observation from figure~\ref{fig:neutral_curves} is the overlapping of the absolute and relative growth thresholds when $\tilde \omega$ is large.
At small $\tilde \omega$, the flow can be stable for absolute growth but unstable for relatively growth, as anticipated earlier.
This is understandable as, since the film thickness decreases with $\tilde r$, it is more stringent for a flow to be stable in the sense of relative growth.

\subsection{Wavelength and phase speed for linear waves}
\label{sec:wavelength_celerity_linearWave}
According to Eq.~\eqref{eq:hhat}, the phase of a linear wave is $\phi = \omega t - A r$, where $A$ is a function of $r$.
Therefore, following a fixed phase with $d \phi = \omega d t - [d(A r)/dr] dr = 0$, the phase speed is calculated as
\begin{equation}
c = \left[ \frac{d r}{dt} \right]_{d\phi=0} =  \frac{\omega}{ d (Ar)/dr} .
\label{eq:phaseSpeed}
\end{equation}

\begin{figure}
	\centering
	\input{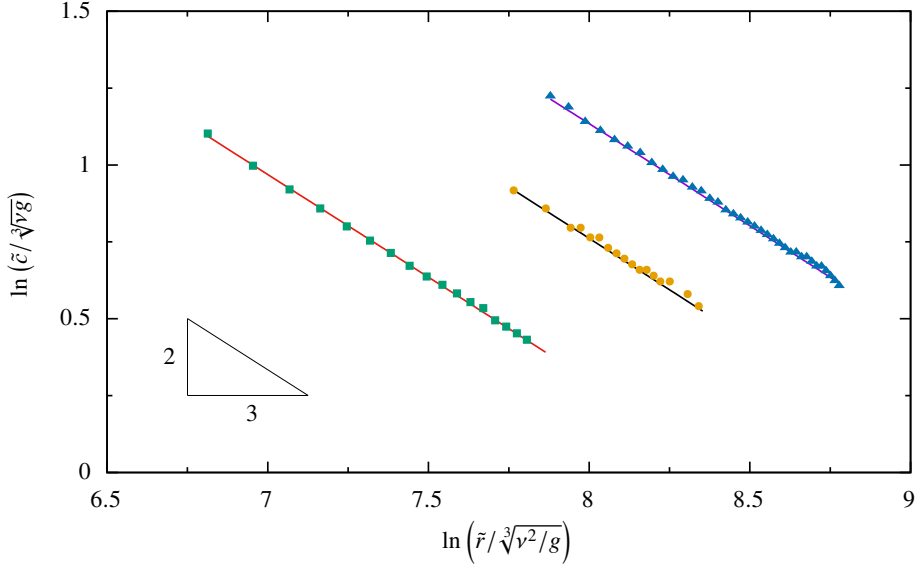}
	\caption{Phase speed of linear waves versus distance from cone apex; $\tilde \beta = 150^\circ$, $\Kap_\perp = 242$.
	Lines: Eq.~\eqref{eq:phaseSpeed};
	symbols: $h$-$q$ model (section~\ref{sec:model}).
	Squares: $N_{q\perp} = 851.5$, $N_{\omega\perp} = 0.108$; 
	circles: $N_{q\perp} = 1703$, $N_{\omega\perp} = 0.128$; 
	triangles: $N_{q\perp} = 3000$, $N_{\omega\perp} = 0.148$. 
	}
    \label{fig:linear_wave_characteristic}
\end{figure}

The lines in figure~\ref{fig:linear_wave_characteristic} show $c$ as functions of $\tilde r$, computed with Eq.~\eqref{eq:phaseSpeed}.
The polar angle is $\tilde \beta = 150^\circ$.
The Kapitza number is $\Kap_\perp = 242$.
From left to right, the dimensionless volumetric flow rates are $N_{q\perp} = 851.5$, 1703, and 3000, respectively.
The corresponding dimensionless disturbance angular frequencies, defined as $N_{\omega\perp} = \tilde \omega (\nu/g^2)^{1/3}$, are 0.108, 0.128, and 0.148.
A remarkable characteristic is that the lines are nearly straight with a slope of $-2/3$ in the logarithm coordinates.
Indeed, inserting the leading-order part of Eq.~\eqref{eq:A} into Eq.~\eqref{eq:phaseSpeed} yields
\begin{equation} \label{eq:celerity-leading}
    c^{(0)}  = \frac{3 q_s^{2/3}}{K r^{2/3} \sin\beta} \propto r^{-2/3}.
\end{equation}
The leading-order wavelength can be calculated accordingly following $\lambda^{(0)} = 2 \pi c^{(0)}/\omega$, which shares the same power law.
The value of $c^{(0)}$ is $3/5$ times the phase speed in \citet{zollars1974linear} (Eq.~(\uppercase\expandafter{\romannumeral6}-17)), if the same scaling is used.
This difference arises because the latter used $A$ instead of $d(Ar)/dr$ as the local wavenumber.

In addition, it would be useful to nondimensionalize the phase speed with the Nusselt scaling~\citep{2011kalliadasis} and compare the results with the flat-plate situation.
In that scaling, the characteristic velocity is chosen as $\tilde{h}_{N\mathrm{2D}}^2 /(l_\nu t_\nu)$, where $ \tilde{h}_{N\mathrm{2D}}$ is the Nusselt film thickness defined in Eq.~\eqref{eq:NusseltThickness}.
The viscous-gravity length and time scales are $l_\nu=[\nu^2/(g\sin\tilde\varphi)]^{1/3}$ and $t_\nu = [\nu/(g\sin\tilde\varphi)^2]^{1/3}$, respectively.
For the current conical scenario, we apply a modified Nusselt scaling by replacing $\tilde{h}_{N\mathrm{2D}}$ with the leading-order approximation to the local base-flow film thickness $\tilde h_{N\ell}$, see Eq.~\eqref{eq:local_hN}.
In this way, the velocity scale is written as 
$\tilde{h}_{N\ell}^2 / (l_\nu t_\nu)$,
with the length and time scales
\begin{equation}
l_\nu = \left( \frac{\nu^2}{-g \cos\tilde\beta} \right)^{1/3}, \quad \mathrm{and} \quad t_\nu =  \left( \frac{\nu}{g^2 \cos^2\tilde\beta} \right)^{1/3},
\label{eq:v_g_scaling}
\end{equation}
respectively (the relation $\tilde\beta = \pi/2 + \tilde\varphi$ has been used).
With this nondimensionalization, Eq.~\eqref{eq:celerity-leading} yields a leading-order phase speed of
\begin{equation}
c^{\prime (0)} = \frac{ \tilde c^{(0)} (\tilde r)  l_\nu t_\nu }{ \tilde h^2_{N\ell} (\tilde r) } \equiv 1,
\end{equation}
which is irrelevant to $\tilde r$.
This value is identical to the leading-order flat-plate result.
The same conclusion, however, does not hold for the first- and second-order components, in which the present conical results contain $\tilde r$-dependent terms.
That being said, when taking $\tilde r \rightarrow \infty$, the present expressions exactly reduce to the flat-plate ones obtained with spatial-mode stability analysis.
In particular, the first-order correction is $\tilde c^{(1)}_\mathrm{2D} = 0$, while the second-order component, in dimensional form, is
\begin{equation}
    \tilde{c}^{(2)}_\mathrm{2D} = - \frac{2 \nu \tilde{\omega}^{2} \sin^{2}\tilde{\beta}}{9 g \cos^{3}\tilde{\beta}} + \frac{\nu \tilde{\omega}^{2}}{g \cos\tilde{\beta}} - \frac{2 \tilde{\omega}^{2} \tilde{Q}_{\mathrm{2D}} \sin\tilde{\beta}}{35 g \cos^{2}\tilde{\beta}} + \frac{44 \tilde{\omega}^{2} \tilde{Q}_{\mathrm{2D}}^{2}}{175 g \nu \cos\tilde{\beta}}.
\end{equation}

\section{Second-order $h$-$q$ model} \label{sec:model}

Unlike the flat-plate case, the dependence of the flow field on the streamwise coordinate prohibits the use of periodic boundary conditions on a single wave unit.
An alternative to efficiently simulate the nonlinear waves is the low-dimensional models.
Recently, \citet{zhou2022hydraulic} derived a ``first-order'' $h$-$q$ model 
via the Galerkin method~\citep[see, e.g.,][]{ruyer2000improved, scheid2006wave} to study hydraulic jumps.
Notably, their model is not purely first order, as it includes the second-order contribution of the streamwise curvature, see Eq. (3.21) in \citet{zhou2022hydraulic}.
The Galerkin method has proven to be the most efficient in obtaining the ``optimal" low-dimensional model for liquid film flow on a flat plate~\citep{ruyer2000improved}.
Here we extend the model in \citet{zhou2022hydraulic} to full second order in $\varepsilon$ for more accurate surface wave simulations.

Starting from the truncated cross-stream momentum equation \eqref{eq:theta-f}, we integrate from $\alpha = 0$ to $\psi$, invoke the continuity equation \eqref{eq:nondim-continuity} and boundary conditions \eqref{eq:noSlipBC_dimless} and \eqref{bc:normal-nondim}, obtaining the first-order pressure distribution 
\begin{equation} \label{eq:p-expr}
	p = r \left(\alpha - \psi\right) \sin\beta + \frac{1}{2} \varepsilon r \left(\alpha^2 - \psi^2\right) \cos\beta - \varepsilon \frac{\partial u}{\partial r} - \varepsilon \left.\left(\frac{\partial u}{\partial r} + 2 \frac{u}{r}\right)\right|_\psi - \left(\frac{\varepsilon}{\Bon}\right) \mathcal{C}.
\end{equation}
Inserting Eq.~\eqref{eq:p-expr} into the streamwise momentum equation \eqref{eq:r-f} yields the second-order boundary-layer equation
\begin{equation} \label{eq:ble}
	\begin{split}
		& \varepsilon \Rey \left(\frac{\partial u}{\partial t} + u \frac{\partial u}{\partial r} + \frac{v}{r} \frac{\partial u}{\partial \alpha}\right) = \frac{1}{r^2} \frac{\partial^2 u}{\partial \alpha^2} + \varepsilon \frac{\cot\beta}{r^2} \frac{\partial u}{\partial \alpha} + \varepsilon^2 \left(2 \frac{\partial^2 u}{\partial r^2} + \frac{4}{r} \frac{\partial u}{\partial r} + \frac{2}{r^2} u\right) \\
		& - \varepsilon^2 \frac{\alpha}{r^2 \sin^2\beta} \frac{\partial u}{\partial \alpha} - \cos\beta + \varepsilon \sin\beta \left(\psi + r \frac{\partial \psi}{\partial r}\right) + \varepsilon \left(\frac{\varepsilon}{\Bon}\right) \frac{\partial \mathcal{C}}{\partial r} + \frac{1}{2} \varepsilon^2 \cos\beta \psi^2 \\
		& + \varepsilon^2 r \cos\beta \psi \frac{\partial \psi}{\partial r} + \varepsilon^2 \frac{\partial}{\partial r}\left[\left.\left(\frac{\partial u}{\partial r} + 2 \frac{u}{r}\right)\right|_\psi\right].
	\end{split}
\end{equation}

Next, we apply the Galerkin method to Eq.~\eqref{eq:ble}.
Guided by the leading-order profile Eq.~\eqref{eq:u0}, the trial function for the streamwise velocity is constructed as a semi-parabolic function of the cross-stream coordinate:
\begin{equation} \label{eq:u-testfunction}
	u^* \equiv m F = m \left[2 \frac{\alpha}{\psi} - \left(\frac{\alpha}{\psi}\right)^2\right].
\end{equation}
The coefficient $m$ is irrelevant to $\alpha$, but may depend on the streamwise coordinate $r$ and time $t$.
It is determined by substituting Eq.~\eqref{eq:u-testfunction} into the dimensionless form of Eq.~\eqref{eq:q-definition}.
The result, retained up to $O(\varepsilon^2)$, reads
\begin{equation}
	m = - \frac{3}{2} \frac{q}{r^2 \psi} \frac{1}{\sin \beta} \left[1 - \varepsilon \frac{5}{8} \psi \cot \beta + \varepsilon^2 \left(\frac{9}{40} \psi^2 + \frac{25}{64} \psi^2 \cot^2 \beta\right) \right].
\end{equation}
Eq.~\eqref{eq:ble} can be formally written as $\mathcal{L} (u) = 0$.
With the trial function $u^\ast$, the residual across the film is
\begin{equation} \label{eq:R0}
    \mathcal{R} = \int_0^\psi F \mathcal{L} \left( u^* \right) d\alpha,
\end{equation}
where $F$ defined in Eq.~\eqref{eq:u-testfunction} serves as the test function.

It is noteworthy that the boundary condition in Eq.~\eqref{eq:bc-tangential-nondim} is non-homogeneous, whereas $\partial^2 u^* /\partial \alpha^2 \equiv 0$.
Therefore, $u^*$ satisfies Eq.~\eqref{eq:bc-tangential-nondim} only to the order of $O(\varepsilon)$.
To recover the missing $O(\varepsilon^2)$ constraint, we embed Eq.~\eqref{eq:bc-tangential-nondim} through the second-order $\alpha$-derivative in Eq.~\eqref{eq:R0}.
In detail, the integral of the term $F \partial^2 u^*/\partial \alpha^2 $ is re-expressed via Green's second identity (or, equivalently, by applying the integration by parts twice) as
\begin{equation}
	\begin{split}
	    \int_{0}^{\psi} F \frac{\partial^2 u^*}{\partial \alpha^2} d\alpha 
	    = & \left.\left(F \frac{\partial u^*}{\partial \alpha}\right)\right|_0^\psi - \left.\left(\frac{\partial F}{\partial \alpha} u^*\right)\right|_0^\psi + \int_{0}^{\psi} \frac{\partial^2 F}{\partial \alpha^2} u^* d\alpha.
	\end{split}
	\label{eq:tau}
\end{equation}
The value $[\partial u^* /\partial \alpha]_\psi$ in the first term of the RHS is replaced by Eq.~\eqref{eq:bc-tangential-nondim}, thus the latter is involved in the derivation.
A similar method is applied to develop the second-order low-dimensional model for the flat-plate case  \citep{ruyer2000improved}.

Inserting Eq.~\eqref{eq:tau}, eliminating $\partial h / \partial t$ with Eq.~\eqref{eq:h-nondim}, and requiring $\mathcal{R}=0$ yields:
\begin{equation} \label{eq:q}
    \begin{split}
        & \varepsilon \Rey \left(1 + \varepsilon \frac{5 h \cot\beta}{8 r} \right) \frac{\partial q}{\partial t} + \frac{\varepsilon \Rey}{r \sin\beta} \left( \frac{17 q}{7 h} \frac{\partial q}{\partial r} - \frac{9 q^{2}}{7 h^{2}} \frac{\partial h}{\partial r} - \frac{9 q^{2}}{7 r h}\right) \\
        & + \frac{\varepsilon^2 \Rey \cos\beta}{r^2 \sin^2\beta} \left( \frac{5167 q}{1792} \frac{\partial q}{\partial r} - \frac{1557 q^{2}}{1792 h} \frac{\partial h}{\partial r} - \frac{4203 q^{2}}{1792 r}\right)\\
        = & - \frac{5 r h \sin\beta \cos\beta}{6} - \frac{5 q}{2 h^{2}} - \varepsilon \left(\frac{5 r h \sin^{2}\beta}{6} \frac{\partial h}{\partial r} + \frac{5 q \cot\beta}{2 r h}\right) + \varepsilon \left(\frac{\varepsilon}{\Bon}\right) \frac{5 r h \sin\beta}{6} \frac{\partial \mathcal{C}}{\partial r} \\
        & + \varepsilon^{2} \left[ \frac{5 h^{2} \sin\beta \cos\beta}{6} \frac{\partial h}{\partial r} + \frac{9}{2} \frac{\partial^{2} q}{\partial r^{2}} - \frac{6 q}{h} \frac{\partial^{2} h}{\partial r^{2}} - \frac{9}{2 h} \frac{\partial h}{\partial r} \frac{\partial q}{\partial r} + \frac{4 q}{h^{2}} \left(\frac{\partial h}{\partial r}\right)^{2} \right.\\
        &\left. - \frac{5  h^{3} \sin\beta \cos\beta}{12 r} - \frac{9}{2 r} \frac{\partial q}{\partial r} + \frac{19 q}{4 r h} \frac{\partial h}{\partial r} - \frac{7 q}{4 r^{2}} - \frac{2 q}{ r^{2} \sin^{2}\beta}\right].
    \end{split}
\end{equation}
Eqs.~\eqref{eq:h-nondim} and \eqref{eq:q} describe the evolution of the film thickness $h$ and volumetric flow rate $q$, respectively.
These two partial differential equations constitute the second-order low-dimensional model (referred to as the $h$-$q$ model hereafter) for the liquid film flow on a cone.
The model recovers to the ``first-order'' one proposed by \citet{zhou2022hydraulic} by dropping all $O (\varepsilon^2)$ terms except for those representing the streamwise surface tension contribution.
On the other hand, by substituting $q_\mathrm{2D}=q/(r\sin\beta)$, replacing $\beta$ with $\pi/2 + \varphi$, and taking the limit $r \rightarrow \infty$, Eqs.~\eqref{eq:h-nondim} and \eqref{eq:q} become
\begin{equation}
    \frac{\partial h}{\partial t} = - \frac{\partial q_\mathrm{2D}}{\partial x},
\label{eq:h_2d}
\end{equation}
and
\begin{equation}
    \begin{split}
        \Rey \frac{\partial q_\mathrm{2D}}{\partial t} = &- \frac{17 \Rey q_\mathrm{2D}}{7 h} \frac{\partial q_\mathrm{2D}}{\partial x} + \frac{9 \Rey q_\mathrm{2D}^{2}}{7 h^{2}} \frac{\partial h}{\partial x} + \frac{5 h \sin\varphi}{6} - \frac{5 q_\mathrm{2D}}{2 h^{2}} - \varepsilon \frac{5 h \cos\varphi}{6} \frac{\partial h}{\partial x} \\
        + & \varepsilon^2 \left[\frac{9}{2} \frac{\partial^{2} q_\mathrm{2D}}{\partial x^{2}} - \frac{6 q_\mathrm{2D}}{h} \frac{\partial^2 h}{\partial x^2} - \frac{9}{2 h} \frac{\partial h}{\partial x} \frac{\partial q_\mathrm{2D}}{\partial x} + \frac{4  q_\mathrm{2D}}{h^{2}} \left(\frac{\partial h}{\partial x}\right)^2 + \left(\frac{\varepsilon}{\Bon}\right) \frac{5 h}{6} \frac{\partial^3 h}{\partial x^3}\right],
    \end{split}
\label{eq:q_2d}
\end{equation}
respectively, where the coordinate $r$ is replaced by $x$ for consistency.
Eqs.~\eqref{eq:h_2d} and \eqref{eq:q_2d} are identical to the simplified second-order model for the thin liquid film flow over an inclined plate \citep{ruyer2000improved}.

The coupled equations (2.21) and (4.7) can be readily advanced in time on a uniform one-dimensional mesh in the radial direction.
For time stepping, we use the Bogacki–Shampine method~\citep{1989Bogacki}, a second/third-order adaptive Runge–Kutta method.
The timestep is automatically adjusted to control the estimated local error below the specified tolerance.
With the (normalized) values of $h$ and $q$ being of order $\gtrsim O(1)$ in our simulations, we set an absolute tolerance of $10^{-6}$.
The spatial derivatives are discretized with central finite difference schemes accurate at least to second order.
The film thickness $h_{\mathrm{in}}$ and volumetric flow rate $q_{\mathrm{in}}$ are specified at the inlet $r_{\mathrm{in}}$ of the computational domain.
In order to facilitate the transition of the flow towards a fully developed state, for a given $q_{\mathrm{in}}$ and $r_{\mathrm{in}}$, $h_{\mathrm{in}}$ is estimated from Eq.~(3.5).
Zero-gradient conditions for $h$ and $q$ are imposed at the outlet, which is sufficiently far from the region of interest.

As a validation of both the model and code, we compare the numerical results with those obtained from the linear theory of section~\ref{sec:linear}.
A periodic disturbance is imposed at the inlet on $q_{\mathrm{in}}$:
\begin{equation} \label{eq:bc-h-simul}
	q = q_{\mathrm{in}} \left[ 1 + \zeta \cos \left(2 \pi f t \right) \right],
\end{equation}
where $f$ and $\zeta$ are the frequency and relative amplitude of the disturbance, respectively.
We set $\zeta = 10^{-7}$ to create small-amplitude surface waves within the linear regime.
The phase speed $c$ at different $r$ is extracted from the wave profiles.
In detail, the oscillatory component $\hat h$ is determined by subtracting $\bar h$ from the computed film thickness, see Eq.~\eqref{eq:h_disturbed}.
Then, the local phase speed is calculated as the speed of the wave peaks in $\hat h (r,t)$.
The result (solid symbols in figure~\ref{fig:linear_wave_characteristic}) agree closely with the theoretical prediction (solid lines) from Eq.~\eqref{eq:phaseSpeed}.

\begin{figure}
    \centering
    \input{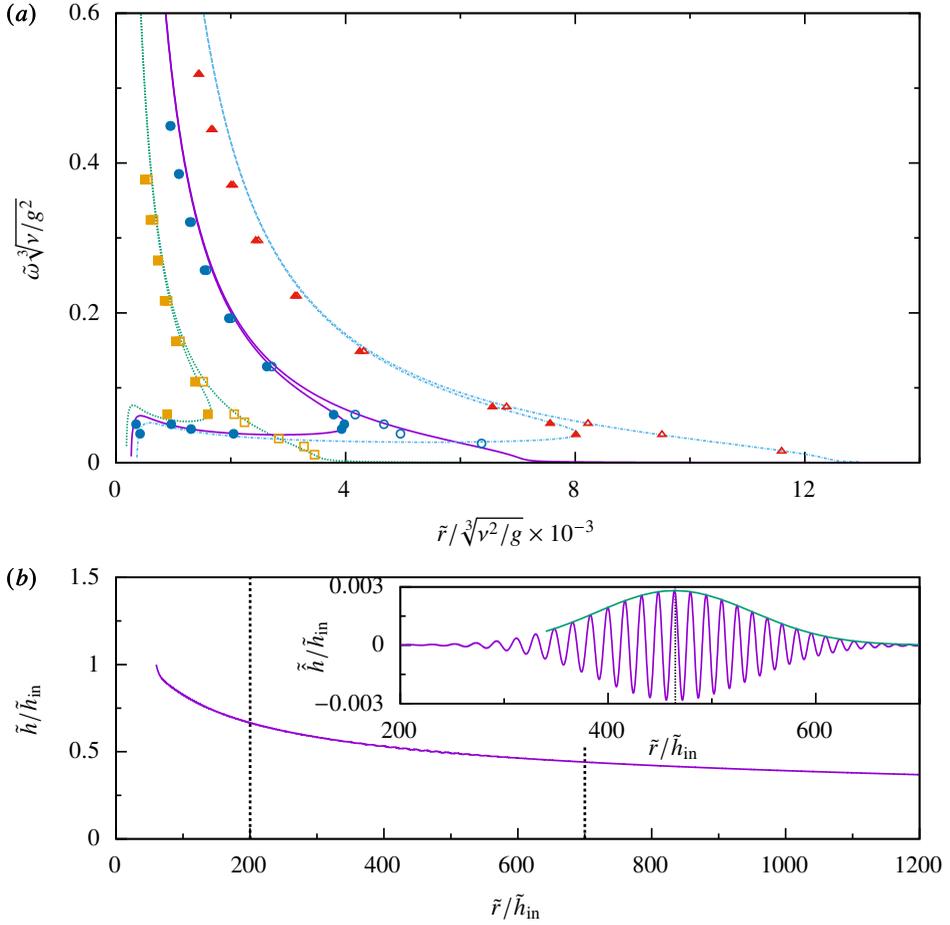}
    \caption{(a) Neutral stability curves; $\tilde \beta = 150^\circ$ and $\Kap_\perp = 242$.
    The volumetric flow rates for dashed, solid, and dash-dotted lines are $N_{q\perp} = 851.5$, 1703, and 3000, respectively.
    For each case, the curves on the left and right correspond to absolute and relative growth, respectively.
    Symbols (solid: absolute growth; open: relative growth) mark transition positions extracted from $h$-$q$ model results. 
    (b) Instantaneous film thickness of a case in panel (a).
    Inset illustrates how the neutral stability position is determined.
    } 
    \label{fig:neutral_curve_with_transition_points}
\end{figure}

The ability of the $h$-$q$ model in predicting neutral stability curves is examined in figure~\ref{fig:neutral_curve_with_transition_points}(a), where it can be seen that, for both absolute and relative growth, the numerical results of the $h$-$q$ model (symbols) have a good consistency with the linear-theory prediction (lines). 

For reference, the shape of the film surface for a typical case is illustrated in figure~\ref{fig:neutral_curve_with_transition_points}(b), where the small-amplitude linear waves can be barely observed at this scale.
The inset shows the isolated $\hat h$ following Eq.~\eqref{eq:h_disturbed}.
The position where the flow is neutrally stable is determined by seeking the local maximum (marked by the vertical dashed line) or minimum of the oscillations' envelope.

\begin{figure}
    \centering
    \input{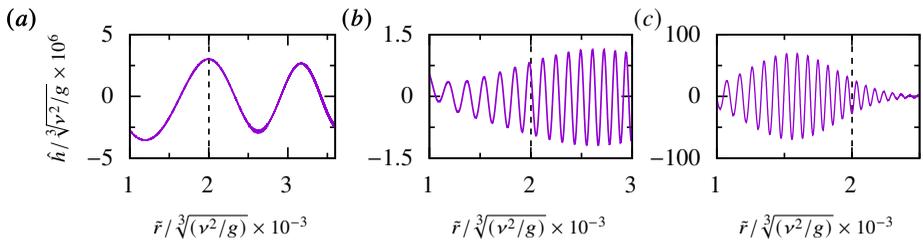}
    \caption{
    Oscillatory part of the film thickness simulated using the $h$-$q$ model.
     $\tilde \beta = 150^\circ$, $\Kap_\perp = 242$, and $N_{q\perp} = 1703$.
    From (a) to (c), $N_{\omega\perp} = 0.026$, 0.128, and 0.257, respectively.
    }
    \label{fig:hhat}
\end{figure}

Moreover, we display the linear wave profiles near a fixed radial distance with different frequencies in figure~\ref{fig:hhat}.
From panels (a) to (c), the normalized disturbance angular frequencies are $N_{\omega\perp}= 0.026$, 0.128, and 0.257, respectively.
With the same radial distance $\tilde{r} / \sqrt[3]{(\nu^2/g)} \times 10^{-3} = 2$, the locations of the three selected situations in the $\tilde \omega$-$\tilde r$ space are marked by the three symbols in figure~\ref{fig:neutral_curves}.
According to figure~\ref{fig:neutral_curves}, at this radius, the flow is stable at $N_{\omega\perp}\equiv\tilde{\omega} \sqrt[3]{\nu/g^2} = 0.026$, unstable at $N_{\omega\perp}=0.128$, and is stable again at $N_{\omega\perp}= 0.257$.
This is indeed true as can be observed from the growth/decay tendency of the wave envelops depicted in figure~\ref{fig:hhat}.

The $h$-$q$ model is further validated in terms of its performance in simulating large-amplitude nonlinear waves.
The results of the model are compared against those of DNS, which is conducted using the interFoam solver embedded in the OpenFOAM-v2412 software.
By employing a volume-of-fluid method, the liquid film is fully resolved in both the streamwise and cross-stream directions.
For the present axisymmetric simulation, the computational domain is a wedge with an angle of $3^\circ$ and one grid layer in the circumferential direction.
The number of cells in the cross-stream direction is typically 48.
The cell aspect ratio is set to less than three.
Therefore, depending on the length of the computational domain, the cell number in the streamwise direction ranges from 5000 to 16000.
The Crank–Nicolson method is used for time marching.
The linear-upwind and van Leer schemes are used for spatial discretization of the convective terms in the momentum equations and the equation for phase fraction evolution, respectively.
The computation is second-order accurate in both time and space.

\begin{figure}
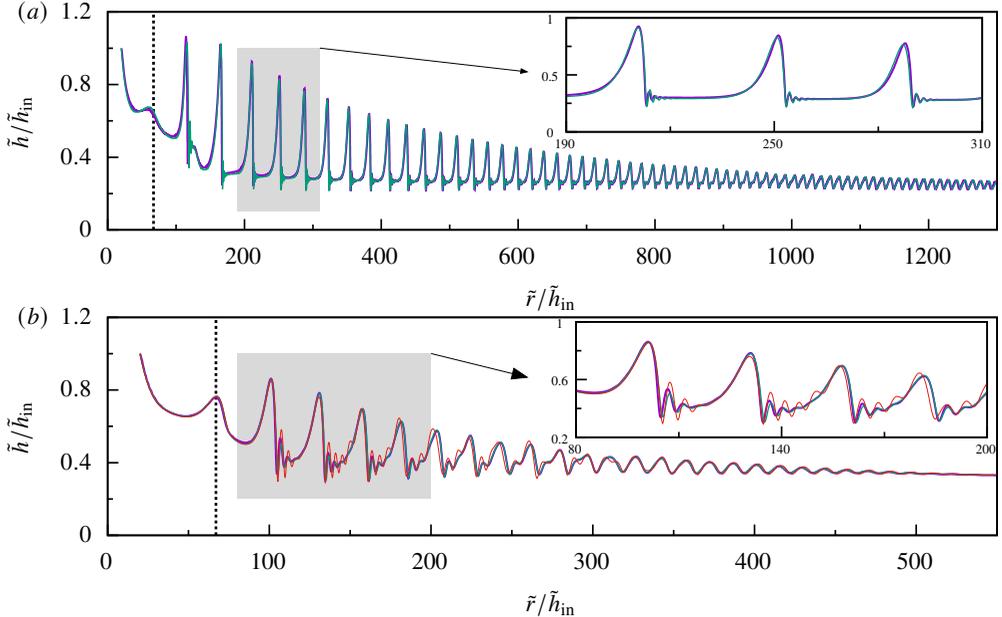

	\centering
    \input{h24.3}
    \input{h6.0}
	\caption{Film thickness versus distance from cone apex at $\Kap_\perp = 242$. 
	(a) $\tilde \beta = 135^\circ$, $N_{q\perp} = 17490$, $N_{\omega\perp} = 0.197$; 
	(b) $\tilde \beta = 110^\circ$, $N_{q\perp} = 11242$, $N_{\omega\perp} = 0.172$.
	Purple lines: DNS;
	green lines: second-order $h$-$q$ model.
	The red line in panel (b) is obtained with the ``first-order'' model of \citet{zhou2022hydraulic}.
    The vertical dashed lines mark the positions where $\tilde{h}_{N\ell} / \tilde{r} = 0.01$.
	}
	\label{fig:validation_wave}
\end{figure}

Figure~\ref{fig:validation_wave} shows the instantaneous shape of the free surface for two cases with large-amplitude waves.
The horizontal coordinate has been highly compressed to more clearly display the surface deformation.
The results of DNS and the $h$-$q$ model are denoted by purple and green lines, respectively.
The waves are generated by setting $\zeta$ in Eq.~\eqref{eq:bc-h-simul} to a relatively large value, typically 0.05.
For the case shown in panel (a),  $\tilde \beta = 135^\circ$, $N_{q\perp} = 17490$, and $N_{\omega\perp} = 0.197$.
For panel (b), $\tilde \beta = 110^\circ$, $N_{q\perp} = 11242$, and $N_{\omega\perp} = 0.172$.
In both cases, $\Kap_\perp = 242$.

Figure~\ref{fig:validation_wave}(a) illustrates a representative case in which solitary waves occur near the inlet.
These waves are characterized by a large main hump and several preceding small capillary ripples within a wave unit.
The steeper slopes render these solitary waves more challenging in numerical simulations than small-amplitude smooth waves.
For this case, a mesh-converged solution requires the number of cells being around 10000 for the $h$-$q$ model.
With this mesh resolution, there are about 350 mesh cells within a single wavelength near the inlet.
In particular, the capillary ripples ahead of the big hump are resolved with 10-15 cells.
As $\tilde r$ increases, the waves convert into sinusoidal-shaped waves whose amplitudes are much smaller.
Throughout the entire computational domain, the result of the $h$-$q$ model is in good agreement with the DNS data.
Notably, for this case, the computational time of the model on a single CPU core is less than 10\% of the DNS simulation, which was conducted in parallel using 64 CPU cores, showing a speedup of two to three orders of magnitude with comparable computational resources.

Figure~\ref{fig:validation_wave}(b) presents a different case in which the main humps of the solitary waves are milder in amplitude.
The result of the ``first-order'' model developed in \citet{zhou2022hydraulic} is also shown (red line).
Both models perform well in the sinusoidal-wave region.
However, in the solitary-wave region, the first-order model incorrectly predicts more capillary waves with larger amplitudes.
\citet[][p. 72]{2011kalliadasis} reported a similar improvement of the second-order model over the first-order one in predicting capillary ripples within the flat-plate scenario.
They attributed this difference to the inclusion of second-order viscous terms in the second-order model.
As a matter of fact, the first-order model developed in \citet{zhou2022hydraulic} is numerically unstable for the case shown in figure~\ref{fig:validation_wave}(a), unless sufficient numerical viscosity is artificially added \citep{zhou2022hydraulic}.
However, for all cases shown in this paper, no special treatment is needed for the second-order model developed in section~\ref{sec:model}.

The dimensionless parameters for the case in figure~\ref{fig:validation_wave}(a) can describe a film flow of a 45\% (w/w) aqueous glycerol solution with $\rho = 1113 \, \mathrm{kg} / \mathrm{m^3}$, $\nu = 5.77 \times 10^{-6} \, \mathrm{m^2} / \mathrm{s}$, and $\sigma = 59.7 \times 10^{-3}\, \mathrm{N} / \mathrm{m}$.
These liquid properties are identical to those in Case C of \citet{Denner2018JFM}.
The volumetric flow rate is $\tilde{Q} = 9.53 \times 10^{-5} \, \mathrm{m^3} / \mathrm{s}$.
The disturbance frequency is $\tilde{f} = 8 \, \mathrm{Hz}$.
The positions of the inlet and outlet of the computational domain, 
when converted into dimensional values, are $\tilde r_{\mathrm{in}} \approx 0.026 \, \mathrm{m}$ and $\tilde r_{\mathrm{out}} \approx 1.66 \, \mathrm{m}$.
The steady-state film thicknesses, based on Eq.~\eqref{eq:local_hN}, are $\tilde h_{N\ell}(\tilde r_{\mathrm{in}}) = 1.28 \, \mathrm{mm}$ and $\tilde h_{N\ell}(\tilde r_{\mathrm{out}}) = 0.32 \, \mathrm{mm}$, respectively.
Correspondingly, the local Reynolds number $\Rey_\ell$, according to Eq.~\eqref{eq:defRel}, decreases from 145 at $\tilde r = \tilde r_{\mathrm{in}}$ to 2.2 at $\tilde r = \tilde r_{\mathrm{out}}$.
Figure~\ref{fig:validation_wave} demonstrates that, although the $h$-$q$ model is formulated under the assumption that $\Rey \sim O(1)$, its accuracy remains acceptable at significantly higher local Reynolds numbers up to the order of $O(10)$-$O(100)$.

\begin{figure}
    \centering
    \input{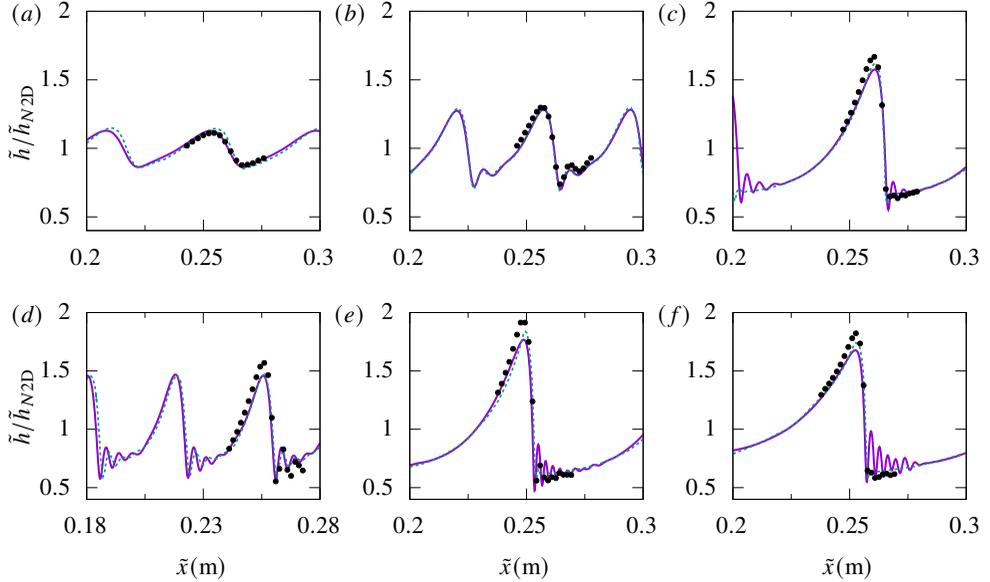}
    \caption{
    Film thickness versus streamwise coordinate.
    The experimental data (symbols) and DNS results (green dashed lines) are extracted from the corresponding panels of figure~9 in \citet{Denner2018JFM};
    the purple solid lines show the results obtained with the present conical $h$-$q$ model in the large-$r$ limit.
    For each panel, horizontal translation is applied for the best match between the present results and those from the original figure.
    From (a) to (f), $\Rey_\mathrm{2D} = $
    5.2, 7.5, 20.6, 12.4, 45.1, and 77.0, respectively.
    }
    \label{fig:denner}
\end{figure}

It would be beneficial to compare the $h$-$q$ model results with experimental data.
As the literature on wavy liquid film flow along conical surfaces are quite limited.
We evaluate the conical model's performance at the flat-plate limit, which requires $r \rightarrow \infty$.
In practice, this is achieved by placing the computational domain in a region where $r$ is extremely large (taken as $\tilde{r}/\tilde{h}_\mathrm{in} > 10^9$ in our simulations).
The wave shapes obtained in this way are compared with experimental results in figure~\ref{fig:denner}, where the parameters of the six flat-plate cases are the same as those shown in figure~9 of \citet{Denner2018JFM}.
The DNS results reported in the same paper are also extracted and plotted for reference. 
As can be seen, within the range of $\Rey_\mathrm{2D} \sim O(10)$ (panels (a), (b), and (d)), the model results are consistent with the experimental data as well as the DNS results. 
For higher Reynolds numbers ($\Rey_\mathrm{2D}$ approaching $O(100)$), the $h$-$q$ model can still reasonably predict the shape of the main humps, although it fails to accurately capture the capillary ripples.

\section{Spatio-temporal evolution of surface waves} \label{sec:nonlinear}

In this section, the spatio-temporal evolution of surface waves is further investigated by using the $h$-$q$ model developed in section~\ref{sec:model} (Eqs.~\eqref{eq:h-nondim} and \eqref{eq:q}), which has been shown to be much more efficient than DNS while retaining comparable accuracy.

\subsection{Morphology of surface waves} \label{sec:wave-evolution}

The instantaneous profiles in figure~\ref{fig:validation_wave} reveal some important characteristics of the waves.
Similar to the linear-theory prediction in section~\ref{sec:wavelength_celerity_linearWave}, the wavelength decreases with the distance from the cone apex.
This is accompanied with the damping of the wave amplitude.
The waves become shorter and weaker, eventually diminishing at large $\tilde r$.
This phenomenon is also consistent with the linear result in section~\ref{sec:linear}, where it is shown that the film flow is stable at large radial distances, see figure~\ref{fig:neutral_curves}.
For small $\tilde r$, the waves are solitary;
as $\tilde r$ increases, a transition to sinusoidal-shaped waves is observed.

\begin{figure}
    \centering
    \input{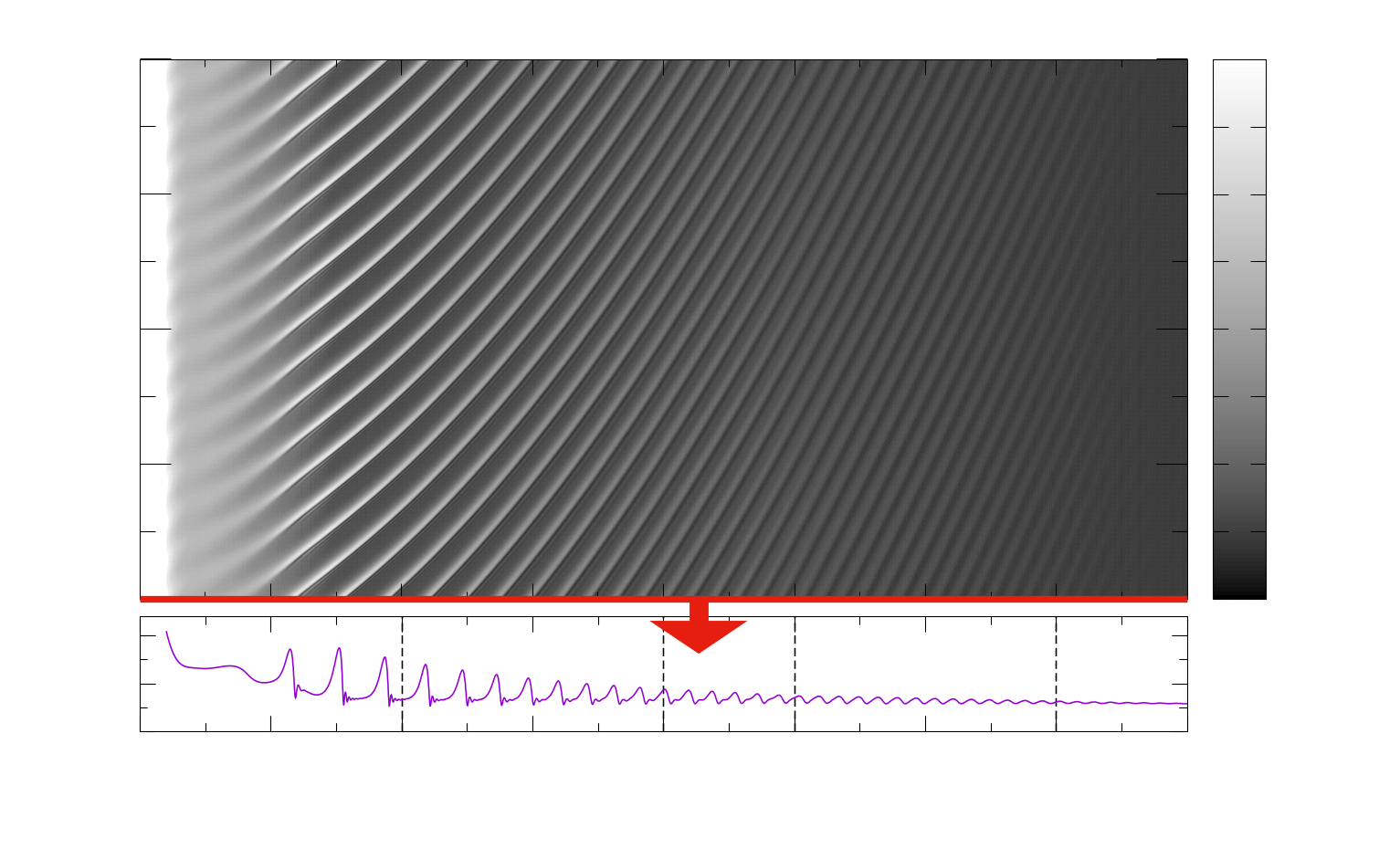}
    \caption{
    Contour map of the spatio-temporal evolution of surface waves;
    $\tilde \beta = 120^\circ$, $\Kap_\perp = 240$, $N_{q\perp} = 12084$, and $N_{\omega\perp}=0.177$.
    The inlet is located at $\tilde r_\mathrm{in}/\tilde h_\mathrm{in}=20$.
    Brightness indicates local film thickness.
    The instantaneous free surface profile at $\tilde t g \tilde h_\mathrm{in}/\nu = 0$ is presented on the bottom.
    The vertical dashed lines show the positions where the data in the four panels of figure~\ref{fig:FFT} are extracted.
    }
    \label{fig:wave_evolution}
\end{figure}

The contour map in figure~\ref{fig:wave_evolution} shows the spatio-temporal evolution of the waves for a representative case we have simulated.
The film thickness is indicated by brightness, so that the wave peaks can be traced following the bright lines in the figure.
The instantaneous shape of the free surface corresponding to $\tilde t = 0$ is presented on the bottom.

The phase speed of the waves, as can be deduced from the slopes of the bright lines in figure~\ref{fig:wave_evolution}, is also a decreasing function of $\tilde r$.
This trend is also the same as that of linear waves, see figure~\ref{fig:linear_wave_characteristic}.
The more curved bright lines at smaller $\tilde r$ indicate larger decreasing rates of the phase speed.
This is because the liquid film spreads out and thins more significantly due to the conical geometry in small-$\tilde r$ regions.
At sufficiently large radial distance, the variation of flow characteristics with $\tilde r$ becomes mild; the film flow approaches the flat-plate limit described by Eqs.~\eqref{eq:h_2d} and \eqref{eq:q_2d}.

\begin{figure}
    \centering
    \input{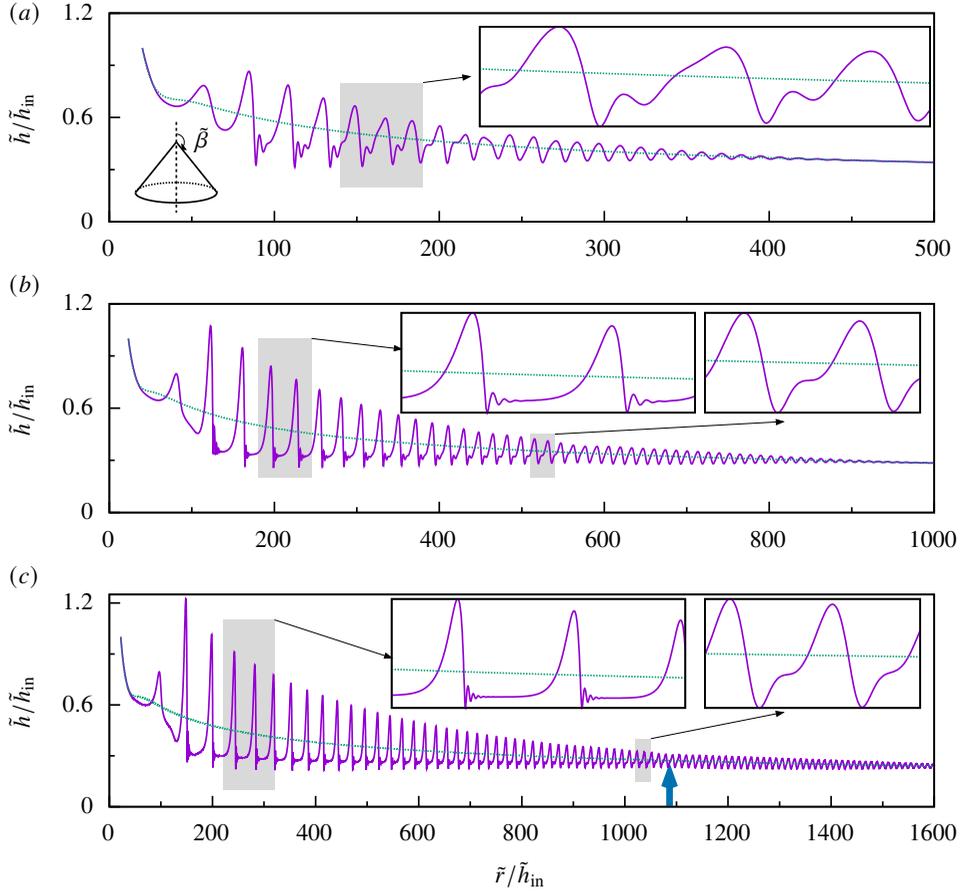}
    \caption{Instantaneous wave profiles for $\tilde \beta = 110^\circ$ (a), $130^\circ$ (b), and $150^\circ$ (c).
    $\Kap_\perp = 242$, $N_{q\perp} = 11242$, and $N_{\omega\perp} = 0.221$.
    Dashed lines show undisturbed free surfaces obtained with the $h$-$q$ model.
    The blue arrow in panel (c) marks the position of the turning point in figure~\ref{fig:hp}.
    A depiction of $\tilde{\beta}$ is displayed in panel (a).
    }
    \label{fig:wave-profile}
\end{figure}

For a more systematic examination, we conducted simulations with $\tilde \beta $ ranging from $100^\circ$ to $150^\circ$ with an increment of $5^\circ$.
The other parameters are set to $\Kap_\perp = 242$, $N_{q\perp} = 11242$, and $N_{\omega\perp} = 0.221$.
These values are selected to be moderate, allowing observation of different types of surface waves as $\tilde{\beta}$ is varied.
Three representative cases are selected from the results with successively increasing angles $\tilde \beta = 110^\circ$, $130^\circ$, and $150^\circ$, see figure~\ref{fig:wave-profile}.
The green dashed lines represent undisturbed free surfaces.
Both the radial and cross-stream coordinates are nondimensionalized by the film thickness $\tilde h_{\mathrm{in}}$ at the inlet.
As can be seen, steeper cones produce larger wave amplitude near the axis, with more solitary waves observed before the waves become sinusoidal.
The number of capillary ripples ahead of the main hump also increases (see insets).
An interesting observation is that, although the steady-state film thickness decreases with $\tilde r$, the envelop of the wave troughs is nearly parallel to the cone surface.

\begin{figure}
    \centering
    \input{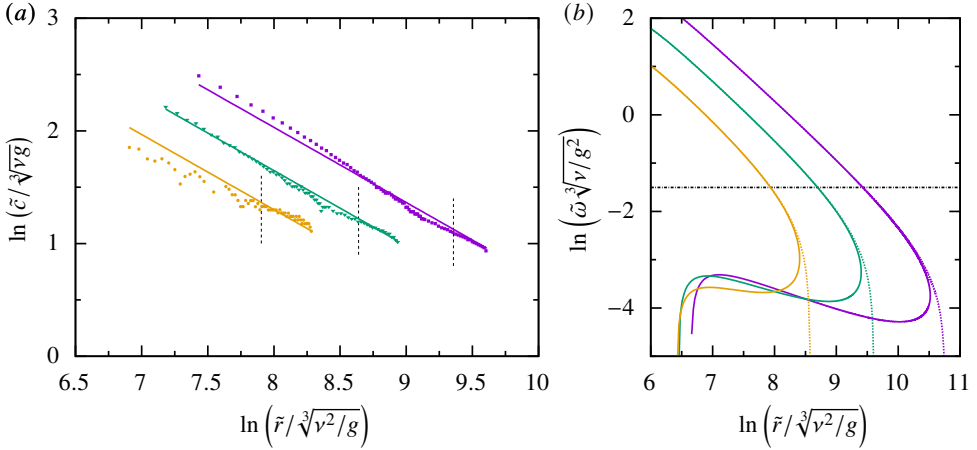}
    \caption{(a) Phase speed versus radial distance.
    Yellow, green, and purple symbols represent the results of $h$-$q$ model for cases in figures~\ref{fig:wave-profile}(a), \ref{fig:wave-profile}(b), and \ref{fig:wave-profile}(c), respectively.
    Straight lines of the same color display the leading-order phase speed predicted by linear theory.
    Vertical dashed lines mark positions where the normalized fundamental-frequency amplitude exceeds $90\%$ (section~\ref{sec:transition}).
    (b) Neutral curves of absolute (solid) and relative (dashed) growth.
    The color scheme is the same as in panel (a).
    The horizontal dash-dotted line shows the disturbance angular frequency $N_{\omega\perp} = 0.221$.
    }
    \label{fig:vel-vs-r}
\end{figure}

The phase speeds of the surface waves for the three cases shown in figure~\ref{fig:wave-profile} are presented in figure~\ref{fig:vel-vs-r}(a).
The symbols are extracted from the results of $h$-$q$ model at discrete radial positions.
The straight lines show the leading-order predictions from linear theory, Eq.~\eqref{eq:celerity-leading}, whose dimensional form reads
\begin{equation}
    \tilde{c}^{(0)} = \left(\frac{3 \tilde{Q}}{2 \pi \tilde{r} \sin\tilde\beta}\right)^{2/3} \left(- \frac{g \cos \tilde \beta}{\nu}\right)^{1/3}.
\label{eq:Nc-linear}
\end{equation}

For all three cases, the phase speed agrees with the linear theory at sufficiently large $\tilde r$, coinciding with the observation in figure~\ref{fig:wave-profile} that the waves convert to small-amplitude linear waves eventually.
In the region where $\tilde r$ is small, however, different behaviors are observed: the waves in figure~\ref{fig:wave-profile}(a) have more pronounced depression, with phase speeds smaller than the leading-order solution of linear waves.
These are the characteristics of the $\gamma_1$-type waves, or slow waves, defined by \citet{chang1993construction, chang1993nonlinear}.
On the contrary, the waves close to the inlet in figure~\ref{fig:wave-profile}(c) are elevation dominant, with phase speeds larger than those of linear waves.
These are $\gamma_2$-type waves, or fast waves, according to \citet{chang1993construction, chang1993nonlinear}.

The occurrence of $\gamma_1$ and $\gamma_2$ waves in different cases can be interpreted with the neutral curves in figure~\ref{fig:vel-vs-r}(b), where the disturbance frequency for the cases in figure~\ref{fig:wave-profile} is shown by the horizontal dash-dotted line.
\citet{liu1993measurements} found that on a flat plate, $\gamma_1$ waves appear when the forcing frequency is just below the cut-off frequency; whereas $\gamma_2$ waves emerge at still lower frequencies.
As illustrated in figure~\ref{fig:vel-vs-r}(b), at a given $\tilde r$, the cut-off angular frequency $\tilde \omega$ increases with $\tilde \beta$.
Therefore, $\gamma_2$ waves are more likely to be observed when $\tilde \beta$ is large.

In the flat-plate situation, $\gamma_1$ waves are unstable \citep[][p.~271]{2011kalliadasis} and easily switch to $\gamma_2$ waves with presence of forced inlet disturbances \citep{liu1994solitary}.
However, this secondary instability can be suppressed on a conical surface: figure~\ref{fig:vel-vs-r}(b) shows that as $\tilde r$ increases, the range of unstable $\tilde \omega$ narrows.
Therefore, disturbances capable of triggering the secondary instabilities may not have sufficient time to amplify before they are damped.
Consequently, for small $\tilde \beta$, the entire free surface can be dominated by $\gamma_1$ waves (figure~\ref{fig:wave-profile}a), whereas for large $\tilde \beta$, a transition from $\gamma_2$ to $\gamma_1$ waves may occur (figure~\ref{fig:wave-profile}c, also see the purple lines and symbols in figure~\ref{fig:vel-vs-r}a).

\subsection{Transition from solitary to sinusoidal waves}
\label{sec:transition}

As shown in section~\ref{sec:wave-evolution}, the surface waves undergo a transition from solitary to sinusoidal-like profiles.
Considering the importance of wave shapes in pertinent heat- and mass-transfer processes~\citep[see, e.g.,][]{Miyara1999, miyara2000numerical, 2022Aktershev}, we elaborate on this transition in detail.

\begin{figure}
	\centering
	\input{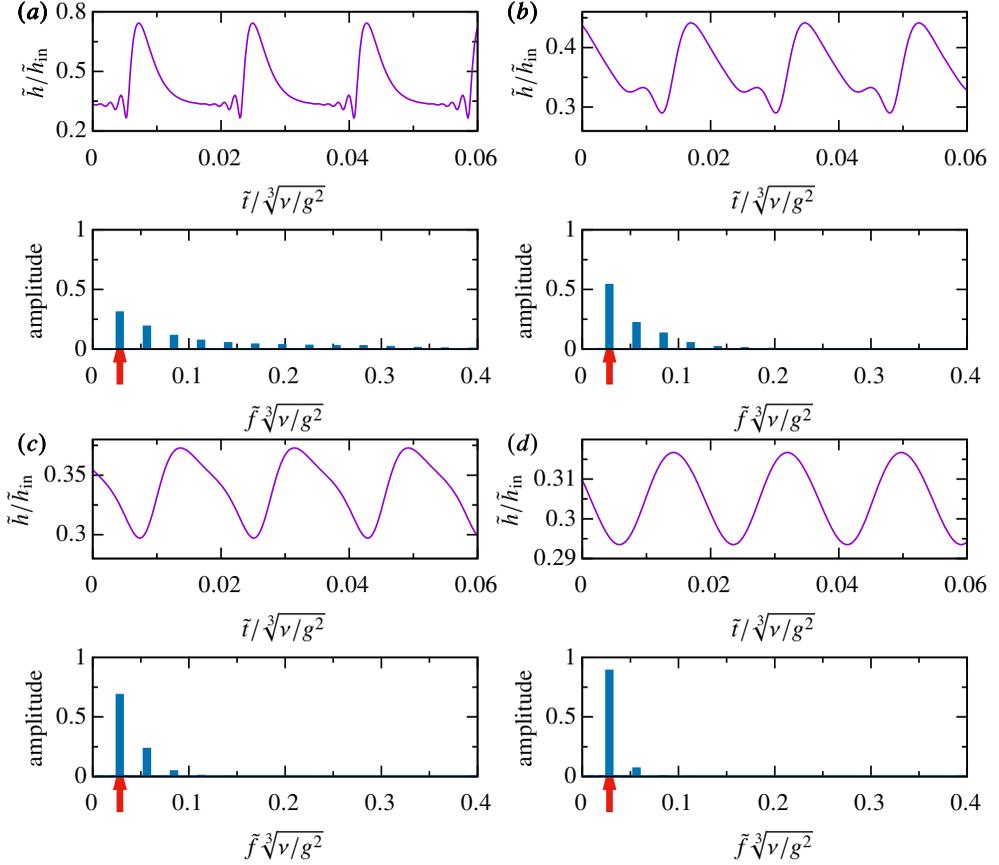}
	\caption{
	Top panels show temporal evolution of film thickness at four radial positions for the case in figure~\ref{fig:wave_evolution}.
	From (a) to (d), $\tilde r/\tilde h_\mathrm{in}=200$, 400, 500, and 700, respectively.
	Bottom panels: corresponding normalized spectra (DC component removed).
	Red arrows indicate frequency of imposed disturbance at inlet.
    }
	\label{fig:FFT}
\end{figure}

The top panels of figures~\ref{fig:FFT}(a-d) present the instantaneous film thickness versus time at four fixed downstream positions with $\tilde r/\tilde h_\mathrm{in}=200$, 400, 500, and 700, respectively.
The case is identical to that shown in figure~\ref{fig:wave_evolution}, in which the four positions are indicated by vertical dashed lines.
The bottom panels show the corresponding Fast Fourier Transform (FFT) spectra based on the periodic time series.
The amplitudes are normalized by their sum for all non-zero frequency components.
The red arrows mark the disturbance frequency imposed at the inlet.
At $\tilde{r} / \tilde h_{\mathrm{in}} = 200$, the waves are solitary, producing multiple spectral peaks in the FFT results.
Further downstream, the wave shape smooths with fewer fluctuations in time.
Correspondingly, the number of visible components in the FFT results decreases, with the fundamental frequency becoming more pronounced.
At $\tilde{r} / \tilde h_{\mathrm{in}}=700$, the temporary wave evolution is almost sinusoidal.
The spectrum is dominated by the fundamental frequency, with most high-order harmonics diminishing.

\begin{figure}
	\centering
	\input{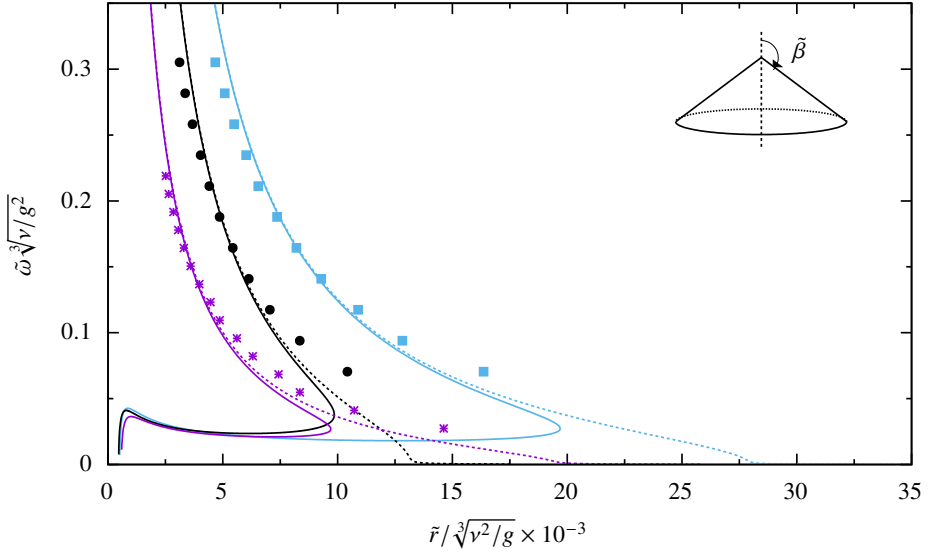}
	\caption{Dimensionless disturbance angular frequency versus radial distance.
	Lines represent neutral curves for absolute (solid) and relative (dashed) growth from linear theory.
	Purple: $\tilde \beta=135^\circ$, $\Kap_\perp=3388$, $N_{q\perp}=11507$;
	black: $\tilde \beta=135^\circ$, $\Kap_\perp=332$, $N_{q\perp}=7762$;
	blue: $\tilde \beta=150^\circ$, $\Kap_\perp=332$, $N_{q\perp}=6722$.
	Symbols of the same colors mark the transition positions where the fundamental frequency amplitude exceeds $90\%$.
	A depiction of $\tilde{\beta}$ is also displayed.
	}
	\label{fig:solitary_to_sinusoidal}
\end{figure}

The transition from solitary to sinusoidal waves is found to exhibit a close correlation with the linear stability threshold.
Figure~\ref{fig:solitary_to_sinusoidal} shows the neutral curves of absolute (solid lines) and relative (dashed lines) growth for cases with different apex angles and liquid properties.
The symbols of the same colors mark the positions where the normalized fundamental spectral amplitude exceeds $90\%$ at given disturbance frequencies.
As can be observed, all symbols roughly align with the neutral curve for relative growth, indicating that the position where sinusoidal waves appear on a conical surface--if a 90\%-amplitude criterion is employed--can be estimated by the position from which the flow becomes stable in the relative sense.
Consequently, when other parameters are fixed, a higher forcing frequency promotes the transition from solitary to sinusoidal waves.

In figure~\ref{fig:vel-vs-r}(a), the radial positions of the $90\%$-amplitude threshold are indicated by short vertical dashed lines.
Clearly, the phase speeds coincide with the linear predictions to the right of these dashed lines.
Therefore, the solitary-sinusoidal wave transition also signals the transition from nonlinear to linear waves.
Indeed, the typical amplitude of the sinusoidal waves observed in our numerical simulations is as small as about $0.2 \%$ of the local wavelength.

\begin{figure}
    \centering
    \resizebox{0.8\linewidth}{!}{\input{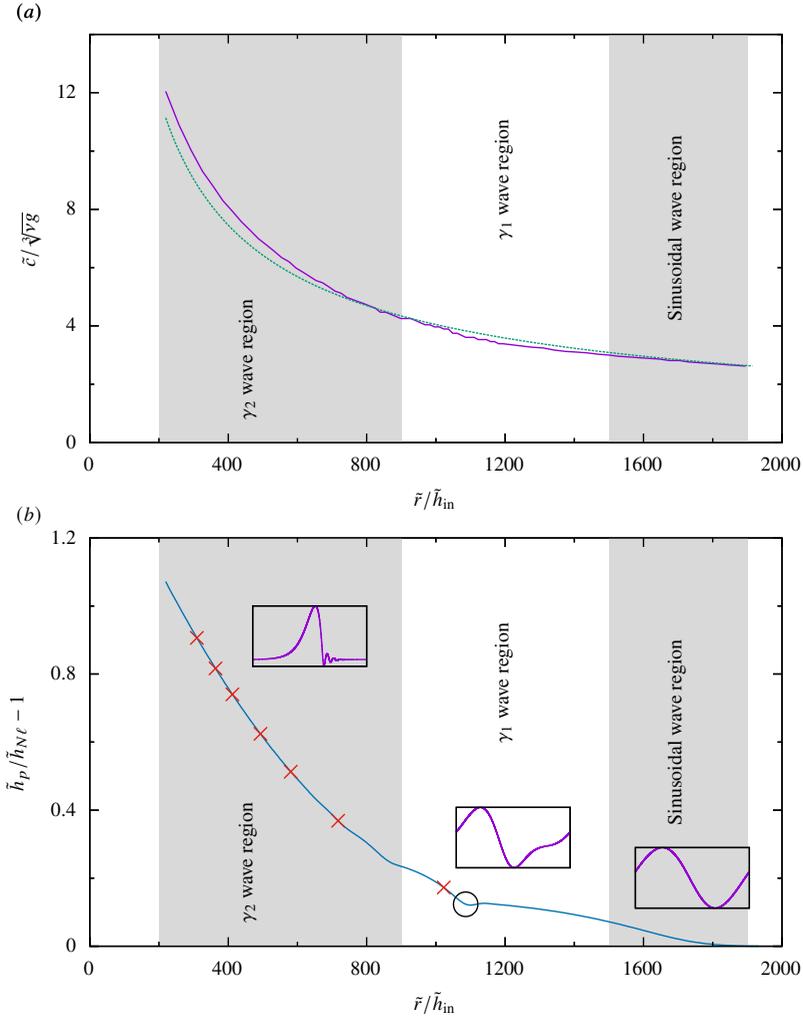}}
    \caption{
    Phase speed (a) and wave peak height (b) versus radial distance for the case in figure~\ref{fig:wave-profile}(c).
    Dashed line in panel (a): leading-order phase speed from linear theory.
    Crosses in panel (b): positions where the number of capillary ripples decreases.
    The black circle highlights the turning point of the curve.
    The insets show typical wave profiles in different regions.
    }
    \label{fig:hp}
\end{figure}

Figure~\ref{fig:hp} shows the phase speed (a) and wave peak height (b) versus $\tilde r$ for the case in figure~\ref{fig:wave-profile}(c).
The peak height is nondimensionalized as $\tilde{h}_p / \tilde{h}_{N\ell} - 1$, where $\tilde{h}_{N\ell}$ is the local Nusselt film thickness given by Eq.~\eqref{eq:local_hN}.
For this representative case, the entire domain can be divided into three distinct regions that are characterized by presence of $\gamma_2$, $\gamma_1$, and sinusoidal waves, respectively.
The typical wave profiles in each region are displayed in the insets of panel (b).
Near the inlet, the forcing frequency lies well below the local cut-off value.
$\gamma_2$ waves develop and travel downstream, with their phase speeds larger than the leading-order linear solution (dashed line in panel a).
At a larger $\tilde r$, the computed and linear phase speeds cross, signaling the transition from $\gamma_2$ to $\gamma_1$ waves.
Further downstream, the $90\%$-amplitude criterion divides the solitary and sinusoidal wave regions.
The position, based on the FFT result, is $\tilde{r} /\tilde h_{\mathrm{in}} \approx 1500$.
The linear theory predicts a relative growth threshold of $\tilde{r} / \tilde h_{\mathrm{in}} \approx 1618$.
The relative error between the two values is less than $8\%$.

\begin{figure}
	\centering
    \input{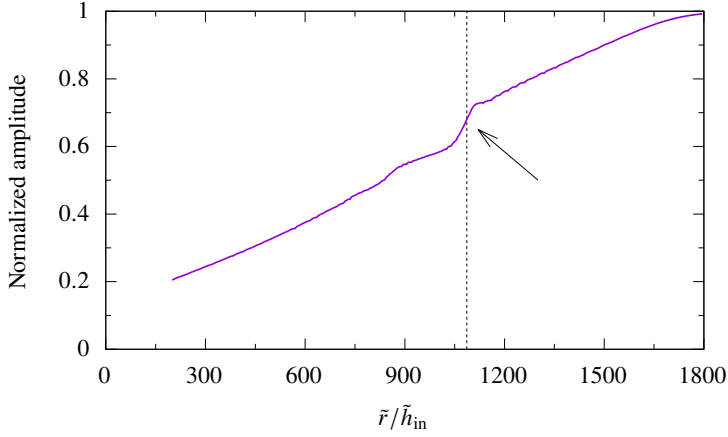}
	\caption{Normalized amplitude of the fundamental frequency versus radial distance for the case in figure~\ref{fig:wave-profile}(c).
	Vertical dashed line shows the position of turning point identified in figure~\ref{fig:hp}(b).
	}
	\label{fig:turningPoint}
\end{figure}

During the entire evolution process, the capillary ripples preceding the main hump diminish successively.
The positions where their number drops are marked by crosses in figure~\ref{fig:hp}(b).
Right after the last capillary wave disappears, the trend of the peak height variation experiences a sudden change at about $\tilde{r}/\tilde h_{\mathrm{in}}= 1086$ (black circle).
This turning point is also marked in figure~\ref{fig:wave-profile}(c) (blue arrow).
Figure~\ref{fig:turningPoint} presents the normalized amplitude of the fundamental frequency as a function of $\tilde r$.
It experiences a significant shift around the turning point (dashed line), as indicated by the black arrow.
Considering the correspondence between the fundamental-frequency amplitude and wave profiles, the turning point can be regarded as the onset of solitary-to-sinusoidal transition.
This will be further evidenced in section~\ref{sec:nonlinear-correlations}.

\subsection{Peak height and phase speed of surface waves} \label{sec:nonlinear-correlations}
For nonlinear two-dimensional solitary waves on a falling liquid film along a vertical plate, \citet{nosoko1996characteristics} proposed empirical correlations for the wave peak height $\tilde h_p$ and phase speed $\tilde c$, which are
\begin{equation}
N_{hp\perp} = 0.49 \Kap_\perp^{-0.132} N_{\lambda\perp}^{0.39} \Rey_{\mathrm{2D}\perp}^{0.46},
\label{eq:Nhp}
\end{equation}
and
\begin{equation}
N_{c\perp} = 1.13 \Kap_\perp^{-0.06} N_{\lambda\perp}^{0.31} \Rey_{\mathrm{2D}\perp}^{0.37},
\label{eq:Nc}
\end{equation}
respectively.
In Eqs.~\eqref{eq:Nhp} and \eqref{eq:Nc}, the vertical version of the viscous-gravity length and time scales mentioned earlier, $l_{\nu\perp} = (\nu^2/g)^{1/3}$ and $t_{\nu\perp} = (\nu/g^2)^{1/3}$, are employed for nondimensionalization, i.e.,
\begin{equation}
N_{hp\perp} = \tilde{h}_p/l_{\nu\perp}, \quad N_{\lambda\perp} = \tilde{\lambda}/l_{\nu\perp}, \quad N_{c\perp} = \tilde{c} t_{\nu\perp}/ l_{\nu\perp} .
\label{eq:nonDim_v_g_scale}
\end{equation}
The Kapitza number $\Kap_\perp = \sigma / ( \rho \nu^{4/3}  g^{1/3} )$ has been defined in section~\ref{sec:linear}.
The Reynolds number $\Rey_{\mathrm{2D}\perp}$ takes the form of Eq.~\eqref{eq:defineRef} with $\tilde \varphi = \pi/2$ in this vertical-plate situation.

Eqs.~\eqref{eq:Nhp} and \eqref{eq:Nc} have shown good consistency with DNS results \citep[see, e.g.,][]{miyara2000numericalsim,2020zhou}.
To apply these correlations to the conical geometry, we (i) replace $g$ with its streamwise component $-g \cos\tilde\beta$, so that $l_\nu$ and $t_\nu$ in Eq.~\eqref{eq:v_g_scaling} are used instead of $l_{\nu\perp}$ and $t_{\nu\perp}$ in Eq.~\eqref{eq:nonDim_v_g_scale}, and $\Kap_\perp$ is changed to $\Kap = \sigma / [ \rho \nu^{4/3}  (-g\cos\tilde\beta)^{1/3} ]$; 
(ii) use the local Reynolds number $\Rey_\ell$ (Eq.~\eqref{eq:defRel}) instead of the constant $\Rey_\mathrm{2D}$.
This leads to
\begin{equation}
N_{hp} = 0.49 \Kap^{-0.132} N_{\lambda}^{0.39} \Rey_\ell^{0.46},
\label{eq:Nhp_adapt}
\end{equation}
and
\begin{equation}
N_{c} = 1.13 \Kap^{-0.06} N_{\lambda}^{0.31} \Rey_\ell^{0.37}.
\label{eq:Nc_adapt}
\end{equation}

\begin{figure}
    \centering
    \input{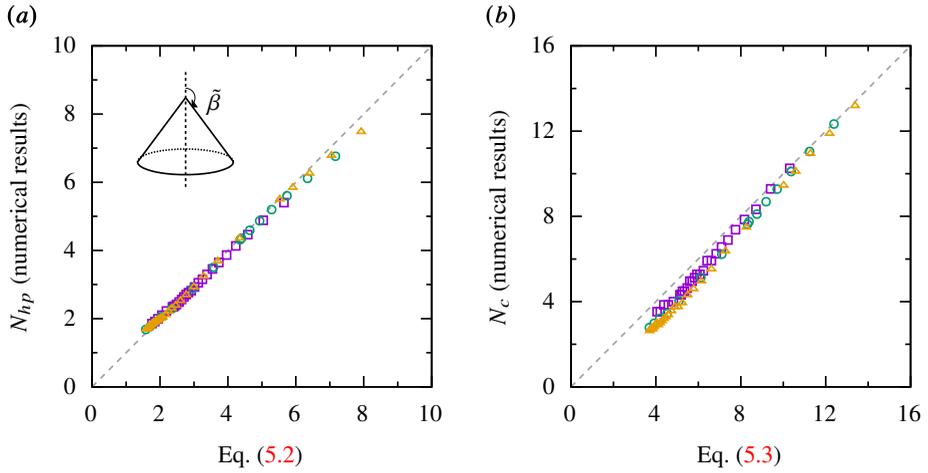}
    \caption{Comparison of peak height (a) and phase speed (b) between the $h$-$q$ model results and Eqs.~\eqref{eq:Nhp_adapt} and \eqref{eq:Nc_adapt};
    $\Kap_{\perp} = 240$ and $N_{\omega\perp} = 0.177$ for all cases.
    Purple squares: $N_{q\perp} = 12084$, $\tilde \beta = 120^\circ$;
    green circles: $N_{q\perp} = 13954$, $\tilde \beta = 135^\circ$;
    yellow triangles: $N_{q\perp} = 12084$, $\tilde \beta = 150^\circ$.
    The dashed diagonal lines represent the points where the values on the horizontal and vertical axes are equal.
    A depiction of $\tilde{\beta}$ is displayed in panel (a).
    }
    \label{fig:Cone_Nosoko}
\end{figure}

Figure~\ref{fig:Cone_Nosoko} compares Eqs.~\eqref{eq:Nhp_adapt} and \eqref{eq:Nc_adapt} with the results of $h$-$q$ model.
The data are extracted at different discrete radial coordinates within the fully developed region, covering a relatively wide range of $N_{hp}$ and $N_{c}$.
In figure~\ref{fig:Cone_Nosoko}(a), the wave peak height $N_{hp}$ calculated with Eq.~\eqref{eq:Nhp_adapt} agrees well with the numerical result, with the largest relative error less than $6\%$.
The consistency is also acceptable for the phase speed $N_c$ as shown in figure~\ref{fig:Cone_Nosoko}(b), especially in the high-$N_c$ region.
According to figure~\ref{fig:hp}(a), the waves in this region are solitary waves.

The slight discrepancy at low $N_c$ therefore arises partly because Nosoko \textit{et al.}'s correlation was obtained in the solitary-wave regime; whereas the phase speed in solitary and sinusoidal regions exhibits different behaviors (figure~\ref{fig:vel-vs-r}a).
A similar mismatch was also reported by \citet{miyara2000numerical} for high-frequency, non-solitary wave cases.
In addition, Nosoko \textit{et al.}'s experiments were conducted for $14 \le \Rey_\ell \le 90$, whereas the data in figure~\ref{fig:Cone_Nosoko} fall in the range $\Rey_\ell \lesssim 15$.
In any case, it is convenient to refit Eq.~\eqref{eq:Nc_adapt} with the present numerical data.
Since it has been shown in section~\ref{sec:transition} that the wave profiles before and after the turning point have difference trends, only the data before the turning point are used for fitting.
The resultant correlation is
\begin{equation} \label{eq:Nc_refit}
    N_c = 0.9 \Kap^{-0.052} N_\lambda^{0.26} \Rey_\ell^{0.58}.
\end{equation}

In the comparison shown in figure~\ref{fig:Cone_Nosoko}, the wavelength $\tilde{\lambda} = \tilde{\lambda} (\tilde{r})$ is extracted from the numerical results by measuring the distance between two neighboring wave peaks.
This $\tilde{r}$-dependent wavelength is unknown a prior in practice, which renders the correlations descriptive rather than predictive. 
Therefore, it is useful to replace the wavelength with the wave frequency, which is constant and known in the current scenario.
We define $N_f = \tilde{f} t_\nu$ as the dimensionless forcing frequency.
By using the relation $\tilde c(\tilde r) = \tilde f \tilde{\lambda}(\tilde r)$, whose dimensionless form is $N_c = N_f N_\lambda$, $N_\lambda$ in Eq.~\eqref{eq:Nc_refit} can be eliminated.
The result is
\begin{equation}
N_c = 0.87 \Kap^{-0.07} N_f^{-0.35} \Rey_\ell^{0.78}.
\label{eq:Nc_Nf}
\end{equation}
Accordingly, with Eq.~\eqref{eq:Nc_Nf}, Eq.~\eqref{eq:Nhp_adapt} can be rewritten as
\begin{equation}
N_{hp} = 0.46 \Kap^{-0.16} N_f^{-0.53} \Rey_\ell^{0.76}.
\label{eq:Nhp_Nf}
\end{equation}
Since $\Rey_\ell \propto 1/\tilde r$ (see Eq.~\eqref{eq:defRel}), $N_c$ and $N_{hp}$ also decay with $\tilde r$ following $N_c \propto \tilde r^{-0.78}$ and $N_{hp} \propto \tilde r^{-0.76}$.

\begin{figure}
    \centering
    \input{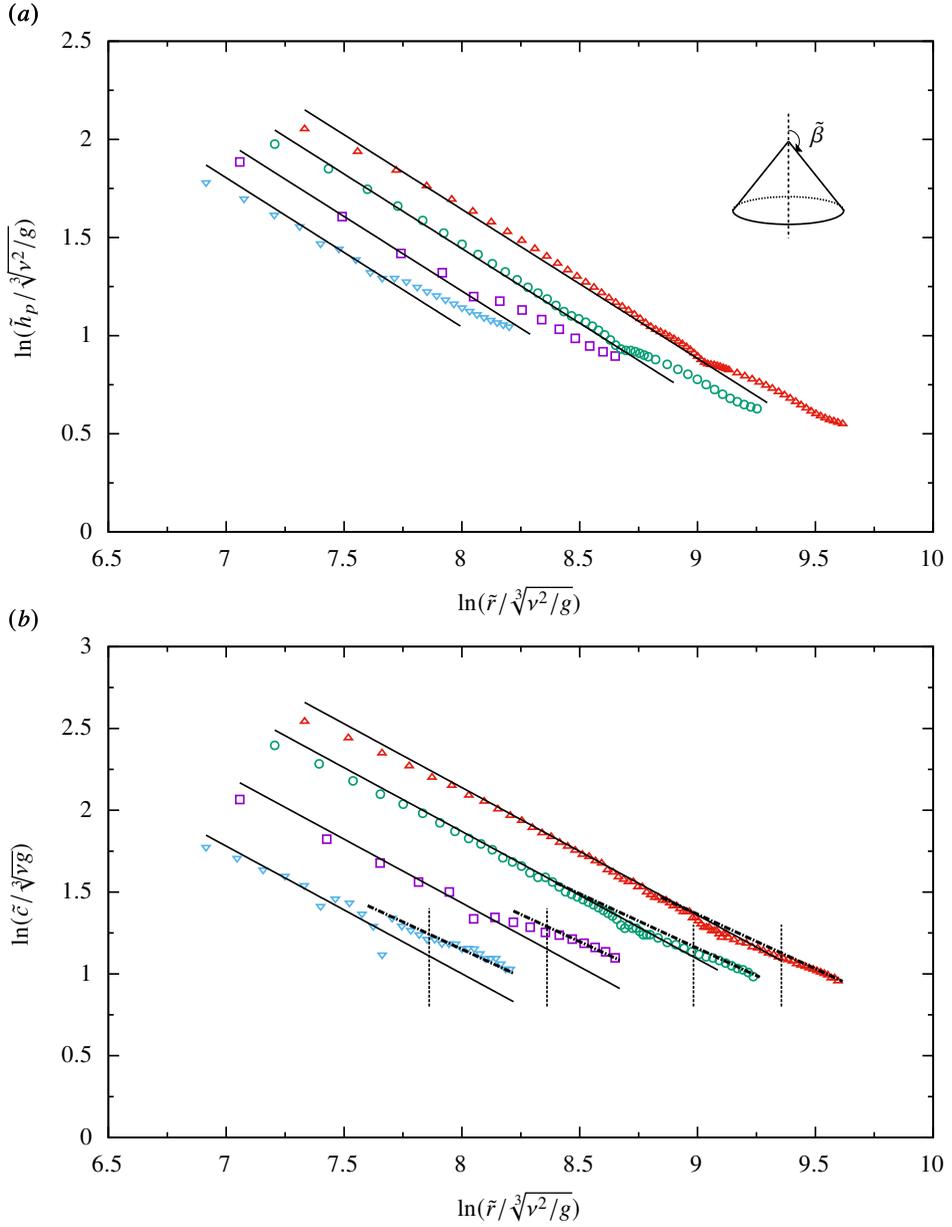}
    \caption{
    Peak height (a) and phase speed (b) versus radial distance.
    Symbols: $h$-$q$ model.
    Solid lines: Eq.~\eqref{eq:Nhp_Nf} and Eq.~\eqref{eq:Nc_Nf}.
    Blue inverted triangles: 
    $\tilde \beta = 110^{\circ}$, $\Kap_{\perp} = 240$, $N_{q\perp} = 8969$, $N_{\omega \perp} = 0.177$;
    purple squares: $\tilde \beta = 120^{\circ}$, $\Kap_{\perp} = 240$, $N_{q\perp} = 12084$, $N_{\omega \perp} = 0.228$;
    green circles: $\tilde \beta = 140^{\circ}$, $\Kap_{\perp} = 242$, $N_{q\perp} = 11242$, $N_{\omega \perp} = 0.221$;
    red triangles: $\tilde \beta = 150^{\circ}$, $\Kap_{\perp} = 242$, $N_{q\perp} = 11242$, $N_{\omega \perp} = 0.221$.
    Dash-dotted lines in panel (b): Eq.~\eqref{eq:Nc-linear};
    vertical dashed lines mark positions where the fundamental frequency amplitude exceeds $90\%$, see section~\ref{sec:transition}.
    A depiction of $\tilde{\beta}$ is displayed in panel (a).
    }
    \label{fig:Nosoko-predict}
\end{figure}

Figure~\ref{fig:Nosoko-predict} shows the peak height (a) and phase speed (b) versus $\tilde r$ for four cases in log-log coordinates.
Good agreement between the results of $h$-$q$ model (symbols) and predictions of Eqs.~\eqref{eq:Nc_Nf} and \eqref{eq:Nhp_Nf} (solid lines) is observed up to the turning points that can be easily located in the figures.
The result of Eq.~\eqref{eq:Nc-linear} is shown with dashed lines in figure~\ref{fig:Nosoko-predict}(b).
The $90\%$-amplitude positions mentioned in section~\ref{sec:transition} are marked with short vertical dashed lines for each case.
As in figure~\ref{fig:vel-vs-r}(a), the linear phase speed coincides well with the numerical results in the sinusoidal-wave region.

In conclusion, the waves ahead of the turning point are solitary and carry preceding capillary ripple(s); their peak height and phase speed can be described by correlations \eqref{eq:Nc_Nf} and \eqref{eq:Nhp_Nf}.
Beyond the $90\%$-amplitude location, the waves are small-amplitude and sinusoidal, and their phase speed can be estimated by linear theory.
The waves between these two positions exhibit transient behavior.

\section{Liquid film flow on the upper surface of an inverted cone} \label{sec:funnel}

\begin{figure}
\centering
\includegraphics[width=7cm]{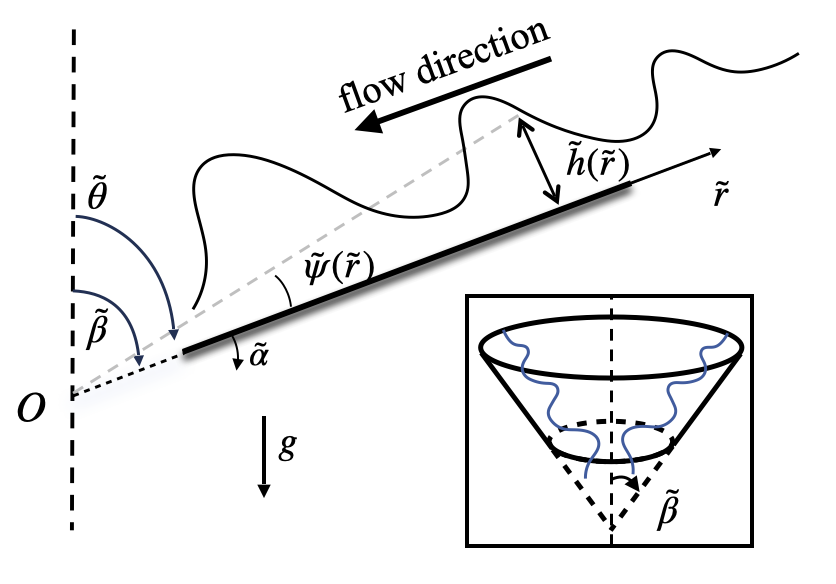}
\caption{
Geometric configuration for liquid film flow along the upper surface of an inverted cone.
The bottom of the cone is open to allow the drainage of liquid.
}
\label{fig:geometry-funnel}
\end{figure}

\begin{figure}
    \centering
    \input{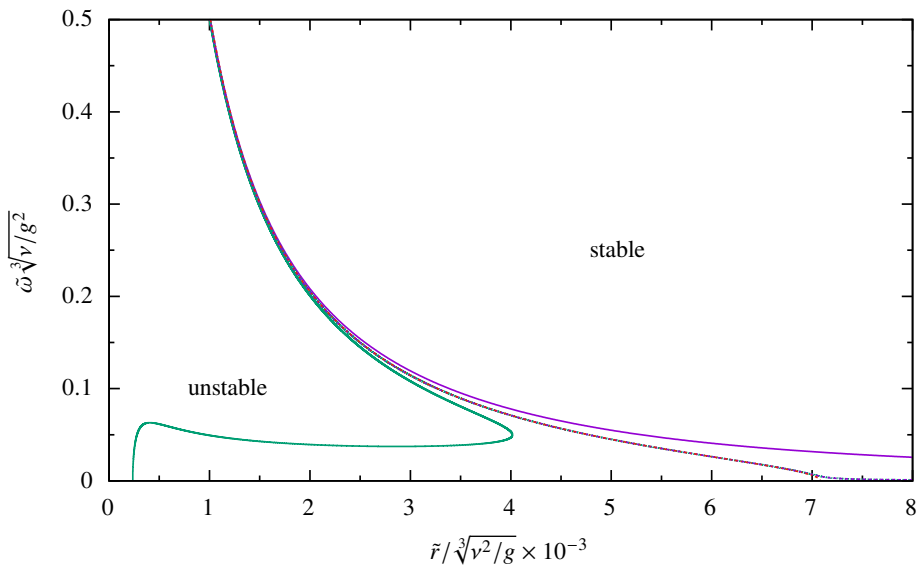}
    \caption{
    Neutral stability curves for absolute (solid lines) and relative (dashed lines) growth.
    Green: cone; purple: inverted cone; red: flat plate.
    The values of $\tilde \beta$, $\Kap_{\perp}$, and $N_{q\perp}$ are the same as in figure~\ref{fig:neutral_curves}.
    }
    \label{fig:funnel_neutral_curve}
\end{figure}

In some industrial applications, such as spinning cone column distillation \citep[see, e.g.,][]{2001Makarytchev, puglisi2022evaluation}, the liquid film exists on the upper surface of a downward-pointing inverted cone.
The geometric configuration is depicted in figure~\ref{fig:geometry-funnel}, with the polar angle $\tilde \beta$ ranging from $0^\circ$ to $90^\circ$.
In this situation, the liquid driven by gravity flows toward the direction in which $\tilde{r}$ decreases, so the volumetric flow rate defined in Eq.~\eqref{eq:q-definition} is negative.
Because spatial stability is evaluated in the flow direction, the amplification factors, Eqs.~\eqref{eq:Ga} and \eqref{eq:Gr}, are replaced by \begin{equation}
    G_a^\prime = - G_a,\quad  G_r^\prime = - G_r,
\end{equation}
respectively.
Using the simplified expressions, Eqs.~\eqref{eq:Ga1st_fullCurv} and \eqref{eq:Gr1st_fullCurv}, one finds
\begin{equation}
    G_r^\prime (-q_s,\pi-\beta) \equiv G_r (q_s,\beta).
\end{equation}
Thus, at the same inclination, the relative-growth stability thresholds for divergent ($\beta>\pi/2$) and convergence ($\beta<\pi/2$) configurations are identical.
As an intermediate situation, the flat-plate liquid film flow is expected to behave similarly (noting that there is no difference between concepts of absolute and relative growth in that situation).
This is confirmed in figure~\ref{fig:funnel_neutral_curve}, where the relative growth neutral curves for the cone (green dashed line), inverted cone (purple dashed line), and flat plate (red dashed line) are indistinguishable when the inclination angles are the same.
Indeed, by substituting $\tilde q_s = \tilde q_\mathrm{2D} \tilde r \sin\tilde\beta$, $G_r$ in Eq.~\eqref{eq:Gr1st_fullCurv} reduces to the flat-plate growth rate plus some $O(\tilde r^{-2})$ terms that decay rapidly as $\tilde r$ increases.

In contrast, the absolute-growth neutral curves differ markedly between the situations of the right cone (green solid line) and inverted cone (purple solid line).
The latter decreases monotonically with $\tilde r$ and lies slightly above the relative-growth curves.
As expected, for the inverted cone on which the liquid converges towards the apex with film thickening, it is more stringent for the flow to be stable in the sense of absolute growth than relative growth.

\begin{figure}
    \centering
    \input{wave-funnel}
    \input{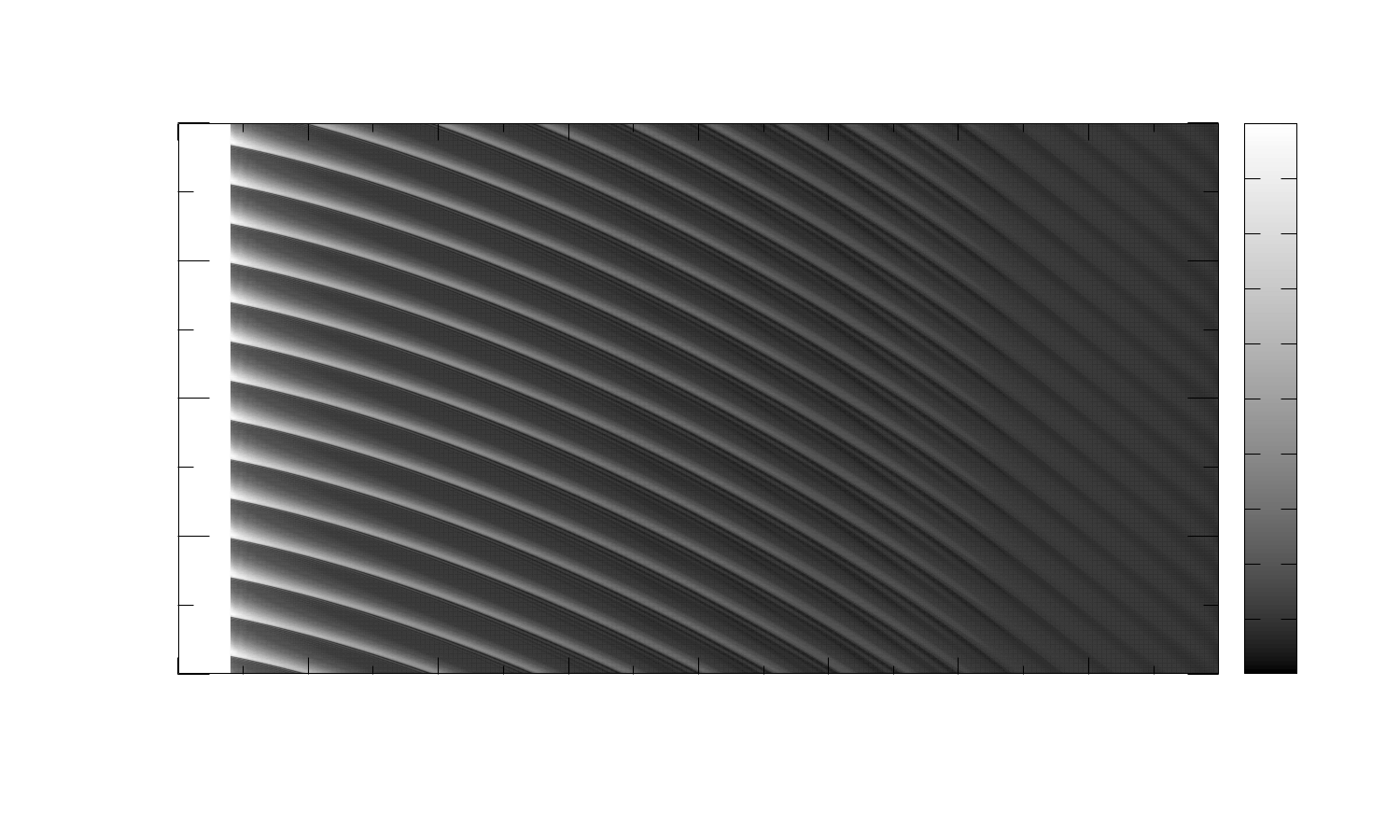}
    \caption{(a) Shape of free surface for liquid film flow on the upper surface of an inverted cone ($\tilde \beta = 60^\circ$).
    (b) Contour map of spatio-temporal wave evolution for the same case.
    Brightness indicates local film thickness.
    Liquid properties and disturbance frequency are the same as in figure~\ref{fig:wave_evolution}.}
    \label{fig:wave_evolution_funnel}
\end{figure}

The characteristics of large-amplitude surface waves obtained with the $h$-$q$ model are illustrated in figure~\ref{fig:wave_evolution_funnel}.
Panels (a) and (b) present the instantaneous wave profile and spatio-temporal evolution, respectively.
The liquid enters the computational domain at $\tilde r_\mathrm{in} / \tilde h_{\mathrm{out}} = 400$ and leaves at $\tilde r_\mathrm{out} / \tilde h_{\mathrm{out}} =20$ ($\tilde r_\mathrm{out}$ and $\tilde h_{\mathrm{out}}$ are related through Eq.~\eqref{eq:hbar-leading}, or its dimensional form, Eq.~\eqref{eq:local_hN}).
The wavelength, peak height, and phase speed all increase with decreasing $\tilde r$, mirroring the trends observed in figure~\ref{fig:wave_evolution}.

\begin{figure}
    \centering
    \input{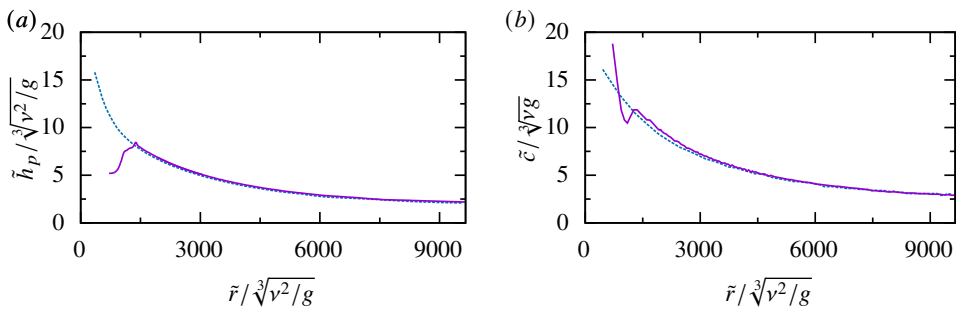}
    \caption{
    Peak height (a) and phase speed (b) of surface waves on a liquid film along the upper surface of a cone (purple solid lines) and an inverted cone (blue dashed lines);
    $\Kap_{\perp} = 240$, $N_{\omega\perp} = 0.177$, and $N_{q\perp} = 13954$.
    The polar angles for the cone and inverted cone are supplementary, with their values of $\tilde \beta=135^\circ$ and $45^\circ$, respectively.
    }
    \label{fig:funnelCone}
\end{figure}

The two situations with supplementary polar angles are further compared quantitatively in figure~\ref{fig:funnelCone}.
The purple solid lines represent the right-cone case with $\tilde \beta = 135^\circ$.
The blue dashed lines denote the inverted-cone case with $\tilde \beta=180^\circ - 135^\circ = 45^\circ$.
For both the peak height (a) and phase speed (b) of the surface waves, the two lines almost coincide at large $\tilde r$.
The discrepancy at small $\tilde r$ is due to the transient effect close to the inlet of the right-cone situation.

The nearly identical behaviors in the fully developed region echoes figure~\ref{fig:funnel_neutral_curve}:
inverting the cone has negligible influence on wave properties provided that the film remains on the upper surface. 
This feature can be explained by the long-wave nature of the flow field, which means the film thickness, as well as the volumetric flow rate per unit width, changes slowly in the radial direction.
As a consequence, the local flow field on either a right cone or an inverted cone resembles the two-dimensional flat-plate case whose parameters are the same as the local values of the conical ones.

That being said, because of inertial effects, the flow field adjust to the local geometry more slowly than the geometry itself changes.
Therefore, the properties tend to retain upstream information, especially in the small $\tilde r$ region where the divergence/convergence is stronger.
This effect can be somewhat observed from figure~\ref{fig:funnelCone}, where the solid lines are slightly above the dashed lines.

\begin{figure}
    \centering
    \input{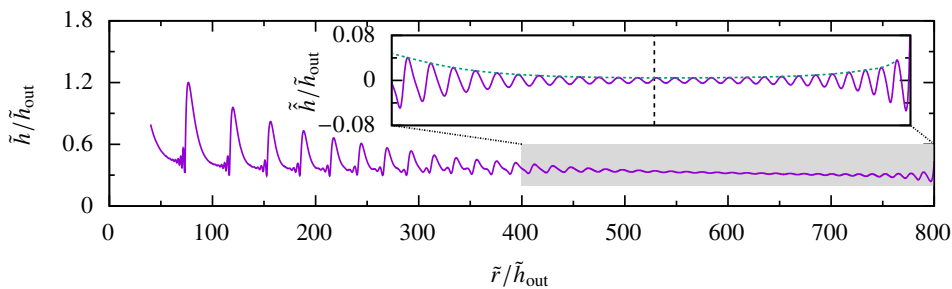}
    \caption{
    Shape of free surface for liquid film flow on the upper surface of an inverted cone.
    The disturbance is imposed at $\tilde{r} / \tilde h_{\mathrm{out}} = 800$.
    Liquid properties and disturbance frequency are the same as in figure~\ref{fig:wave_evolution}.
    Inset: $\tilde {\hat h}$ (defined in Eq.~\eqref{eq:h_disturbed}) as a function of $\tilde r$.
    The transition position is determined as the position where the wave amplitude is minimum.
    }
    \label{fig:funnel_stability}
\end{figure}

An implication of figure~\ref{fig:funnel_neutral_curve} is illustrated in figure~\ref{fig:funnel_stability}: for an inverted cone, an inlet disturbance may first decay and then grow at smaller $\tilde r$ where the flow becomes unstable to the imposed frequency.
The inset locates the minimum-amplitude position at $\tilde{r} / \tilde h_{\mathrm{out}} \approx 614$, within $6\%$ deviation from the linear-theory stability threshold 650.
The free surface around this location is nearly flat.
A reduction of heat- and mass-transfer rate across the film is expected in this region.

\section{Further considerations on the long-wave expansion}
\label{sec:furtherCons}
\subsection{Validity range of the theory}
\label{subsec:valiRang}

Stemming from the long-wave assumption, both the linear stability analysis and the $h$-$q$ model are expected to be valid for $\varepsilon =H/L \ll 1$.
As a conservative estimate, we set $H/L \lesssim 0.01$.
Replacing $L$ with the radial distance $\tilde{r}$ and $H$ with the local film thickness estimated using Eq.~\eqref{eq:local_hN}, one obtains an inequality for $\tilde{r}$, whose solution reads
\begin{equation}
    \tilde{r} \gtrsim 100^{3/4} \left(- \frac{3 \nu \tilde Q}{2 \pi g \sin \tilde \beta \cos \tilde \beta }  \right)^{1/4}.
    \label{eq:rCrit}
\end{equation}
For all data shown in the present paper, the minimal radial distance where they are extracted is taken around the threshold predicted by Eq.~\eqref{eq:rCrit}.
In figures~\ref{fig:validation_wave}(a) and \ref{fig:validation_wave}(b), these thresholds for $\tilde r$ are indicated by vertical dashed lines.

Another assumption that is made during the derivation is the smallness of the Reynolds number, i.e., $\Rey \sim O(1)$.
However, comparisons with DNS (figure~\ref{fig:validation_wave}) and experimental results (figure~\ref{fig:denner}) demonstrate that, the current $h$-$q$ model works consistently well up to $\Rey \lesssim 10$; and major wave properties can still be captured at $\Rey \lesssim 100$.
Such good performance ``beyond expectation'' has been richly documented for the $h$-$q$ model families derived from the long-wave expansion in the flat-plate scenario.

\subsection{A double-$\varepsilon$ expansion strategy} \label{sec:doublee}
In the Navier-Stokes equations \eqref{eq:continuity}-\eqref{eq:momentum-theta}, there are two ways in which the radial distance $\tilde r$ is involved: it may appear alone (e.g., $\tilde r$, $\tilde r^{-1}$, and $\tilde r^{-2}$) or in the denominator of partial derivatives (e.g., $\partial/\partial \tilde r$ and $\partial^2/\partial \tilde r^2$).
Strictly speaking, $\tilde{r}$ in the two scenarios should not be normalized using the same characteristic length: while an individual $\tilde{r}$ reflects the influence of the radial distance itself on the flow, the derivatives of physical quantities with respect to $\tilde{r}$ represent how quickly these quantities vary with $\tilde{r}$.
The length scale across which these variations occur is the local wavelength, yet is irrelevant to $\tilde r$'s absolute value.
Accordingly, an individual $\tilde{r}$ should be normalized by its own scale, whereas $\tilde{r}$ in the bottom of derivatives is more appropriate to be normalized using the wavelength\footnote{An exception is the situation in which the independent variable $\tilde r$ appears in the numerator of the derivative, as the length scale for the change of $\tilde r$ is itself ($L_1$), instead of the wavelength $L_2$.
For example, as is evident, $\partial \tilde r / \partial \tilde r$ should have the order $L_1/L_1 \sim O(1)$ instead of $L_1/L_2 \sim O(\varepsilon_2/\varepsilon_1)$.
Thus, care should be taken when dealing with terms such as $\partial \tilde \psi/\partial \tilde r \approx -\partial \tilde (\tilde  h/ \tilde r)/\partial \tilde r$, in which case the normalization should be performed after expansion, i.e., after $\tilde r$ is removed from the numerator of the partial derivative.
}.
Based on this consideration, the nondimensionalization and long-wave expansion of the Navier-Stokes equations can be refined as follows.

Let $L_1$ and $L_2$ denote the characteristic length scales for the radial distance and wavelength, respectively. 
Two, instead of one, small parameters can be defined as
\begin{equation}
    \varepsilon_1 = \frac{H}{L_1} \ll 1 , \quad \text{and} \quad \varepsilon_2 = \frac{H}{L_2} \ll 1.
\end{equation}
Since $|\tilde \psi| \approx  \tilde h/ \tilde r $, $\tilde \psi$ as well as $\tilde \alpha $ is of order $O(\varepsilon_1)$.
With $u =\tilde u/U$ and $v = \tilde v/V$, the balance of continuity equation leads to $V/U \sim \varepsilon_2$. 
Thus, the following scalings are adopted:
\begin{equation} \label{eq:scaling}
	\tilde{r} = L_1 r, \quad \frac{\partial}{\partial \tilde{r}} = \frac{1}{L_2} \frac{\partial}{\partial r}, \quad \tilde{\alpha} = \varepsilon_1 \alpha,  \quad \tilde{u} = U u, \quad \tilde{v} = \varepsilon_2 U v, \quad \tilde{t} = \frac{L_2 t}{U}, \quad \tilde{p} = \rho g H p.
\end{equation}
On substitution, the dimensionless Navier-Stokes equations read
\begin{equation} \label{eq:cont-eps}
    \frac{\partial u}{\partial r} + \frac{1}{r} \frac{\partial v}{\partial \alpha} + 2 \frac{\varepsilon_1}{\varepsilon_2} \frac{u}{r} + \varepsilon_1 \frac{v}{r} \cot\beta - \varepsilon_1^2 \frac{v}{r} \alpha \csc^2\beta = 0,
\end{equation}
\begin{equation} \label{}
    \begin{aligned}
        & \varepsilon_2 Re \frac{\partial u}{\partial t} + \varepsilon_2 Re u \frac{\partial u}{\partial r} + \varepsilon_2 Re \frac{v}{r} \frac{\partial u}{\partial \alpha}
        = -\cos\beta + \varepsilon_1 \alpha \sin\beta + \frac{1}{2} \varepsilon_1^2 \alpha^2 \cos\beta - \varepsilon_2 \frac{\partial p}{\partial r} \\
        & + \varepsilon_1 \varepsilon_2 \frac{\partial^2 u}{\partial r^2} + \varepsilon_1 \varepsilon_2 \frac{2}{r} \frac{\partial u}{\partial r} + \frac{1}{r^2} \frac{\partial^2 u}{\partial \alpha^2} + \varepsilon_1 \frac{\cot\beta}{r^2} \frac{\partial u}{\partial \alpha} - \varepsilon_1^2 \frac{\alpha \csc^2\beta}{r^2} \frac{\partial u}{\partial \alpha} - \varepsilon_1^2 2 \frac{u}{r^2} - \varepsilon_1 \varepsilon_2 \frac{2}{r^2} \frac{\partial v}{\partial \alpha},
    \end{aligned}
\end{equation}
and
\begin{equation} \label{}
    0 = \sin\beta + \varepsilon_1 \alpha \cos\beta - \frac{1}{r} \frac{\partial p}{\partial \alpha} + \varepsilon_2 \frac{1}{r^2} \frac{\partial^2 v}{\partial \alpha^2} + \varepsilon_1 \frac{2}{r^2} \frac{\partial u}{\partial \alpha}.
\end{equation}
The no-slip boundary condition at the wall is the same as Eq.~\eqref{eq:noSlipBC_dimless}.
And the kinematic boundary condition at the free surface becomes
\begin{equation}
    \frac{\partial h}{\partial t} - \frac{\varepsilon_1}{\varepsilon_2} \frac{h u}{r} + u \frac{\partial h}{\partial r} + v = 0.
\end{equation}
The tangential and normal stress balances at the free surface are given by
\begin{equation}
    \begin{aligned}
        & \frac{1}{r} \frac{\partial u}{\partial \alpha} = 2 \left( \varepsilon_1 \frac{\partial u}{\partial r} - \varepsilon_1 \frac{u}{r} - \varepsilon_2 \frac{1}{r} \frac{\partial v}{\partial \alpha}\right) \left(- \varepsilon_2  \frac{\partial h}{\partial r} + \varepsilon_1 \frac{h}{r}\right) - \varepsilon_1\varepsilon_2 \left( \frac{\partial v}{\partial r} + \frac{v}{r} \right)
    \end{aligned}
\end{equation}
and
\begin{equation}
    \begin{split}
        & - p + 2 \left(\varepsilon_2 \frac{1}{r} \frac{\partial v}{\partial \alpha} + \varepsilon_1 \frac{u}{r}\right) =  \left(\frac{\varepsilon_2}{Bo}\right) \mathcal{C},
    \end{split}
\end{equation}
respectively, where the curvature $\mathcal{C}$ is
\begin{equation} \label{eq:curv-eps}
    \mathcal{C} = \frac{\varepsilon_1}{\varepsilon_2} \frac{1}{r} \cot\beta + \frac{\varepsilon_1^2}{\varepsilon_2} \frac{h}{r^2} \cot^2\beta + \varepsilon_1 \frac{1}{r} \frac{\partial h}{\partial r} + \varepsilon_2 \frac{\partial^2 h}{\partial r^2}.
\end{equation}
It is worthwhile to note that, when $\varepsilon_1 \sim \varepsilon_2 \sim \varepsilon$, which correspond to regions not too far from the cone apex, Eqs.~\eqref{eq:cont-eps}-\eqref{eq:curv-eps} reduce to Eqs.~\eqref{eq:nondim-continuity}-\eqref{eq:curv}.
Whereas for positions at which the radial distance is much larger than the local wavelength, i.e., $\varepsilon_1 \ll \varepsilon_2$, the effect of $\varepsilon_1$ terms in Eqs.~\eqref{eq:cont-eps}-\eqref{eq:curv-eps} diminishes or even vanishes in the present second-order framework. 
In such a scenario, Eqs.~\eqref{eq:nondim-continuity}-\eqref{eq:curv} actually contain extra high-order (higher than second order) $\varepsilon_1$ terms, representing a more conservative approximation.

With the above preparations, we are able to examine the impact of different terms in the expression of curvature, Eq.~\eqref{eq:curv-eps}.
In this equation, the first term is the lateral curvature whose order is $O(\varepsilon_1/\varepsilon_2)$.
The last three terms represent the contribution of the streamwise curvature,
with their orders of magnitude being $O(\varepsilon_1^2/\varepsilon_2)$, $O(\varepsilon_1)$, and $O(\varepsilon_2)$, respectively.
If $\varepsilon_1 \sim \varepsilon_2$, the lateral curvature is dominant.
However, as $\tilde r$ increases (thus $\varepsilon_1$ decreases), its influence weakens, rendering the last term of the streamwise curvature pronounced.
In other words, the present double-$\varepsilon$ expansion approach reveals that, the lateral curvature is a dominant quantity only in the region close to the apex.
It degrades to higher orders as $\tilde r$ increases.
Physically speaking, this behavior reflects the vanishing effect of cone surface curvature far from the apex.

\section{Summary and conclusions} \label{sec:conclusions}

In this study, the linear and nonlinear wave dynamics of axisymmetric thin liquid film flow on a conical surface are systematically investigated.
The linear stability result of \citet{1980Zollars} is improved by including the streamwise curvature of the free surface, which is found to have a non-negligible stabilizing effect.
A second-order $h$-$q$ model is derived using the weighted residual integral boundary-layer approach.
The model is shown to predict the wave dynamics with good accuracy and efficiency at moderate Reynolds numbers.

Unlike the liquid film flow on an inclined flat plate which is irrelevant to the streamwise coordinate, the characteristics of surface waves on a cone exhibit strong dependence on the radial distance $\tilde r$ from the apex.
Together with the progressively thinning of the film due to the divergent geometry, a transition from solitary to sinusoidal waves is observed as $\tilde r$ increases.
The transition position is found to coincide with the stability threshold in the sense of relative growth.
The peak height and phase speed of the surface waves also decrease with $\tilde r$.
In the solitary-wave region, these quantities can be predicted by empirical correlations similar to those proposed by \citet{nosoko1996characteristics};
whereas the phase speed of the sinusoidal waves are well described by the linear theory.

Additionally, results are compared with liquid film flow on the upper surface of an inverted cone whose half-apex angle is supplementary to that of the right cone.
Differences in most linear and nonlinear wave characteristics are small, indicating negligible inertial influence on wave evolution.
This observation implies the applicability of quantitative flat-plate results in the conical scenarios, provided the local film thickness, flow rate, and Reynolds number $\Rey_\ell$ are substituted.
Because $\Rey_\ell$ varies continuously along the conical surface, a single experiment or numerical simulation yields a data set spanning a wide range of $\Rey_\ell$.
This data can then be used to efficiently develop and validate correlations for the flat-plate situation.

In industrial applications, liquid film flow is often accompanied by heat and mass transfer across the film \citep[see, e.g.,][]{tanguy2015concentration, 2020zhou}.
These processes are jointly influenced by film thickness and surface-wave dynamics.
In general, higher heat- and mass-transfer efficiency is favored by thinner films and more pronounced solitary waves, yet the two compete on a conical surface.
In film coating operations \citep[e.g.][]{1999Quere, 2008Benilov, 2024Molefe, 2024Pino}, adjustment of film thickness is sometimes required \citep{2006Lacanette}.
Therefore, research on optimization and control strategies \citep[see, e.g.][]{2020Aniszewski, 2021Mendez, 2024Barreiro} for conical liquid film flow is needed in the future.
Coating flows are also seen in natural environments, such as in the formation of karst landscapes. 
The stability of such flows and the associated surface waves are believed to play a crucial role in shaping speleothem morphology within limestone caves \citep{2012Camporeale, Ledda2021, ledda2022}. 
These speleothems, particularly stalagmites and stalactites, often exhibit elongated conical shapes, whose formation involves low-Reynolds-number flows \citep{2005Short}.
Therefore, the methods and discussions proposed in the current work are potentially useful in hydrodynamic studies related to stalagmite/stalactite growth.

Compared with stationary cones, rotating geometries exhibit even richer phenomena \citep[see, e.g.,][]{kim2024low, stafford2025thin} and have broad industrial applications \citep[e.g.,][]{duruk2023three, shepherd2024general}. 
Linear stability, nonlinear wave evolution, and quantitative wave characteristics on rotating cones warrant further investigation.

The current work focuses on the response of liquid films under axisymmetric disturbances.
In practice, transition to fully three-dimensional flows may occur due to lateral disturbances.
Future work is needed to extend the present analysis and $h$-$q$ model to fully three-dimensional versions.
In that case, the superiority of low-dimensional models over 3D DNS simulations would be more significant.

\backsection[Acknowledgements]{The authors thank Prof. Andrea Prosperetti for insightful discussions during the preparation of this work.}

\backsection[Funding]{The authors acknowledge the support of the National Natural Science Foundation of China (12202441, 12572298) and the Fundamental Research Funds for the Central Universities.}

\backsection[Declaration of interests]{The authors report no conflict of interest.}

\appendix
\section{Derivation of the curvature at the free surface} \label{ap:curv}
The position of the free surface can be expressed by the equation $\tilde \alpha - \tilde \psi(\tilde r,  \tilde t) \equiv 0$.
Accordingly, the unit normal vector is given by
\begin{equation}
    \tilde{\boldsymbol{n}}
    = \frac{\nabla (\tilde \alpha - \tilde \psi)}{\lVert \nabla (\tilde \alpha - \tilde \psi)\rVert} = \frac{1}{\sqrt{1 + (\tilde r \partial_{\tilde r} \tilde{\psi})^2}} 
    \begin{bmatrix}
    -\tilde r \partial_{\tilde r} \tilde \psi, 1, 0
    \end{bmatrix}^T.
\end{equation}
The curvature is calculated as
\begin{equation}
    \begin{aligned}
        \tilde{\mathcal{C}} = \nabla \cdot \tilde{\boldsymbol{n}}
        = & \frac{1}{\tilde{r}^2} \frac{\partial}{\partial \tilde{r}} \left[\tilde{r}^2 \left(- \frac{\tilde{r} \partial_{\tilde{r}} \tilde{\psi}}{\sqrt{1 + (\tilde{r} \partial_{\tilde{r}} \tilde{\psi})^2}}\right)\right] + \frac{1}{\tilde{r} \sin\tilde{\theta}} \frac{\partial}{\partial \tilde{\theta}}\left( \frac{\sin \tilde{\theta}}{\sqrt{1 + (\tilde{r} \partial_{\tilde{r}} \tilde{\psi})^2}}\right).
    \end{aligned}
    \label{eq:curvAppend}
\end{equation}
On substitution of $\tilde{\theta} = \tilde{\beta} + \tilde{\psi}$, Eq.~\eqref{eq:curvAppend} yields
\begin{equation}
    \tilde{\mathcal{C}} = \frac{1}{\tilde{r} \sqrt{1 + (\tilde{r} \partial_{\tilde r} \tilde \psi)^2}} \left[\cot(\tilde{\beta} + \tilde{\psi}) - 2 \tilde r \partial_{\tilde r} \tilde \psi - \frac{\tilde r \left(\partial_{\tilde r} \tilde \psi + \tilde r \partial^2_{\tilde r} \tilde \psi\right)}{1 + (\tilde r \partial_{\tilde r} \tilde \psi)^2}\right],
\end{equation}
whose dimensionless form reads
\begin{equation}
    \mathcal{C} = \frac{1}{r \sqrt{1 + (\varepsilon r \partial_r \psi)^2}} \left[\cot(\beta + \varepsilon \psi) - 2 \varepsilon r \partial_r \psi - \frac{r \left(\varepsilon \partial_r \psi + \varepsilon r \partial^2_r \psi\right)}{1 + (\varepsilon r \partial_r \psi)^2}\right].
    \label{eq:A3}
\end{equation}
On substitution of $\psi \approx - h / r$ into Eq.~\eqref{eq:A3}, applying Taylor expansions, and retaining terms up to $O(\varepsilon)$, one obtains Eq.~\eqref{eq:curv}.

\section{Some results in the linear stability analysis} \label{ap:linear-results}
The results of the asymptotic expansions for $u$, $v$, and $p$ shown in Eq.~\eqref{eq:expansion} are presented here for reader's convenience ($v^{(2)}$ and $p^{(2)}$ are not listed as they are not used in the derivation):
\begin{equation} \label{eq:v0}
    v^{(0)} = - \frac{2 \alpha^{3} r^{2} \cos\beta}{3} - \frac{\alpha^{2} r^{2} \cos\beta}{2} \frac{\partial h}{\partial r} - \frac{3 \alpha^{2} r h \cos\beta}{2},
\end{equation}
\begin{equation} \label{eq:p0}
    p^{(0)} = \alpha r \sin\beta + h \sin\beta - 
    \left( \frac{\varepsilon}{Bo} \right)
    \frac{ \cos\beta}{ r \sin\beta},
\end{equation}
\begin{equation} \label{eq:u1}
    \begin{aligned}
    u^{(1)} = & - \frac{\Rey \alpha^{6} r^{5} \cos^{2}\beta}{180} - \frac{\Rey \alpha^{5} r^{4} h \cos^{2}\beta}{30} + \frac{\Rey \alpha^{4} r^{4} h \cos^{2}\beta}{24} \frac{\partial h}{\partial r} - \frac{\Rey \alpha^{4} r^{3} h^{2} \cos^{2}\beta}{24} \\
    & + \frac{\Rey \alpha^{3} r^{3} h^{2} \cos^{2}\beta}{6} \frac{\partial h}{\partial r} + \frac{\Rey \alpha^{3} r^{2} h^{3} \cos^{2}\beta}{18} - \frac{\Rey \alpha r h^{4} \cos^{2}\beta}{3} \frac{\partial h}{\partial r} - \frac{\Rey \alpha h^{5} \cos^{2}\beta}{5} \\
    & - \frac{\alpha^{3} r^{2} \cos^{2}\beta}{6 \sin\beta} + \frac{\alpha^{2} r^{2} \sin\beta}{2} \frac{\partial h}{\partial r} - \frac{\alpha^{2} r h \cos^{2}\beta}{2 \sin\beta} + \alpha r h \sin\beta \frac{\partial}{\partial r} h - \frac{\alpha h^{2} \cos^{2}\beta}{2 \sin\beta} \\
    & + \left(\frac{\varepsilon}{\Bon}\right) \left( \frac{\alpha^{2} \cos\beta}{2 \sin\beta} + \frac{\alpha h \cos\beta}{r \sin\beta}\right),
    \end{aligned}
\end{equation}
\begin{equation} \label{eq:v1}
    \begin{aligned}
    v^{(1)} = & \frac{\Rey \alpha^{7} r^{5} \cos^{2}\beta}{180} + \frac{\Rey \alpha^{6} r^{5} \cos^{2}\beta}{180} \frac{\partial h}{\partial r} + \frac{\Rey \alpha^{6} r^{4} h \cos^{2}\beta}{30} - \frac{\Rey \alpha^{5} r^{5} h \cos^{2}\beta}{120} \frac{\partial^{2} h}{\partial r^{2}} \\
    & - \frac{\Rey \alpha^{5} r^{5} \cos^{2}\beta}{120} \left(\frac{\partial h}{\partial r}\right)^{2} - \frac{\Rey \alpha^{5} r^{4} h \cos^{2}\beta}{30} \frac{\partial h}{\partial r} + \frac{\Rey \alpha^{5} r^{3} h^{2} \cos^{2}\beta}{24} \\
    & - \frac{Re \alpha^{4} r^{4} h^{2} \cos^{2}\beta}{24} \frac{\partial^{2} h}{\partial r^{2}} - \frac{\Rey \alpha^{4} r^{4} h \cos^{2}\beta}{12} \left(\frac{\partial h}{\partial r}\right)^{2} - \frac{\Rey \alpha^{4} r^{3} h^{2} \cos^{2}\beta}{4} \frac{\partial h}{\partial r} \\
    & - \frac{\Rey \alpha^{4} r^{2} h^{3} \cos^{2}\beta}{18} + \frac{\Rey \alpha^{2} r^{2} h^{4} \cos^{2}\beta}{6} \frac{\partial^{2} h}{\partial r^{2}} + \frac{2 \Rey \alpha^{2} r^{2} h^{3} \cos^{2}\beta}{3} \left(\frac{\partial h}{\partial r}\right)^{2} \\
    & + \Rey \alpha^{2} r h^{4} \cos^{2}\beta \frac{\partial h}{\partial r} + \frac{\Rey \alpha^{2} h^{5} \cos^{2}\beta}{5} + \frac{\alpha^{4} r^{2} \cos^{2}\beta}{3 \sin\beta} - \frac{\alpha^{3} r^{3} \sin\beta}{6} \frac{\partial^{2} h}{\partial r^{2}} \\
    & - \frac{2 \alpha^{3} r^{2} \sin\beta}{3} \frac{\partial h}{\partial r} + \frac{\alpha^{3} r^{2} \cos^{2}\beta}{3 \sin\beta} \frac{\partial h}{\partial r} + \frac{\alpha^{3} r h \cos^{2}\beta}{\sin\beta} - \frac{\alpha^{2} r^{2} h \sin\beta}{2} \frac{\partial^{2} h}{\partial r^{2}} \\
    & - \frac{\alpha^{2} r^{2} \sin\beta}{2} \left(\frac{\partial h}{\partial r}\right)^{2} - \frac{3 \alpha^{2} r h \sin\beta}{2} \frac{\partial h}{\partial r} + \frac{\alpha^{2} r h \cos^{2}\beta}{2 \sin\beta} \frac{\partial h}{\partial r} + \frac{\alpha^{2} h^{2} \cos^{2}\beta}{2 \sin\beta} \\
    & - \left(\frac{\varepsilon}{\Bon}\right) \left(\frac{\alpha^{3} \cos\beta}{3 \sin\beta} + \frac{\alpha^{2} \cos\beta}{2 \sin\beta} \frac{\partial h}{\partial r} + \frac{\alpha^{2} h \cos\beta}{2 r \sin\beta}\right),
    \end{aligned}
\end{equation}
\begin{equation} \label{eq:p1}
    \begin{aligned}
        p^{(1)} = & - \frac{\alpha^{2} r \cos\beta}{2} - \alpha r \cos\beta \frac{\partial}{\partial r} h - \alpha h \cos\beta + h \cos\beta \frac{\partial}{\partial r} h + \frac{h^{2} \cos\beta}{2 r} \\
        & - \left(\frac{\varepsilon}{\Bon}\right) \left(\frac{\partial^{2} h}{\partial r^{2}} + \frac{1}{r} \frac{\partial h}{\partial r} + \frac{h \cos^{2}\beta}{r^{2} \sin^{2}\beta}\right),
    \end{aligned}
\end{equation}
\begin{scriptsize}
\begin{equation} \label{eq:u2}
    \begin{aligned}
    u^{(2)} = & \frac{\Rey^{2} \alpha^{10} r^{8} \cos^{3}\beta}{10800} + \frac{\Rey^{2} \alpha^{9} r^{7} h \cos^{3}\beta}{1080} + \frac{\Rey^{2} \alpha^{8} r^{8} h \cos^{3}\beta}{4480} \frac{\partial^{2} h}{\partial r^{2}} + \frac{\Rey^{2} \alpha^{8} r^{8} \cos^{3}\beta}{4480} \left(\frac{\partial h}{\partial r}\right)^{2} \\
    & - \frac{\Rey^{2} \alpha^{8} r^{7} h \cos^{3}\beta}{1440} \frac{\partial h}{\partial r} + \frac{17 Re^{2} \alpha^{8} r^{6} h^{2} \cos^{3}\beta}{5760} + \frac{Re^{2} \alpha^{7} r^{7} h^{2} \cos^{3}\beta}{560} \frac{\partial^{2} h}{\partial r^{2}}  + \frac{\Rey^{2} \alpha^{7} r^{7} h \cos^{3}\beta}{560} \left(\frac{\partial h}{\partial r}\right)^{2} \\
    & - \frac{\Rey^{2} \alpha^{7} r^{6} h^{2} \cos^{3}\beta}{180} \frac{\partial h}{\partial r} + \frac{\Rey^{2} \alpha^{7} r^{5} h^{3} \cos^{3}\beta}{720} + \frac{\Rey^{2} \alpha^{6} r^{6} h^{3} \cos^{3}\beta}{180} \frac{\partial^{2} h}{\partial r^{2}} + \frac{7 \Rey^{2} \alpha^{6} r^{6} h^{2} \cos^{3}\beta}{720} \left(\frac{\partial h}{\partial r}\right)^{2} \\
    & - \frac{\Rey^{2} \alpha^{6} r^{5} h^{3} \cos^{3}\beta}{135} \frac{\partial h}{\partial r} - \frac{13 \Rey^{2} \alpha^{6} r^{4} h^{4} \cos^{3}\beta}{2160} + \frac{\Rey^{2} \alpha^{5} r^{5} h^{4} \cos^{3}\beta}{120} \frac{\partial^{2} h}{\partial r^{2}} + \frac{\Rey^{2} \alpha^{5} r^{5} h^{3} \cos^{3}\beta}{30} \left(\frac{\partial h}{\partial r}\right)^{2} \\
    & + \frac{\Rey^{2} \alpha^{5} r^{4} h^{4} \cos^{3}\beta}{30} \frac{\partial h}{\partial r} + \frac{\Rey^{2} \alpha^{5} r^{3} h^{5} \cos^{3}\beta}{150} - \frac{\Rey^{2} \alpha^{4} r^{4} h^{5} \cos^{3}\beta}{72} \frac{\partial^{2} h}{\partial r^{2}} - \frac{5 \Rey^{2} \alpha^{4} r^{4} h^{4} \cos^{3}\beta}{72} \left(\frac{\partial h}{\partial r}\right)^{2} \\
    & - \frac{\Rey^{2} \alpha^{4} r^{3} h^{5} \cos^{3}\beta}{45} \frac{\partial h}{\partial r} + \frac{\Rey^{2} \alpha^{4} r^{2} h^{6} \cos^{3}\beta}{40} - \frac{7 \Rey^{2} \alpha^{3} r^{3} h^{6} \cos^{3}\beta}{90} \frac{\partial^{2} h}{\partial r^{2}} - \frac{7 \Rey^{2} \alpha^{3} r^{3} h^{5} \cos^{3}\beta}{15} \left(\frac{\partial h}{\partial r}\right)^{2} \\
    & - \frac{56 \Rey^{2} \alpha^{3} r^{2} h^{6} \cos^{3}\beta}{135} \frac{\partial h}{\partial r} - \frac{\Rey^{2} \alpha^{3} r h^{7} \cos^{3}\beta}{27} + \frac{10 \Rey^{2} \alpha r h^{8} \cos^{3}\beta}{63} \frac{\partial^{2} h}{\partial r^{2}} + \frac{316 \Rey^{2} \alpha r h^{7} \cos^{3}\beta}{315} \left(\frac{\partial h}{\partial r}\right)^{2} \\
    & + \frac{44 \Rey^{2} \alpha h^{8} \cos^{3}\beta}{45} \frac{\partial h}{\partial r} + \frac{4 \Rey^{2} \alpha h^{9} \cos^{3}\beta}{27 r} + \frac{11 \Rey \alpha^{7} r^{5} \cos^{3}\beta}{1260 \sin\beta} + \frac{\Rey \alpha^{6} r^{6} \sin\beta \cos\beta}{360} \frac{\partial^{2} h}{\partial r^{2}} \\
    & - \frac{\Rey \alpha^{6} r^{5} \sin\beta \cos\beta}{90} \frac{\partial h}{\partial r} + \frac{\Rey \alpha^{6} r^{5} \cos^{3}\beta}{180 \sin\beta} \frac{\partial h}{\partial r} + \frac{\Rey \alpha^{6} r^{4} h \cos^{3}\beta}{18 \sin\beta} + \frac{\Rey \alpha^{5} r^{5} h \sin\beta \cos\beta}{60} \frac{\partial^{2} h}{\partial r^{2}} \\
    & - \frac{\Rey \alpha^{5} r^{4} h \sin\beta \cos\beta}{15} \frac{\partial h}{\partial r} - \frac{\Rey \alpha^{5} r^{4} h \cos^{3}\beta}{60 \sin\beta} \frac{\partial h}{\partial r} + \frac{\Rey \alpha^{5} r^{3} h^{2} \cos^{3}\beta}{10 \sin\beta} + \frac{\Rey \alpha^{4} r^{4} h^{2} \sin\beta \cos\beta}{12} \frac{\partial^{2} h}{\partial r^{2}} \\
    & + \frac{\Rey \alpha^{4} r^{4} h \sin\beta \cos\beta}{6} \left(\frac{\partial h}{\partial r}\right)^{2} - \frac{\Rey \alpha^{4} r^{3} h^{2} \sin\beta \cos\beta}{24} \frac{\partial h}{\partial r} - \frac{7 \Rey \alpha^{4} r^{3} h^{2} \cos^{3}\beta}{48 \sin\beta} \frac{\partial h}{\partial r} - \frac{\Rey \alpha^{4} r^{2} h^{3} \sin\beta \cos\beta}{72} \\
    & + \frac{5 \Rey \alpha^{4} r^{2} h^{3} \cos^{3}\beta}{144 \sin\beta} + \frac{2 \Rey \alpha^{3} r^{3} h^{3} \sin\beta \cos\beta}{9} \frac{\partial^{2} h}{\partial r^{2}} + \frac{2 \Rey \alpha^{3} r^{3} h^{2} \sin\beta \cos\beta}{3} \left(\frac{\partial h}{\partial r}\right)^{2} + \frac{5 \Rey \alpha^{3} r^{2} h^{3} \sin\beta \cos\beta}{18} \frac{\partial h}{\partial r} \\
    & - \frac{2 \Rey \alpha^{3} r^{2} h^{3} \cos^{3}\beta}{9 \sin\beta} \frac{\partial h}{\partial r} - \frac{\Rey \alpha^{3} r h^{4} \sin\beta \cos\beta}{18} + \frac{\Rey \alpha^{2} r h^{4} \cos^{3}\beta}{6 \sin\beta} \frac{\partial h}{\partial r} + \frac{\Rey \alpha^{2} h^{5} \cos^{3}\beta}{10 \sin\beta} \\
    & - \frac{2 \Rey \alpha r h^{5} \sin\beta \cos\beta}{5} \frac{\partial^{2} h}{\partial r^{2}} - \frac{4 \Rey \alpha r h^{4} \sin\beta \cos\beta}{3} \left(\frac{\partial h}{\partial r}\right)^{2} - \frac{11 \Rey \alpha h^{5} \sin\beta \cos\beta}{15} \frac{\partial h}{\partial r} + \frac{8 \Rey \alpha h^{5} \cos^{3}\beta}{15 \sin\beta} \frac{\partial h}{\partial r} \\
    & + \frac{\Rey \alpha h^{6} \sin\beta \cos\beta}{9 r} + \frac{\Rey \alpha h^{6} \cos^{3}\beta}{9 r \sin\beta} - \frac{7 \alpha^{4} r^{2} \cos\beta}{12} + \frac{\alpha^{4} r^{2} \cos^{3}\beta}{24 \sin^{2}\beta} + \frac{\alpha^{4} r^{2} \cos\beta}{12 \sin^{2}\beta} - \frac{\alpha^{3} r^{3} \cos\beta}{3} \frac{\partial^{2} h}{\partial r^{2}} - \frac{3 \alpha^{3} r^{2} \cos\beta}{2} \frac{\partial h}{\partial r} \\
    & - \alpha^{3} r h \cos\beta + \frac{\alpha^{3} r h \cos^{3}\beta}{6 \sin^{2}\beta} + \frac{\alpha^{3} r h \cos\beta}{6 \sin^{2}\beta} + \frac{\alpha^{2} r^{2} h \cos\beta}{2} \frac{\partial^{2} h}{\partial r^{2}} + \frac{\alpha^{2} r^{2} \cos\beta}{2} \left(\frac{\partial h}{\partial r}\right)^{2} - \frac{\alpha^{2} h^{2} \cos\beta}{4} + \frac{\alpha^{2} h^{2} \cos^{3}\beta}{4 \sin^{2}\beta} \\
    & + \frac{5 \alpha r h^{2} \cos\beta}{2} \frac{\partial^{2} h}{\partial r^{2}} + 5 \alpha r h \cos\beta \left(\frac{\partial h}{\partial r}\right)^{2} + \frac{7 \alpha h^{2} \cos\beta}{2} \frac{\partial h}{\partial r} - \frac{3 \alpha h^{3} \cos\beta}{2 r} + \frac{\alpha h^{3} \cos^{3}\beta}{6 r \sin^{2}\beta} - \frac{\alpha h^{3} \cos\beta}{6 r \sin^{2}\beta} \\
    & + \left(\frac{\varepsilon}{\Bon}\right) \left(- \frac{\Rey \alpha^{6} r^{3} \cos^{2}\beta}{60 \sin\beta} - \frac{\Rey \alpha^{5} r^{2} h \cos^{2}\beta}{10 \sin\beta} + \frac{\Rey \alpha^{4} r^{2} h \cos^{2}\beta}{12 \sin\beta} \frac{\partial h}{\partial r} - \frac{\Rey \alpha^{4} r h^{2} \cos^{2}\beta}{6 \sin\beta} + \frac{\Rey \alpha^{3} r h^{2} \cos^{2}\beta}{3 \sin\beta} \frac{\partial h}{\partial r} \right.\\
    &\left. - \frac{2 \Rey \alpha h^{4} \cos^{2}\beta}{3 r \sin\beta} \frac{\partial h}{\partial r} - \frac{4 \Rey \alpha h^{5} \cos^{2}\beta}{15 r^{2} \sin\beta} - \frac{\alpha^{3} \cos^{2}\beta}{6 \sin^{2}\beta} - \frac{\alpha^{2} r^{2}}{2} \frac{\partial^{3} h}{\partial r^{3}} - \frac{\alpha^{2} r}{2} \frac{\partial^{2} h}{\partial r^{2}} + \frac{\alpha^{2}}{2} \frac{\partial h}{\partial r} - \frac{\alpha^{2} \cos^{2}\beta}{2 \sin^{2}\beta} \frac{\partial h}{\partial r} \right.\\
    &\left. + \frac{\alpha^{2} h \cos^{2}\beta}{2 r \sin^{2}\beta} - \alpha r h \frac{\partial^{3} h}{\partial r^{3}} - \alpha h \frac{\partial^{2} h}{\partial r^{2}} + \frac{\alpha h}{r} \frac{\partial h}{\partial r} - \frac{\alpha h \cos^{2}\beta}{r \sin^{2}\beta} \frac{\partial h}{\partial r} + \frac{3 \alpha h^{2} \cos^{2}\beta}{2 r^{2} \sin^{2}\beta}\right).
    \end{aligned}
\end{equation}
\end{scriptsize}

The Benney-type equation, Eq.~\eqref{eq:be2}, can be written as
\begin{equation} \label{eq:be2-final}
	\frac{\partial h}{\partial t} = - \frac{1}{r \sin(\beta - \varepsilon h / r)} \frac{\partial}{\partial r} \left(q^{(0)} + \varepsilon q^{(1)} + \varepsilon^2 q^{(2)}\right),
\end{equation}
where
\begin{equation} \label{eq:be2-q0}
	q^{(0)} = \int_\psi^0 r^{2} u^{(0)} \sin\beta d\alpha = - \frac{1}{3} r h^{3} \sin\beta \cos\beta,
\end{equation}
\begin{equation} \label{eq:be2-q1}
	\begin{aligned}
	q^{(1)} = & \int_\psi^0 (\alpha r^{2} u^{(0)} \cos\beta + r^{2} u^{(1)} \sin\beta) d\alpha \\
	= & \frac{2 \Rey h^{6} \sin\beta \cos^{2}\beta}{15} \left(\frac{13 h}{21} + r \frac{\partial h}{\partial r}\right) - \frac{r h^{3} \sin^{2}\beta}{3} \frac{\partial h}{\partial r} + \frac{h^{4} \cos^{2}\beta}{3} - \left(\frac{\varepsilon}{\Bon}\right) \frac{h^{3} \cos\beta}{3 r},
	\end{aligned}
\end{equation}
and
\begin{equation} \label{eq:be2-q2}
	\begin{split}
			q^{(2)} = & \int_\psi^0 \left(- \frac{\alpha^{2} r^{2} u^{(0)} \sin\beta}{2} + \alpha r^{2} u^{(1)} \cos\beta + r^{2} u^{(2)} \sin\beta\right) d\alpha \\
			= & - \frac{4 \Rey^{2} r h^{10} \sin\beta \cos^{3}\beta}{63} \frac{\partial^2 h}{\partial r^2} - \frac{127 \Rey^{2} r h^{9} \sin\beta \cos^{3}\beta}{315} \left(\frac{\partial h}{\partial r}\right)^2 - \frac{13 h^{5} \cos^{3}\beta}{120 r \sin\beta} \\
			& - \frac{5608 \Rey^{2} h^{10} \sin\beta \cos^{3}\beta}{14175} \frac{\partial h}{\partial r} - \frac{9623 \Rey^{2} h^{11} \sin\beta \cos^{3}\beta}{155925 r} - \frac{7 \Rey h^{8} \cos^{3}\beta}{90 r} \\
			& + \frac{10 \Rey r h^{7} \sin^{2}\beta \cos\beta}{63} \frac{\partial^2 h}{\partial r^2} + \frac{8 \Rey r h^{6} \sin^{2}\beta \cos\beta}{15} \left(\frac{\partial h}{\partial r}\right)^2 + \frac{7 h^{5} \sin\beta \cos\beta}{8 r} \\
			& - \frac{67 \Rey h^{7} \cos^{3}\beta}{252} \frac{\partial h}{\partial r} - \frac{2 \Rey h^{8} \sin^{2}\beta \cos\beta}{45 r} - r h^{4} \sin\beta \cos\beta \frac{\partial^2 h}{\partial r^2} \\
			& - \frac{7 r h^{3} \sin\beta \cos\beta}{3} \left(\frac{\partial h}{\partial r}\right)^2 - \frac{7 h^{4} \sin\beta \cos\beta}{6} \frac{\partial h}{\partial r} + \frac{94 \Rey h^{7} \sin^{2}\beta \cos\beta}{315} \frac{\partial h}{\partial r} \\
			& + \frac{7 h^5 \cos\beta}{120 r \sin\beta} + \left(\frac{\varepsilon}{\Bon}\right) \left(\frac{4 \Rey h^{6} \cos^{2}\beta}{15 r} \frac{\partial h}{\partial r} + \frac{4 \Rey h^{7} \cos^{2}\beta}{35 r^{2}} + \frac{r h^{3} \sin\beta}{3} \frac{\partial^{3} h}{\partial r^{3}} \right.\\
			&\left. + \frac{h^{3} \sin\beta}{3} \frac{\partial^{2} h}{\partial r^{2}} - \frac{h^{3} \sin\beta}{3 r} \frac{\partial h}{\partial r} + \frac{h^{3} \cos^{2}\beta}{3 r \sin\beta} \frac{\partial h}{\partial r} - \frac{h^{4} \cos^{2}\beta}{3 r^{2} \sin\beta}\right).
		\end{split}
\end{equation}

The coefficients $S_0$-$S_4$ in the linearized equation, Eq.~\eqref{eq:hhat-linear}, are
\begin{equation} \label{eq:S0}
    \begin{split}
        S_0 = & - \frac{K^{2} q_{s}^{2/3} \cos\beta}{3 r^{5/3}} + \varepsilon \left[- \frac{148 \Rey q_{s}^{2}}{35 r^{4} \sin^{2}\beta}  + \frac{q_{s}}{3 r^{3} \cos\beta} + \frac{3 q_{s} \cos\beta}{r^{3} \sin^{2}\beta} + \left(\frac{\varepsilon}{\Bon}\right) \frac{5 K^{2} q_{s}^{2/3}}{9 r^{11/3} \tan\beta}\right] \\
        & + \varepsilon^2 \left[ - \frac{1486862 K \Rey^{2} q_{s}^{10/3}}{363825 r^{19/3} \sin^{3}\beta} + \frac{356 K \Rey q_{s}^{7/3}}{189 r^{16/3} \sin\beta \cos\beta} - \frac{1093 K \Rey q_{s}^{7/3} \cos\beta}{90 r^{16/3} \sin^{3}\beta} \right.\\
        &\left. + \frac{14 K q_{s}^{4/3} \sin\beta}{27 r^{13/3} \cos^{2}\beta} + \frac{4229 K q_{s}^{4/3}}{216 r^{13/3} \sin\beta} + \frac{1781 K q_{s}^{4/3} \cos^{2}\beta}{360 r^{13/3} \sin^{3}\beta} + \frac{49 K q_{s}^{4/3}}{40 r^{13/3} \sin^{3}\beta} \right.\\
        &\left. - \left(\frac{\varepsilon}{\Bon}\right) \left(\frac{128 \Rey q_{s}^{2}}{35 r^{6} \sin^{3}\beta} + \frac{34 q_{s}}{9 r^{5} \sin\beta \cos\beta} + \frac{15 q_{s} \cos\beta}{r^{5} \sin^{3}\beta}\right) - \left(\frac{\varepsilon}{\Bon}\right)^2 \frac{11 K^{2} q_{s}^{2/3} \cos\beta}{27 r^{17/3} \sin^{2}\beta}\right],
    \end{split}
\end{equation}
\begin{equation}
    \begin{split}
        S_1 = & - \frac{K^{2} q_{s}^{2/3} \cos\beta}{r^{2/3}} + \varepsilon \left[\frac{32 \Rey q_{s}^{2}}{35 r^{3} \sin^{2}\beta} - \frac{q_{s}}{3 r^{2} \cos\beta} + \frac{q_{s} \cos\beta}{r^{2} \sin^{2}\beta} - \left(\frac{\varepsilon}{\Bon}\right) \frac{K^{2} q_{s}^{2/3} \cos\beta}{3 r^{8/3} \sin\beta}\right] \\
        & + \varepsilon^2 \left[ - \frac{312482 K \Rey^{2} q_{s}^{10/3}}{40425 r^{16/3} \sin^{3}\beta} - \frac{136 K \Rey q_{s}^{7/3}}{315 r^{13/3} \sin\beta \cos\beta} + \frac{304 K \Rey q_{s}^{7/3} \cos\beta}{105 r^{13/3} \sin^{3}\beta} \right.\\
        &\left. - \frac{2 K q_{s}^{4/3} \sin\beta}{3 r^{10/3} \cos^{2}\beta} - \frac{451 K q_{s}^{4/3}}{72 r^{10/3} \sin\beta} + \frac{157 K q_{s}^{4/3} \cos^{2}\beta}{120 r^{10/3} \sin^{3}\beta} - \frac{21 K q_{s}^{4/3}}{40 r^{10/3} \sin^{3}\beta} \right.\\
        &\left. + \left(\frac{\varepsilon}{\Bon}\right) \left(\frac{32 \Rey q_{s}^{2}}{35 r^{5} \sin^{3}\beta} + \frac{34 q_{s}}{27 r^{4} \sin\beta \cos\beta} + \frac{19 q_{s} \cos\beta}{3 r^{4} \sin^{3}\beta}\right) + \left(\frac{\varepsilon}{\Bon}\right)^2 \frac{K^{2} q_{s}^{2/3} \cos\beta}{9 r^{14/3} \sin^{2}\beta}\right],
    \end{split}
\end{equation}
\begin{equation}
    \begin{split}
        S_2 = & \varepsilon \left(\frac{6 \Rey q_{s}^{2}}{5 r^{2} \sin^{2}\beta} + \frac{q_{s}}{r \cos\beta}\right) + \varepsilon^2 \left[ \frac{169 K \Rey q_{s}^{7/3} \cos\beta}{140 r^{10/3} \sin^{3}\beta} - \frac{K q_{s}^{4/3}}{6 r^{7/3} \sin\beta} + \frac{K q_{s}^{4/3} \sin\beta}{3 r^{7/3} \cos^{2}\beta} \right.\\
        &\left. - \frac{734 K \Rey^{2} q_{s}^{10/3}}{525 r^{13/3} \sin^{3}\beta} - \frac{206 K \Rey q_{s}^{7/3}}{105 r^{10/3} \sin\beta \cos\beta} + \left(\frac{\varepsilon}{\Bon}\right) \left(\frac{q_{s}}{r^{3} \sin\beta \cos\beta} - \frac{q_{s} \cos\beta}{r^{3} \sin^{3}\beta}\right)\right],
    \end{split}
\end{equation}
\begin{equation}
    S_3 = \varepsilon^2 \left[\frac{12 K \Rey^{2} q_{s}^{10/3}}{7 r^{10/3} \sin^{3}\beta} + \frac{10 K \Rey q_{s}^{7/3}}{7 r^{7/3} \sin\beta \cos\beta} + \frac{3 K q_{s}^{4/3}}{r^{4/3} \sin\beta} - \left(\frac{\varepsilon}{\Bon}\right) \frac{q_{s}}{r^{2} \sin\beta \cos\beta}\right],
\end{equation}
and
\begin{equation} \label{eq:S4}
	S_4 = - \varepsilon^{2} \left(\frac{\varepsilon}{\Bon}\right) \frac{q_s}{r \sin\beta \cos\beta}.
\end{equation}

In Eq.~\eqref{eq:sol-hypo}, the functions $f_i (r)$ ($i=0$, 1, and 2) can be expressed in the form
\begin{equation} \label{eq:f}
	f_i (r) = \Real (f_i) + \Imag (f_i)\, \mathrm{j},
\end{equation}
with
\begin{equation}
	\Real (f_0) = - \frac{1}{3} \ln r ,  \quad \Imag (f_0) = - \frac{K \omega r^{5/3} \sin\beta}{5 q_{s}^{2/3}},
\end{equation}
\begin{equation}
    \begin{split}
        \Real (f_1) = & \omega^{2} \left(- \frac{2 \Rey r}{15 \cos\beta} - \frac{r^{2} \tan^{2}\beta}{18 q_{s}}\right) \\
        & - \frac{4 K \Rey q_{s}^{4/3}}{7 r^{7/3} \sin\beta} + \frac{2 K q_{s}^{1/3}}{9 r^{4/3} \cot\beta} + \frac{2 K q_{s}^{1/3}}{3 r^{4/3} \tan\beta} - \left(\frac{\varepsilon}{\Bon}\right) \frac{1}{3 r^{2} \sin\beta},
    \end{split}
\end{equation}
\begin{equation}
    \Imag (f_1) = \omega \left[- \frac{16 K^{2} \Rey q_{s}^{2/3}}{105 r^{2/3}} - \frac{K^{2} r^{1/3} \sin^{2}\beta}{9 q_{s}^{1/3} \cos\beta} + \frac{K^{2} r^{1/3} \cos\beta}{3 q_{s}^{1/3}} - \left(\frac{\varepsilon}{\Bon}\right) \frac{K}{3 q_{s}^{2/3} r^{1/3}}\right],
\end{equation}
and
\begin{equation}
    \begin{split}
        \Real (f_2) = & - \left(\frac{\varepsilon}{\Bon}\right) \frac{K^{2} \omega^{4} r^{10/3} \sin\beta}{270 q_{s}^{7/3} \cot^{2}\beta} + \omega^{2} \left[- \frac{242 K \Rey^{2} q_{s}^{4/3}}{3150 r^{4/3} \sin(2 \beta)} - \frac{32 K \Rey \sqrt[3]{q_{s}} \sin\beta}{105 \sqrt[3]{r} \cos^{2}\beta} \right.\\
        &\left. + \frac{K \Rey \sqrt[3]{q_{s}}}{420 \sqrt[3]{r} \sin\beta} - \frac{5 K r^{2/3}}{54 q_{s}^{2/3} \cot^{3}\beta} - \frac{11 K r^{2/3}}{36 q_{s}^{2/3} \cot\beta} + \left(\frac{\varepsilon}{\Bon}\right) \left(- \frac{4 \Rey}{15 r \sin(2 \beta)} \right.\right.\\
        &\left.\left. + \frac{13 \ln r \sin\beta}{81 q_{s} \cos^{2}\beta} + \frac{\ln r}{9 q_{s} \sin\beta}\right)\right] - \frac{26248 K^{2} \Rey^{2} q_{s}^{8/3}}{40425 r^{14/3} \sin^{2}\beta} - \frac{152 K^{2} \Rey q_{s}^{5/3}}{315 r^{11/3} \cos\beta} \\
        & - \frac{169 K^{2} \Rey q_{s}^{5/3} \cos\beta}{210 r^{11/3} \sin^{2}\beta} + \frac{19 K^{2} q_{s}^{2/3}}{81 r^{8/3} \cot^{2}\beta} + \frac{571 K^{2} q_{s}^{2/3}}{216 r^{8/3}} + \frac{163 K^{2} q_{s}^{2/3}}{360 r^{8/3} \tan^{2}\beta} \\
        & + \frac{7 K^{2} q_{s}^{2/3}}{40 r^{8/3} \sin^{2}\beta} - \left(\frac{\varepsilon}{\Bon}\right) \left(\frac{8 K \Rey q_{s}^{4/3}}{21 r^{13/3} \sin^{2}\beta} + \frac{74 K \sqrt[3]{q_{s}}}{81 r^{10/3} \cos\beta} + \frac{16 K \sqrt[3]{q_{s}} \cos\beta}{9 r^{10/3} \sin^{2}\beta}\right) \\
        & +\left(\frac{\varepsilon}{\Bon}\right)^2 \frac{1}{6 r^{4} \sin^{2}\beta},
    \end{split}
\end{equation}
\begin{equation} \label{eq:mIm}
    \begin{split}
        \Imag (f_2) = & \omega^{3} \left[\frac{44 K^{2} \Rey^{2} q_{s}^{2/3} \sqrt[3]{r}}{525 \cos\beta} - \frac{K^{2} \Rey r^{4/3}}{210 \sqrt[3]{q_{s}} \cot^{2}\beta} - \frac{2 K^{2} r^{7/3} \sin^{4}\beta}{189 q_{s}^{4/3} \cos^{3}\beta} + \frac{K^{2} r^{7/3} \sin^{2}\beta}{21 q_{s}^{4/3} \cos\beta} \right.\\
        &\left.- \left(\frac{\varepsilon}{\Bon}\right) \frac{11 K r^{5/3} \tan^{2}\beta}{135 q_{s}^{5/3}}\right] + \omega \left[- \frac{8606 \Rey^{2} q_{s}^{2}}{24255 r^{3} \sin^{2}\beta \cos\beta} + \frac{152 \Rey q_{s}}{315 r^{2} \cos^{2}\beta} + \frac{4 \Rey q_{s}}{35 r^{2} \sin^{2}\beta} \right.\\
        &\left. - \frac{13 \sin^{2}\beta}{27 r \cos^{3}\beta} - \frac{145}{72 r \cos\beta} + \frac{13 \cos\beta}{40 r \sin^{2}\beta} - \frac{7}{40 r \sin^{2}\beta \cos\beta} \right.\\
        &\left. - \left(\frac{\varepsilon}{\Bon}\right) \left(\frac{71 K^{2} \Rey q_{s}^{2/3}}{630 r^{8/3} \sin\beta} + \frac{26 K^{2} \tan\beta}{81 \sqrt[3]{q_{s}} r^{5/3}} + \frac{17 K^{2} \cot\beta}{45 \sqrt[3]{q_{s}} r^{5/3}}\right) + \left(\frac{\varepsilon}{\Bon}\right)^2 \frac{2 K}{63 q_{s}^{2/3} r^{7/3} \sin\beta}\right].
    \end{split}
\end{equation}

\bibliographystyle{jfm}
%\bibliography{jfm}

\end{document}